\documentclass[a4paper,11pt]{article}
\pdfoutput=1
\usepackage[english]{babel}
\usepackage{jheppub}
\usepackage{subfigure}
\usepackage{xspace}
\usepackage{amsmath,amssymb,textcomp,enumerate,alltt,xspace,multirow}
\usepackage{slashed}
\usepackage{multicol}
\usepackage{boxedminipage}

\usepackage{graphicx}
\usepackage{relsize}
\usepackage{xcolor}

\newcommand{\QCD}{\mathrm{QCD}}
\newcommand{\EW}{\mathrm{EW}}

\newcommand{\NLO}{\mathrm{NLO}}

\newcommand{\GeV}{\text{GeV}\xspace}
\newcommand{\TeV}{\text{TeV}\xspace}

\newcommand{\QCDEW}{QCDxEW }

\newcommand{\fb}{\mathrm{fb}}

\newcommand{\LV}{\mathrm{LV}}
\newcommand{\LM}{\mathrm{LM}}
\newcommand{\fin}{\mathrm{fin}}

\newcommand{\litwo}{\rm Li_2}

\newcommand{\calC}{\mathcal{C}}

\newcommand{\calG}{\mathcal{G}}

\newcommand{\ord}{\mathcal{O}}
\newcommand{\calO}{\ord}

\renewcommand{\it}[1]{\textit{#1}}

\newcommand{\ie}{i.e.~}

\newcommand{\lb}{\left(}
\newcommand{\rb}{\right)}
\newcommand{\lsb}{\left[}
\newcommand{\rsb}{\right]}

\newcommand\Dphirec[1]{\Delta\phi_{\mathrm{rec},X}}

\newcommand{\OpenLoops}{{\rmfamily\scshape OpenLoops}\xspace}
\newcommand{\OLL}{\mbox{\OpenLoops\hspace{-0.5mm}2}\xspace}
\newcommand{\OLtwo}{\OLL}

\def\bit{\begin{itemize}}
\def\eit{\end{itemize}}

\def\be{\begin{equation}}
\def\ee{\end{equation}}

\def\bea{\begin{eqnarray}}

\def\eea{\end{eqnarray}}

\def\bce{\begin{center}}
\def\ece{\end{center}}

\def\refeq#1{\mbox{(\ref{#1})}}
\def\refeqs#1#2{\mbox{(\ref{#1})--(\ref{#2})}}
\def\reffi#1{\mbox{Fig.~\ref{#1}}}

\def\refta#1{\mbox{Table~\ref{#1}}}

\newcommand{\beq}{\begin{eqnarray}}
\newcommand{\eeq}{\end{eqnarray}}
\newcommand{\flm}{F_{\rm{LM}}}

\newcommand{\ep}{\epsilon}
\newcommand{\eps}{\epsilon}

\newcommand{\la}{\big\langle}
\newcommand{\ra}{\big\rangle}

\newcommand{\nn}{\nonumber}
\newcommand{\barq}{{\bar{q}}}

\def\plotboxrescale{2}
\def\plotboxrescaletwo{1.75}

\def\plots{figure}

\newcommand{\getmainP}[2]{
\includegraphics[scale=0.33,trim=0 24 0 0,clip]{\plots/#1/#2.pdf}
}

\newcommand{\getMratio}[2]{%
\includegraphics[scale=0.33,trim=0 24 0 15,clip]{\plots/#1/#2.pdf}
}

\newcommand{\getBratio}[2]{%
\includegraphics[scale=0.33,trim=0 0 0 15,clip]{\plots/#1/#2.pdf}
}

\newcommand{\makeplotA}[4]{
  \begin{minipage}{0.29\textwidth}
  \getmainP{#1}{#4}  \vspace*{-1mm} \\
  \getMratio{#2}{#4} \vspace*{-1mm} \\
  \getBratio{#3}{#4}
  \end{minipage}
}

\newcommand{\makeplotB}[4]{
  \begin{minipage}{0.25\textwidth}
  \getmainP{#1}{#4}  \vspace*{-6mm} \\
  \getMratio{#2}{#4} \vspace*{-6mm} \\
  \getBratio{#3}{#4}
  \end{minipage}
}

\newcommand{\plotobs}[1]{
\centering
\scalebox{\plotboxrescale}{\makeplotA{main}{ratio-1}{ratio-2}{#1}}
\vspace*{-1.ex}
}
\newcommand{\showobs}[1]{\plotobs{#1}}

\newcommand{\plottwoobs}[2]{
\begin{tabular}{cc}
\centering
\scalebox{\plotboxrescaletwo}{\makeplotB{main}{ratio-1}{ratio-2}{#1}}
\scalebox{\plotboxrescaletwo}{\makeplotB{main}{ratio-1}{ratio-2}{#2}}
\end{tabular}
\vspace*{-6.ex}
}
\newcommand{\showtwoobs}[2]{\plottwoobs{#1}{#2}}

\allowdisplaybreaks

\graphicspath{{figure/}}

\def\OXF{Rudolf Peierls Centre for Theoretical Physics, University of Oxford,
Clarendon Laboratory, Parks Road, Oxford OX1 3PU, UK}
\def\WDH{Wadham College, Oxford OX1 3PN, UK}
\def\MITP{PRISMA$^+$ Cluster of Excellence, Institut f{\"u}r Kernphysik, Johannes Gutenberg Universit{\"a}t, 55099 Mainz, Deutschland}
\def\MSU{Department of Physics and Astronomy, Michigan State University,
East Lansing, Michigan 48824, USA}
\def\KITA{Institute for Theoretical Particle Physics, KIT, Karlsruhe, Germany}
\def\KITB{Institut f{\"u}r Astroteilchenphysik, Karlsruher Institut f{\"u}r Technologie (KIT), D-76021 Karlsruhe, Germany}
\def\TIF{Tif Lab, Dipartimento di Fisica, Universit\'{a} di Milano and
INFN, Sezione di Milano, Via Celoria 16, I-20133 Milano, Italy}

\preprint{
\begin{flushright}
TIF-UNIMI-2022-5,
TTP22-015,
P3H-22-028\\
OUTP-22-02P, 
MSUHEP-22-012
\end{flushright}
}

\title{Mixed QCD-electroweak corrections to dilepton production at the
  LHC in the high invariant mass region}

\author[a]{Federico Buccioni,}
\author[a,b]{Fabrizio Caola,}
\author[a]{Herschel A. Chawdhry,}
\author[a]{Federica Devoto,}
\author[c]{Matthias Heller,}
\author[d]{Andreas von Manteuffel,}
\author[e]{Kirill Melnikov,}
\author[f]{Raoul R{\"o}ntsch,}
\author[e,g]{Chiara Signorile-Signorile}

\emailAdd{federico.buccioni@physics.ox.ac.uk}
\emailAdd{fabrizio.caola@physics.ox.ac.uk}
\emailAdd{herschel.chawdhry@physics.ox.ac.uk}
\emailAdd{federica.devoto@physics.ox.ac.uk}
\emailAdd{maheller@students.uni-mainz.de}
\emailAdd{vmante@msu.edu}
\emailAdd{kirill.melnikov@kit.edu}
\emailAdd{raoul.rontsch@unimi.it}
\emailAdd{chiara.signorile-signorile@kit.edu}

\affiliation[a]{\OXF}
\affiliation[b]{\WDH}
\affiliation[c]{\MITP}
\affiliation[d]{\MSU}
\affiliation[e]{\KITA}
\affiliation[f]{\TIF}
\affiliation[g]{\KITB}

\abstract{ We compute mixed QCD-electroweak corrections to the
  neutral-current Drell-Yan production of a pair of massless leptons
  in the high invariant mass region.  Our computation is fully
  differential with respect to the final state particles. At
  relatively low values of the dilepton invariant mass, $m_{\ell \ell}
  \sim 200$ GeV, we find unexpectedly large mixed QCD-electroweak
  corrections at the level of -1\%.  At higher invariant masses,
  $m_{\ell \ell} \sim 1$ TeV, we observe that these corrections can be
  well approximated by the product of QCD and electroweak
  corrections. Hence, thanks to the well-known Sudakov enhancement of
  the latter, they increase at large invariant mass and reach e.g.
  -3\% at $m_{\ell \ell} = 3$ TeV.
  Finally, we note that the inclusion of mixed corrections reduces the
  theoretical uncertainty related to the choice of electroweak input
  parameters to below the percent level.  }


\begin{document}
\maketitle

\flushbottom

\section{Introduction}

The production of lepton pairs in hadron collisions, commonly referred
to as the Drell-Yan (DY) process~\cite{Drell:1970wh}, continues to
play an important role in testing the Standard Model (SM) of particle
physics and searching for physics beyond it.  In particular, many
recent studies of the DY process \cite{CMS:2021ctt, ATLAS:2017fih}
have focused on the dilepton high invariant mass region, where
high-precision experimental results are becoming available.

Interest in the high invariant mass region stems from the fact that
many extensions of the SM contain weakly-coupled states which can
decay to lepton pairs.  Even if such states are too heavy to be
directly produced at the LHC, their presence can still be detected
through searches for shape distortions in kinematic distributions of
SM signatures.  Such a strategy was explored to improve on the mass
reach of direct searches for heavy neutral gauge bosons in
Ref.~\cite{Alioli:2017nzr}.
More generally, studies of dileptons with high invariant masses can be
used to constrain heavy New Physics in a model-independent way, using
the Standard Model Effective Field Theory
(SMEFT)~\cite{Buchmuller:1985jz, Grzadkowski:2010es}.  In particular,
the dilepton invariant mass distribution is affected by SMEFT
operators that also impact the so-called oblique
  parameters~\cite{Peskin:1990zt} constrained with a few per mille
precision using LEP data \cite{Falkowski:2015krw}.  Since studies of
the DY process in the high invariant-mass region are expected to reach
only percent-level precision at the LHC, it may seem surprising that
the LHC data could help to improve constraints on SMEFT
operators. However, since such contributions are generated by
dimension-6 operators, they grow quadratically with energy.
Effectively, the higher energy of the LHC compensates for the limited
precision, since the enhancement factor for $\sqrt{s} \simeq 1$~TeV is
around 150~\cite{Farina:2016rws, Dawson:2021ofa} when compared to
studies at $\sqrt{s} = m_Z$.
Investigations of dilepton pairs with high invariant mass may also help to
elucidate the physical origin of flavour anomalies
~\cite{LHCb:2017avl,LHCb:2019hip,LHCb:2021trn,Bifani:2018zmi}.
Indeed, by looking at the difference between dimuon and dielectron
production at high invariant masses, one can set appropriate bounds on
the corresponding models~\cite{Greljo:2017vvb}.

To achieve these goals, high-precision theoretical predictions within
the SM are needed; in fact, to constrain the Wilson coefficients of
SMEFT operators, percent precision is required.  Since the strong
coupling constant $\alpha_s$ is about $0.1$, QCD corrections have to
be accounted for through, at least, next-to-next-to leading order
(NNLO).  At this perturbative order, both inclusive and
fully-differential results are
available~\cite{Hamberg:1990np,vanNeerven:1991gh,Harlander:2002wh,Anastasiou:2003yy,Anastasiou:2003ds,Melnikov:2006kv,
  Catani:2009sm,Gavin:2010az}.  Very recently, N$^3$LO QCD corrections
to DY processes were
calculated~\cite{Duhr:2020seh,Duhr:2020sdp,Duhr:2021vwj,Chen:2021vtu,Camarda:2021ict,Chen:2022cgv}
and found to be close to a percent, motivating their inclusion at this
level of precision.
In addition to QCD corrections, electroweak (EW) contributions also
need to be accounted for to achieve percent-level precision.  The NLO
EW corrections were calculated long
ago~\cite{Dittmaier:2001ay,Baur:2001ze,Baur:2004ig,Arbuzov:2005dd,Zykunov:2005tc,CarloniCalame:2006zq,Zykunov:2006yb,CarloniCalame:2007cd,Arbuzov:2007db,Dittmaier:2009cr},
and found to be small, close to one percent for moderate values
of the dilepton invariant mass.  However, it was also found that EW
corrections are significantly enhanced at large invariant masses
$\sqrt{s} \gg m_Z$ and can reach tens of percent in this region
because of so-called electroweak Sudakov
logarithms~\cite{Kuhn:1999nn,Ciafaloni:2001vu,Denner:2000jv,Denner:2001gw}.

The enhancement of EW corrections at large dilepton invariant masses
and the fact that QCD corrections can be as large as twenty percent at
$\sqrt{s} \gg m_Z$ raise the question of the magnitude of mixed \QCDEW
corrections and make it plausible that these corrections can reach
$\calO(1\%)$ at high invariant masses. If so, they become relevant for
the many interesting phenomenological studies that were mentioned
earlier.  Although the impact of QCD and electroweak radiation has
been studied using parton
showers~\cite{Barze:2013fru,Frederix:2018nkq}, it is important to
obtain predictions for the exact mixed \QCDEW corrections to the DY
process, and their explicit computation is the goal of this paper.

We note that mixed \QCDEW corrections have already been studied for
\emph{resonant} production of $Z$ and $W$
bosons~\cite{Delto:2019ewv,Buccioni:2020cfi,Bonciani:2020tvf,Behring:2020cqi,Behring:2021adr,Bonciani:2021iis}
and were found to be small, close to one per mille.  Although one may
think that calculations of these corrections in the resonance 
and high invariant mass regions are technically similar, this is
actually not the case. Indeed, in the resonance region, all
contributions that connect initial and final states are suppressed by
the ratio of the vector boson width to its mass $\Gamma_V/m_V
\sim 10^{-2}$ and can be
neglected~\cite{Dittmaier:2014qza,Dittmaier:2015rxo}.  Hence, when
computing mixed \QCDEW corrections in such a case, it is
sufficient to only consider corrections to the
subprocesses $q \bar q' \to V$ and $V\to \ell_1 \bar\ell_2$.
However, in the high invariant mass region this is no longer the case
and corrections to the full $q\bar q'\to \ell_1\bar\ell_2$ process
need to be considered.

This leads to two significant complications with respect to the resonant
case.  First, one has to deal with the full $q \bar q' \to \ell_1 \bar
\ell_2$ two-loop amplitude and compute Feynman integrals that include
e.g. two-loop four-point functions with various internal and external
masses. Fortunately, the relevant integrals and helicity amplitudes
have been computed recently in
Refs.~\cite{Bonciani:2016ypc,vonManteuffel:2017myy,Heller:2019gkq,Heller:2020owb,Hasan:2020vwn,Armadillo:2022bgm}
and can be used to describe the mixed \QCDEW virtual corrections
in the high invariant mass region.
Second, computing fully differential second-order corrections to the
$q\bar q' \to \ell_1 \ell_2$ process requires properly extracting soft
and collinear singularities arising from real emission of partons off
the initial \emph{and} final state. 

In this paper, we develop the nested soft-collinear subtraction scheme
of Ref.~\cite{Caola:2017dug} to deal with infrared singularities
originating from QCD and EW emissions. In particular, we extend our
previous results~\cite{Buccioni:2020cfi,Behring:2020cqi} to cope with
parton radiation off both initial and final states. This, combined
with the availability of the two-loop
amplitudes~\cite{Heller:2020owb}, allows us to obtain mixed \QCDEW
corrections to neutral-current DY at high invariant mass in a robust
and efficient way. As a consequence, we are able to perform an
in-depth phenomenological study of high-mass dilepton production at
the LHC that accounts for both NNLO QCD and mixed \QCDEW corrections.

We note that an independent calculation of the mixed \QCDEW
corrections to the production of massive dileptons was performed
recently~\cite{Bonciani:2021zzf}.\footnote{A similar calculation for
lepton-neutrino production also exists~\cite{Buonocore:2021rxx}, albeit
with approximate virtual corrections.} In the high invariant-mass
region, Ref.~\cite{Bonciani:2021zzf} observed percent-level effects. A
direct comparison of our results with the ones in
Ref.~\cite{Bonciani:2021zzf} is not possible because this reference
performed studies in the so-called ``bare lepton'' setup (i.e. without
recombining leptons and photons). However, our analysis qualitatively
confirms these findings.

The rest of the paper is organised as follows. In Section~\ref{sec2}
we review the nested soft-collinear subtraction scheme and explain how
to apply it to the computation of mixed \QCDEW corrections to dilepton
production.  In Section~\ref{sec:VV} we provide a
brief summary of the relevant virtual amplitudes and
discuss the adaptation of the two-loop amplitudes of
Refs.~\cite{Heller:2019gkq,Heller:2020owb} for our numerical code.
Phenomenological results are reported in Section~\ref{sec:pheno}.
We conclude in Section~\ref{sec:conclusions}. Useful formulas are collected
in several appendices. 

\section{Subtraction scheme for mixed \QCDEW corrections}
\label{sec2}
The goal of this section is to review the theoretical framework that
we employ for calculating mixed \QCDEW corrections to the Drell-Yan
process.  We begin by describing the main obstacles in performing
perturbative computations at higher orders and discuss how these
obstacles manifest themselves when computing mixed \QCDEW corrections.

\subsection{General considerations}
\label{sec:generalities_subtra}
Higher-order computations in quantum field theory suffer from
ultraviolet and infrared divergences that need to be regularized and
extracted. While ultraviolet divergences are removed once measurable
quantities are used as input parameters in perturbative computations,
the situation with infrared divergences is more subtle.  Indeed,
although these are present both in virtual and real
corrections to physical observables, they manifest themselves in
different ways.
Virtual corrections to scattering amplitudes contain explicit 
poles in $\ep$ that arise once the integration over loop momenta is
performed.\footnote{For all computations employed in this paper, we
use dimensional regularization and work in $d = 4-2\eps$ space-time
dimensions.}  On the other hand, since real emission contributions
represent a distinct physical process, they are regular in the bulk of
the phase space, but develop singularities if one or several emitted
partons become soft or collinear to other partons in the process.
When integrated over energies and emission angles of soft and
collinear partons, these singularities turn into poles in $\ep$ which
cancel against similar poles in virtual corrections for \emph{infrared
safe} observables.

However, the integration over unresolved phase space of soft and collinear
partons has to be performed in a manner that preserves the
fully-differential nature of a particular calculation.  This can be
achieved in two different ways. One possibility is to restrict such
integrations to regions of phase space where unresolved partons are
either soft or collinear, ensuring that they do not affect the kinematic
features of hard observable partons.  This method is usually referred
to as \emph{slicing}.
Another option is to subtract suitably-defined expressions from the
full matrix element so that the difference is integrable throughout
the entire phase space. One has then to add back the subtracted terms
and ensure that they are observable-independent, so that they can be
integrated to produce explicit poles in $\ep$. This procedure defines  a
\emph{subtraction} scheme which can be used to perform
fully-differential computations at higher orders in perturbation
theory.  In recent years, both subtraction and slicing schemes have
been developed and used to compute NNLO QCD corrections to various
processes at the LHC and beyond, see e.g.
Refs.~\cite{Heinrich:2020ybq,TorresBobadilla:2020ekr} for a review.

In this paper we use the so-called \emph{nested soft-collinear
subtraction
scheme}~\cite{Caola:2017dug,Caola:2018pxp,Caola:2019nzf,Delto:2019asp} for
NNLO QCD calculations.
It is designed by exploiting two properties of scattering amplitudes.
The first one is their factorization in the soft and collinear limits
into a product of universal kernels and lower-multiplicity on-shell
amplitudes \cite{Altarelli:1977zs, Catani:1999ss,
  Bern:1999ry,Catani:2000pi, Kosower:1999rx}.  The second one is QCD
color coherence \cite{Ermolaev:1981cm, Bassetto:1983mvz,
  Dokshitzer:1987nm}, which implies that soft and collinear limits of
on-shell amplitudes are not entangled. One can use these features to
set up an iterative subtraction procedure that starts with the
subtraction of soft divergences. To regulate the remaining collinear
singularities, one introduces a partitioning of the phase space
to deal with the minimal number of collinear configurations at a time.
This allows one to subtract collinear divergences in a relatively
simple and modular way. This method was
developed for NLO QCD computations in Ref.~\cite{Frixione:1995ms} and
then extended to NNLO in Refs.~\cite{Czakon:2010td,Czakon:2011ve}.

The nested soft-collinear subtraction scheme can also be used to
compute mixed \QCDEW
corrections~\cite{Buccioni:2020cfi,Behring:2020cqi}. In fact, in this
case, significant simplifications can be expected since gluons and
photons do not interact with each other. As a result, NNLO soft limits
are described by a product of two NLO eikonal functions and no
singularities are present when a photon and a gluon become collinear to
each other.  However, triple-collinear limits remain complicated and
their integration over unresolved phase space is non-trivial.  The
integration of the triple-collinear subtraction terms for the QCD and mixed
\QCDEW cases was performed in Refs.~\cite{Delto:2019asp} and
\cite{Behring:2020cqi} respectively.

We note that the particular features of mixed \QCDEW corrections to DY
production which can be used to simplify the subtraction of infrared
divergences do not depend on whether the vector boson that decays into
a lepton pair is produced on the mass shell or not.  However,
computations in the latter case require more care because, since
radiation off initial and final states has to be considered
simultaneously, more singular limits need to be considered with
respect to the on-shell case.  Nevertheless, this complication does
not affect the overall structure of the subtraction and it can easily
be addressed by adapting singular kernels and phase space partitions
used in NNLO QCD computations.

Despite significant similarities between this calculation of mixed
\QCDEW corrections to DY production and the earlier ones with on-shell
vector bosons~\cite{Buccioni:2020cfi,Behring:2020cqi}, we describe the
subtraction of infrared singularities in detail in this paper, both to
make it self-contained and to highlight the differences with respect
to the on-shell case. We do this in the next two sections, starting
with the calculation of QCD and EW corrections at next-to-leading
order and continuing with the discussion of mixed \QCDEW corrections.

\subsection{Computation of EW and QCD corrections at next-to-leading order}
\label{sec:NLO}
We begin the discussion of NLO corrections by considering the real
emission process
\begin{equation}
f_1(p_1) + f_2(p_2)
\rightarrow \ell^-(p_3) +\ell^+(p_4) + f_5(p_5)\; ,
\label{eq:NLO_process}
\end{equation}
where the label $f_i = \{\gamma, g, q, \bar{q}\}$ specifies the parton
that participates in the hard scattering and $p_i$ is the four-momentum
of the parton $i$.  Following Ref.~\cite{Caola:2017dug} we define the
function
\begin{equation}
\flm(1_{f_1},2_{f_2}, 3, 4| 5_{f_5})
=\mathcal{N} 
\sum_{\rm col, \, pol}
\int {\rm dLips}_{34} \, 
(2\pi)^{d} \, 
\delta^{(d)}\Big(p_{12}-\sum_{j=3}^{5} p_j \Big) \, 
\big| {\cal M}(p_1 \dots p_5) \big|^2 \; ,
\label{eq2}
\end{equation}
where $p_{12}=p_1+p_2$, ${\cal M}$ is the matrix element of the
process in Eq.~\eqref{eq:NLO_process}, ${\rm dLips}_{34}$ is the
Lorentz-invariant phase space of the two leptons and ${\cal N}$ is a
quantity that includes spin- and color-averaging factors, if required.

The partonic cross section of the process Eq.~\eqref{eq:NLO_process} is
obtained by integrating Eq.~(\ref{eq2}) over the phase space of parton
$f_5$
\begin{equation}
2 s\cdot d\hat{\sigma}^{f_1 f_2}_{\rm r}
=  \int [dp_5] \, 
\flm(1_{f_1},2_{f_2}, 3, 4| 5_{f_5})
\equiv
\big\langle
\flm(1_{f_1},2_{f_2}, 3, 4| 5_{f_5})
\big\rangle \; ,
\label{eq:xsect_r_NLO}
\end{equation}
where $s=2 p_1 \cdot p_2$ is the partonic center-of-mass energy
squared. In the nested soft-collinear subtraction framework, the phase
space element $[dp]$ is assumed to include an upper bound on the
parton energy $E_{\rm max}$ \cite{Caola:2017dug}
\begin{equation}
[dp] = \frac{d^{d-1}p}{(2\pi)^{d-1} 2E_p} \, \theta \big( E_{\rm
  max}-E_p \big) \, .
\label{eq:PS_element}
\end{equation}
We note that any $E_{\rm max}$ can be chosen as long as it exceeds the
maximal energy that parton $f_5$ can reach in the process
Eq.~(\ref{eq:NLO_process}).  The reason for introducing $E_{\rm max}$
will become clear momentarily.

For the sake of concreteness, we will now focus on NLO electroweak
corrections to the $q \bar q$ production channel; then $f_1=q,
\,f_2=\bar{q} \,, f_5 = \gamma$. The matrix element in Eq.~(\ref{eq2})
develops singularities when the photon becomes either soft or
collinear to one of the four charged partons; we need to regulate
these singularities and extract them without integrating over
resolved parts of the photon's phase space.
To accomplish this, we follow Ref.~\cite{Caola:2017dug} and introduce
operators $S_5$ and $C_{5i}$, $i=1,2,3,4$, that extract leading
singularities of the function $\flm$ in the soft $p_5 \to 0$ and
collinear $\vec p_5 || \vec p_i$ limits, respectively.  These singular
limits can be written as products of universal functions and
lower-multiplicity matrix elements.  More specifically, we have
\begin{equation}
S_5 \, \flm(1_{q},2_{\bar{q}},3,4 |5_{\gamma})=
- 2 \, 
e^2 
\sum_{j>i  = 1}^{4}  \, \lambda_{ij} \; Q_iQ_j \; 
\frac{p_i\cdot p_j}{p_i \cdot p_5 \; p_j \cdot p_5} \, 
\flm(1_{q},2_{\bar{q}},3,4) \, ,
\label{eqa5}
\end{equation}
where $e = \sqrt{4 \pi \alpha} $ is the electric charge of the positron,
$Q_i$ is the physical electric charge of parton $i$ in units
of the positron charge, and $\lambda_{ij}$ is equal to $+1$ if $i,j$
are both incoming or outgoing and $-1$ otherwise. For our process,
$Q_1 = -Q_2 = Q_q$ and $Q_3=-Q_4=Q_{e^-} \equiv Q_e$.

To describe collinear limits, we need to distinguish between cases
where the photon is collinear to an incoming QCD parton or to an
outgoing lepton.  The corresponding formulas read
 \begin{equation}
C_{5i} \, F_{\rm LM} (1_{q},2_{\bar{q}},3,4|5_{\gamma}) = e^2 \, Q_i^2
\; \frac{ P_{qq} (z)}{p_i\cdot p_5} \cdot
\begin{cases}
  \frac{F_{\rm LM}(\dots z \cdot i \dots )}{z}, \;\;\;\;\;\;\;\;\;\;
z=\frac{E_i-E_5}{E_i},\; i \in\{1,2\}, \\ \;\;\flm\left( \dots
\frac{i}{z} \dots \right), \; z = \frac{E_i}{E_i+E_5}, \; i
\in\{3,4\},
\end{cases}
\label{eq2.6}
\end{equation}
where $P_{qq}(z)$ is the color-stripped quark splitting function
\begin{equation}
P_{qq}(z)=\frac{1+z^2}{1-z}-\epsilon(1-z),
\end{equation}
and the notation $z \cdot i$ in Eq.~(\ref{eq2.6}) implies that the
function $\flm$ has to be computed with the momentum of the parton $i$ set
to $z p_i$.

We can use these soft and collinear operators to construct expressions that
are finite in the corresponding limits. We start with the soft
operator $S_5$ and write
\begin{equation}
\la \flm(1_{q},2_{\bar{q}},3,4 | 5_{\gamma})\ra
=
\la S_5 \, 
\flm(1_{q},2_{\bar{q}},3,4|5_{\gamma})\ra
+\la [I-S_5]\, \flm(1_{q},2_{\bar{q}},3,4 |5_{\gamma})\ra \, ,
\label{eq:subtra_step1_NLO}
\end{equation}
where $I$ is the identity operator.  The two terms in
Eq.~(\ref{eq:subtra_step1_NLO}) have very different
properties. Indeed, according to Eq.~(\ref{eqa5}), in the first term
of Eq.~(\ref{eq:subtra_step1_NLO}) the four-momentum $p_5$ factorizes
from the function $\flm$. Hence, we can analytically integrate over
$p_5$ without affecting the kinematics of other particles. We
note that the integration over the photon energy becomes UV divergent
once the soft limit is taken; this potential divergence is regulated  by
$E_{\rm max}$.  The result of such integration is well-known (see
e.g.~\cite{Caola:2018pxp}) and can be written as follows
\begin{equation} 
\label{eq:soft_integ_count_NLO}
\begin{split}
& \la S_5 \, 
\flm(1_{q},2_{\bar{q}},3,4|5_{\gamma})\ra
 = \\ 
& -\frac{2 [\alpha]}{\eps^2} \big(2 E_{\rm max}\big)^{-2\eps} 
\sum_{j>i = 1}^4  
\lambda_{ij} \; Q_iQ_j  \, 
\la 
\eta_{ij}^{-\eps} \;  \mathcal{F}(\eta_{ij})  \; 
\flm(1_{q},2_{\bar{q}},3,4)\ra \, .
\end{split}
\end{equation}
In Eq.~\eqref{eq:soft_integ_count_NLO} we introduced
\be
\eta_{ij} = \frac{\rho_{ij}}{2}=\frac{1-\cos \theta_{ij}}{2},
\label{eqa91}
\ee
where $\theta_{ij}$ is the relative angle between the directions
of partons $i$ and $j$, 
and
\begin{equation}
\mathcal{F}(\eta) = \frac{\Gamma^2(1-\eps)}{\Gamma(1-2\eps)} \; 
\eta^{1+\eps} \;
{}_2F_1 (1, 1, 1-\eps; 1-\eta).
\label{eqa9}
\end{equation}
We have also introduced the coupling $[\alpha]$,  that
reads\footnote{A similar
definition is implied, \textit{mutatis mutandis}, for the
strong coupling $[\alpha_s]$ in the case of QCD corrections.}
\begin{equation} 
[\alpha] = \frac{e^2}{8\pi^2}\frac{(4\pi)^{\eps}}{\Gamma(1-\eps)} \; .
\end{equation}

The second term on the r.h.s. of Eq.~(\ref{eq:subtra_step1_NLO}) is
regular in the soft $p_5 \to 0$ limit but it still contains collinear
singularities that arise when the emitted photon is collinear to
quarks or leptons.  Since we would like to deal with one collinear
singularity at a time, we introduce
a partition of unity 
\begin{equation}
  1 = \omega^{51}+ \omega^{52}+ \omega^{53}+\omega^{54} \;,
  \label{eq:part1}
\end{equation}
where the partition functions $\omega^{5i}$ are designed to
have the following property 
\begin{equation}
C_{5i} \, \omega^{5j} = \delta_{ij}.
\label{eq2:12}
\end{equation}
This implies that the function $\omega^{5i}
[I-S_5]\flm(1_{q},2_{\bar{q}},3,4|5_{\gamma})$ is only singular in the
limit $\vec p_5 || \vec p_i$, while all other collinear singularities
are damped. Our choice of partition functions reads
\begin{equation}
\omega^{5i} = \frac{1/\rho_{5i}}{\sum_{j=1}^4 1/\rho_{5j}} \;,
\label{eq2:13}
\end{equation}
with $\rho_{ij}$ defined in Eq.~(\ref{eqa91}). It is straightforward
to check that with this choice Eqs.~(\ref{eq:part1}, \ref{eq2:12}) are
satisfied.
We can use Eqs.~(\ref{eq2:12}, \ref{eq2:13}) to extract collinear
singularities from the soft-regulated contribution in
Eq.~(\ref{eq:subtra_step1_NLO}). We arrive at
\begin{equation}
\label{eqa13}
\begin{split} 
&\la \flm(1_{q},2_{\bar{q}},3,4| 5_{\gamma})\ra = \la S_5 \,
  \flm(1_{q},2_{\bar{q}},3,4|5_{\gamma})\ra
  \\ &+\displaystyle\sum_{i=1}^4\la{[I-S_5]\, C_{5i} \,
    \flm(1_{q},2_{\bar{q}},3,4|5_{\gamma})}\ra + \, \la
  \mathcal{O}^\gamma_{\rm{nlo}} \,
  \flm(1_{q},2_{\bar{q}},3,4|5_{\gamma}) \ra \;,
\end{split} 
\end{equation}
 where
 \begin{equation}
 \la \mathcal{O}^\gamma_{\rm{nlo}} \,
 \flm(1_{q},2_{\bar{q}},3,4|5_{\gamma}) \ra = \la[I -S_5] \,
 \sum_{i=1}^4 [I- C_{5i}] \, \omega^{5i}
 \flm(1_{q},2_{\bar{q}},3,4|5_{\gamma}) \ra
 \label{eq:onloa}
\end{equation}
is fully regulated and can be numerically computed in four dimensions
with any infrared safe restriction on the phase space.

The only ingredients that we still require to compute the function
$\langle \flm(1_q,2_{\bar q},3,4|5_\gamma) \rangle $ in
Eq.~\eqref{eqa13} are the hard-collinear subtraction terms $ (I - S_5)
C_{5i} \flm$, with $i=1,..,4$.  They were calculated  in
Refs.~\cite{Caola:2017dug,Caola:2019pfz} and can be borrowed from
there.  The results read
\begin{equation}
\la{[1-S_5]\, C_{5i} \, \flm(1_{q},2_{\bar{q}},3,4|5_{\gamma})}\ra
 = \frac{[\alpha]\,Q_i^2}{\eps}  \,
\frac{\Gamma^2(1-\eps)}{\Gamma(1-2\eps)} \, 
 \,   
\big(2 E_i\big)^{-2\eps} \mathcal{H}\mathcal{C}_i(L_i) \; ,
\label{eq:hc_integ_count_NLO}
\end{equation} 
where $L_i =  \log  \left(E_{\rm max}/ E_i\right)$ and
\begin{equation} 
\mathcal{H}\mathcal{C}_i(L_i)
=
\begin{cases}
- \int \limits_0^1 dz \, 
\la
 P_{qq}^{\rm NLO}(z, L_i) \, 
F_{\rm LM}^{(i)}(1_{q},2_{\bar{q}},3,4; z) \ra \, ,
 \quad i \in\{1,2\}  \, ,
 \\[3pt]
 \la
P_{qq}^{\rm NLO}(L_i)\; 
\flm(1_{q},2_{\bar{q}},3,4)
\ra
\, ,
 \hspace{22mm} i\in\{3,4\}  \, .
 \end{cases}
\label{eq2.17}
\end{equation}
Following Ref.~\cite{Behring:2020cqi} we have used the notation
\begin{equation}
\label{eq:boosted_FLM}
F_{\rm LM}^{(i)}(1_{q},2_{\bar{q}},3,4; z)
= 
\begin{cases}
F_{\rm LM}(z \cdot 1_q, 2_\barq, 3, 4)/ z \; , \qquad i = 1 \; ,
 \\
F_{\rm LM}(1_q, z \cdot 2_\barq, 3, 4)/ z \; , \qquad i = 2 \; .
\end{cases}
\end{equation}
The splitting functions $P_{qq}^{\rm NLO}$ are related to the
Altarelli-Parisi splitting functions and their integrals. Their
explicit expressions can be found in Eq.~\eqref{eq:PqqNLO}. It is 
important to emphasize that the integration over $z$ in
Eq.~(\ref{eq2.17}) does not introduce additional $1/\ep$
singularities.

The explicit $1/\eps$ poles that appear in
Eqs.~(\ref{eq:soft_integ_count_NLO}, \ref{eq:hc_integ_count_NLO}) have
to cancel against similar poles in the one-loop EW corrections to
$q\bar{q} \rightarrow \ell^+ \ell^-$ and in contributions that
describe collinear renormalization of parton distribution functions
(PDFs).  Infrared divergences that appear in one-loop virtual
corrections can be written in a process-independent way, see
e.g.~\cite{Catani:1998bh}.  To do so, we introduce the function
$F_{\rm LV}$ that describes the contributions of virtual electroweak
corrections to the DY cross section and write
\begin{equation}
\begin{split} 
  &   2 s \cdot d\hat{\sigma}^{q\bar q}_{\rm v} =
 \la F_{\rm LV} (1_q,2_{\barq},3,4) \ra = \\
  & 
\mathcal{N} 
\sum_{\rm col, \, pol}
\int {\rm d Lips}_{34} \, 
(2\pi)^{d} \, 
\delta^{(d)}\Big(p_{12}-\sum_{j=3}^{4} p_j \Big) \,
2 {\rm Re} \Big[
\, {\cal M}^{\rm 1loop}(p_1\dots p_4)
 {\cal M}^\dagger(p_1 \dots p_4)
\Big] \, . 
\end{split} 
\end{equation}
This function can be written as a sum of divergent and finite terms
\begin{equation}
\la F_{\rm LV} (1_q,2_{\barq},3,4) \ra
=
[\alpha]\,I^{(1)}_{EW}  \, 
\langle F_{\rm LM} (1_q,2_{\barq},3,4)  \rangle
+\la F_{\rm LV}^{\rm fin} (1_q,2_{\barq},3,4)  \ra \, ,
\label{eqa19}
\end{equation}
with Catani's operator $I^{(1)}_{ EW}$ defined as follows \cite{Catani:1998bh}
\begin{equation}
I^{(1)}_{EW} = 
\Big( \frac1{\eps^2}+\frac32 \frac1\eps 
\Big) \; 
\sum_{j>i = 1}^4 2 \, 
 \tilde{\lambda}_{ij} \; Q_iQ_j \; 
\bigg(\frac{\mu^2}{s_{ij}}\bigg)^\eps,
\label{eq2.40a}
\end{equation}
where $s_{ij} = 2p_i\cdot p_j$ and $\tilde{\lambda}_{ij} = \cos(\pi
\eps)$ if $i,j$ are both incoming or outgoing and
$\tilde{\lambda}_{ij} = -1$ otherwise.  We note that the finite
remainder of the virtual corrections $\la F_{\rm LV}^{\rm fin}
(1_q,2_{\barq},3,4) \ra$ can only be obtained through a dedicated
computation; for the current discussion the only important point is
that it contains no divergences, either explicit or implicit.

Finally, we note that collinear singularities related to the photon
emission by incoming quarks are removed by re-defining parton
distribution functions.  The corresponding contribution to the cross
section in the $\overline{\rm MS}$ scheme reads (see
e.g. \cite{Behring:2020cqi})
\begin{equation}
2s \cdot d\hat{\sigma}^{q\bar q}_{\rm pdf} = [\alpha]\frac{Q_q^2 }{\eps}
\frac{\Gamma(1-\eps)}{ \, \mu^{2\ep} e^{\eps \gamma_E}} \; \sum
\limits_{i=1}^{2} \; \int \limits_{0}^{1}dz \, \bar{P}_{qq}^{\rm AP,
  0} (z) \; \la F_{\rm LM}^{(i)}(1_{q},2_{\bar{q}},3,4; z) \ra \; ,
\label{eqa20}
\end{equation}
where we used the fact that the absolute values of electric charges of
the incoming quark and anti-quark are equal.  We also note that
$\bar{P}_{qq}^{\rm AP, 0}$ in Eq.~(\ref{eqa20}) is the color-stripped
leading order Altarelli-Parisi splitting function; it reads
\begin{equation}
  \bar{P}_{qq}^{\rm AP, 0}(z)
  = 
  2 \mathcal{D}_0 (z) -(1+z) +\frac32 \delta(1-z) \, ,
  \qquad
  \mathcal{D}_0 (z) = \Big[\frac{1}{1-z}\Big]_+ \; .
  \label{eq:PqqAP0}
\end{equation}

To compute the NLO EW contribution to the partonic cross section $q
\bar q \to \ell^+ \ell^-$, we need to combine
Eqs.~(\ref{eqa13}, \ref{eqa19}, \ref{eqa20}) and expand the result up to
$\mathcal{O}(\eps^0)$. Working in the partonic center-of-mass frame
and choosing $E_{\rm max} = \sqrt{s}/2$, we obtain the following
result
\begin{equation}
  \label{eq:subtra_finite_NLO}
  \begin{split} 
    & 2s  \cdot d\hat{\sigma}_{\rm nlo,EW}^{q\bar{q}}
    = 
    \la \mathcal{O}_{\rm nlo}^{\gamma} \, 
    F_{\rm LM}(1_{q},2_{\bar{q}},3,4|5_{\gamma}) \ra
    +\la F_{\rm LV}^{\rm fin} (1_{q},2_{\bar{q}},3,4) \ra
    \\ 
    & 
    + 
    \frac{\alpha}{2 \pi}
    \bigg\{
    Q_q^2 \, \sum_{i=1}^2
    \int_{0}^{1}dz
    \Big[ \bar{P}_{qq}^{\rm AP, 0}(z) \log \Big( \frac{s}{\mu^2}\Big) + P^{'}_{qq}(z)
      \Big]  
    \la F_{\rm LM}^{(i)}(1_{q},2_{\bar{q}},3,4; z)
    \ra
    \\
    & \hspace{15mm}
    +
    \la \, \calG_{\EW} \,  
    F_{\rm LM}(1_{q},2_{\bar{q}},3,4) \ra
    \bigg\} \; ,
  \end{split}
\end{equation}
where we have defined
\begin{equation}
\label{eq:Gewk}
 \calG_{\EW} = Q_q^2 \, \frac{2 \pi^2}{3} + Q_{e}^2 \Big( 13 - \frac{2
   \pi^2}{3} \Big) + 2 \, Q_{q}Q_{e} \bigg[ 3 \log \Big(
   \frac{\eta_{13}}{\eta_{23}} \Big) + 2\, {\rm Li}_2(1-\eta_{13}) -
   2\, {\rm Li}_2(1-\eta_{23}) \bigg] \, ,
\end{equation}
and
\begin{equation}
  P^{'}_{qq}(z) = 4 \, \mathcal{D}_1(z) +(1-z) -2 \,(1+z)\, \log (1-z),
  ~~~~\mathcal D_n(z) = \left[\frac{\ln^n(1-z)}{1-z}\right]_+
\, .
\end{equation}
We note  that in the chosen reference frame, the momentum-conserving
delta function included in $F_{\rm LM}(1_{q},2_{\bar{q}},3,4)$ forces
$\eta_{13} = \eta_{24}$ and $\eta_{23} = \eta_{14}$; we have used this fact to simplify the
appearance of Eq.~(\ref{eq:Gewk}).

Before moving to the discussion of NNLO mixed \QCDEW corrections, we
note that NLO QCD corrections to the $q \bar q$ channel can easily be
obtained from the above formulas by replacing electric charges with
QCD charges, $\alpha\to\alpha_s$, $Q_e \to 0$ and $Q_q^2 \to C_F$, and
restricting the collinear subtractions in $\cal{O}_{\rm nlo}^{\gamma}$
to incoming partons only.  We also note that the computations
described above can easily be extended to other partonic channels and
for this reason we do not consider them here. Their discussion in a
similar case can be found in Ref.~\cite{Behring:2020cqi}.

\subsection{Computation of mixed \QCDEW corrections}
\label{sec:NNLOsubtraction}

We continue with mixed \QCDEW corrections and focus on the $q \bar q$
partonic channel. To obtain a finite partonic cross section in this
case, we need to combine the following contributions
\begin{equation}
\label{eq2.24}
\begin{split}
d\hat{\sigma}_{\rm mix}^{q \bar{q}}
 = 
d\hat{\sigma}_{\rm vv}^{q \bar{q}}
+d\hat{\sigma}_{{\rm rv},\gamma}^{q \bar{q}}
+d\hat{\sigma}_{{\rm rv},g}^{q \bar{q}}
+d\hat{\sigma}_{{\rm rr},g\gamma}^{q \bar{q}}
+d\hat{\sigma}_{{\rm rr},q\bar q}^{q \bar{q}}
+d\hat{\sigma}_{\rm pdf}^{q \bar{q}}  \; ,
\end{split} 
\end{equation}
where $d\hat{\sigma}_{\rm vv}$ is the double-virtual correction to the
elastic process $q \barq \rightarrow \ell^-\ell^+$,
$d\hat{\sigma}_{{\rm rv},\gamma}$ describes the one-loop QCD
correction to the process with an additional photon in the final
state, $d\hat{\sigma}_{{\rm rv},g}$ is the one-loop EW correction to the
process with an additional gluon, $d\hat{\sigma}_{{\rm rr},ij}$
represents the tree-level double-real emission of partons $i$ and $j$,
and $d\hat{\sigma}_{\rm pdf}$ describes the collinear renormalization of
parton distribution functions.

We note that the singularity structures
of the processes $q\barq\to\ell_1\ell_2+g\gamma$ and $q\barq\to\ell_1\ell_2
+q\bar q$ are very different. Indeed, the latter only contains
triple-collinear singularities, which are removed through PDF
renormalization.  Because of this, we find it convenient to treat
the $g\gamma$ and $q\bar q$ final states separately. Hence, we write 
\begin{equation}
  d\hat{\sigma}_{{\rm mix}}^{q \bar{q}} =
  d\hat{\sigma}_{{\rm mix},g\gamma}^{q \bar{q}} +
  d\hat{\sigma}_{{\rm mix},q\bar q}^{q \bar{q}}\,,
\end{equation}
with
\begin{equation}
  \begin{split}
    &d\hat{\sigma}_{{\rm mix},g\gamma}^{q \bar{q}}
    =
    d\hat{\sigma}_{\rm vv}^{q \bar{q}}
    +d\hat{\sigma}_{{\rm rv},\gamma}^{q \bar{q}}
    +d\hat{\sigma}_{{\rm rv},g}^{q \bar{q}}
    +d\hat{\sigma}_{{\rm rr}, g\gamma}^{q \bar{q}}
    +d\hat{\sigma}_{{\rm pdf},g\gamma}^{q \bar{q}}  \; ,
    \\
    &d\hat{\sigma}_{{\rm mix},q\bar q}^{q \bar{q}}
    =
    d\hat{\sigma}_{{\rm rr}, q\bar q}^{q \bar{q}}
    +d\hat{\sigma}_{{\rm pdf},q\bar q}^{q \bar{q}}  \; .
  \end{split}
  \label{eq:mix_qqb_split}
\end{equation}
In this section, we describe in detail the infrared regularization of
$d\hat{\sigma}_{{\rm mix},g\gamma}^{q \bar{q}}$. Results for the much
simpler contribution $d\hat{\sigma}_{{\rm mix},q\bar q}^{q \bar{q}}$
are reported in Appendix~\ref{sec:finite_parts_qq_qq}.

We begin with the analysis of the double-real emission cross section.
We write it as
\begin{align}
2s\cdot d\hat{\sigma}_{{\rm rr},g\gamma}^{q \bar{q}} 
\equiv 
\int[dp_5] [dp_6] F_{\rm LM}(1_q,2_{\bar{q}},3,4 | 5_g,6_\gamma) \equiv
\la F_{\rm LM}(1_q,2_{\bar{q}},3,4|5_g,6_\gamma)\ra. \; \; 
\label{eq:real_sigma_general}
\end{align}
The phase space elements for the gluon and the photon are defined in
Eq.~\eqref{eq:PS_element} and the meaning of the function $\flm$
should be clear from the discussion in the previous section.
In analogy to the NLO case, we first isolate soft singularities in
Eq.~(\ref{eq:real_sigma_general}).  Since in the case
of mixed \QCDEW corrections they factorize, we can write
\begin{equation}
\label{eq:NNLO_soft_reg}
\begin{split}
\la F_{\rm LM}(1_q,2_{\bar{q}},3,4|5_g,6_\gamma) \ra & = \la S_g \,
S_\gamma \, F_{\rm LM}(1_q,2_{\bar{q}},3,4|5_g,6_\gamma)\ra \\ & +
\la\big[ (I-S_g) \, S_\gamma + (I-S_\gamma) \, S_g \, \big] F_{\rm
  LM}(1_q,2_{\bar{q}},3,4|5_g,6_\gamma)\ra \\ & +\la (I-S_g) \,
(I-S_\gamma) \, F_{\rm LM}(1_q,2_{\bar{q}},3,4|5_g,6_\gamma)\ra \; .
 \end{split} 
\end{equation}
In Eq.~(\ref{eq:NNLO_soft_reg}), $S_g$ and  $S_\gamma$ are operators that
extract the leading soft behavior of the function $F_{\rm LM}$ in the
limits $E_5 \rightarrow 0$ and $E_6 \rightarrow 0$ respectively.  The
first term on the right hand side of Eq.~(\ref{eq:NNLO_soft_reg})
corresponds to the double-soft limit; it is equal to the product of
two NLO soft factors (cf.  Eq.~\eqref{eq:soft_integ_count_NLO})
\begin{equation}
\label{eq:double_soft_integ_count_NNLO}
\begin{split} 
\la S_g \, S_\gamma \, F_{\rm LM}(1_q,2_{\bar{q}},3,4|5_g,6_\gamma)\ra
&=
-4 \, 
\frac{[\alpha_s] [\alpha]}{\eps^4} \, 
C_F \, 
\big(2 E_{\rm max}\big)^{-4\eps} \,
\eta_{12}^{-\eps} \, \mathcal{F}(\eta_{12}) 
\\
&\times 
\sum_{j>i = 1}^4  
\lambda_{ij} \; Q_iQ_j \; 
\la
\eta_{ij}^{-\eps} \, \mathcal{F}(\eta_{ij})  \, 
\flm(1_{q},2_{\bar{q}},3,4)\ra.
\end{split}
\end{equation}

The two contributions in the second line of
Eq.~\eqref{eq:NNLO_soft_reg} correspond to kinematic configurations
where either a gluon or a photon is soft.  These terms still contain
single collinear singularities that need to be regulated. We follow
the discussion in the previous section and write
\begin{equation}
\label{eq:single_soft_term1_NNLO}
\begin{split} 
& \langle S_g \, ( I- S_\gamma )\, F_{\rm
    LM}(1_q,2_{\bar{q}},3,4|5_g,6_\gamma)\rangle = \frac{ 2 \, C_F
    [\alpha_{s}] \big(2 E_{\rm max}\big)^{-2\eps} }{\eps^2} \;
  \Big\langle \eta_{12}^{-\eps} \, \mathcal{F}(\eta_{12}) \\ & \times
  \, \bigg[\mathcal{O}^\gamma_{\rm nlo} \, F_{\rm
      LM}(1_q,2_\barq,3,4|6_\gamma) +\frac{[\alpha]}{\eps}\,
    \frac{\Gamma^2(1-\eps)}{\Gamma(1-2\eps)} \, \sum_{i=1}^4 Q_i^2 \;
    \big(2 E_{i}\big)^{-2\eps} \; \mathcal{H}\mathcal{C}_i(L_i) \bigg]
  \Big\rangle\, ,
\end{split}
\end{equation}
and 
\begin{equation}
\label{eq:single_soft_term2_NNLO}
\begin{split} 
&
\langle S_\gamma \, 
( I- S_g )\, F_{\rm LM}(1_q,2_\barq,3,4|5_g, 6_\gamma)
\rangle  =
\\
&
-2 \, [\alpha]  \, \frac{1}{\eps^2} 
\big(2 E_{\rm max}\big)^{-2\eps}
\Big\langle 
\sum_{j>i = 1}^4  
\lambda_{ij} \; Q_iQ_j \; 
\eta_{ij}^{-\eps} \, \mathcal{F}(\eta_{ij})  \, 
\,  
\bigg[
\mathcal{O}_{\rm nlo}^{g} \, 
 F_{\rm LM}(1_q,2_\barq,3,4|5_g)
 \\
&
-[\alpha_s]  \frac{C_F}\eps \, 
\frac{\Gamma^2(1-\eps)}{\Gamma(1-2\eps)} 
 \sum_{i=1}^2 
 \big(2 E_{i}\big)^{-2\eps}
 \int\limits_0^1
 dz \, P_{qq}^{\rm NLO}(z,L_i) \, 
 F_{\rm LM}^{(i)}(1_{q},2_{\bar{q}},3,4; z)
 \bigg]
\Big\rangle   \; ,
\end{split} 
 \end{equation}
where $\mathcal{O}_{\rm nlo}^{g} = (I-S_g)(I- C_{g1}-C_{g2})$.
Eqs.~(\ref{eq:single_soft_term1_NNLO},
\ref{eq:single_soft_term2_NNLO}) provide formulas for $\langle S_g \,
( I- S_\gamma )\, F_{\rm LM} \rangle$ and $\langle S_\gamma \, ( I-
S_g )\, F_{\rm LM} \rangle$ with all the $1/\ep$ singularities
extracted and no implicitly divergent contributions left.

We now focus on the term in the last line of
Eq.~\eqref{eq:NNLO_soft_reg} which is soft-regulated, but 
still contains multiple collinear singularities that need to be
isolated.  To do this, we partition the phase space in such a way that
for each partition only a subset of kinematic configurations becomes
singular. Using the $\eta_{ij}$ defined in Eq.~\eqref{eqa91}, we
construct the partition functions
\begin{equation} \omega^{\gamma i,gj} =
  \frac{\eta_{gj}^{-1}}{\sum\limits_{k=1}^2
    \eta^{-1}_{gk}}
\times
\frac{ \eta^{-1}_{\gamma i}}{\sum \limits_{k=1}^4
  \eta^{-1}_{\gamma k}} \; ,
 \label{eq:NNLO_sector}
 \end{equation}
 with $ i\in \{1,2, 3, 4\}  \, , \,  j \in \{1,2\}$.
They clearly add up to unity
\begin{equation}
1 =
 \sum_{i=1}^4\sum_{j=1}^2 \omega^{\gamma i,gj} \, .
 \label{eq:partition_NNLO_prel}
\end{equation}
The partition functions $\omega^{\gamma i, g j}$ are designed in
such a way that $\omega^{\gamma i, g j} \, F_{\rm LM}(1,2,3,4|5_g,
6_\gamma)$ is only singular when the photon becomes collinear to
parton $i$ and/or the gluon becomes collinear to parton $j$.
They also satisfy the further relations
\begin{equation}
 \label{eq:sector_NNLO_prop}
\begin{split} 
& C_{g\gamma,i} \, \omega^{\gamma i,gi} = C_{g\gamma,i} \; , \qquad
  \quad \,\;\;\;\;\; i \in \{1,2\} \; , \\ & C_{\gamma i} \, C_{g j}
  \, \omega^{\gamma i,g j} = C_{\gamma i} \, C_{gj} \; , \qquad \,
  i\in \{1,2, 3, 4\} \; , \qquad j\in \{1,2\} \; ,
 \end{split}
\end{equation}
where $C_{g\gamma,i}$ is the projection operator that describes the
triple-collinear limit $\vec p_g || \vec p_\gamma || \vec p_i$.

We note that triple-collinear configurations, which correspond to
the partition functions $\omega^{\gamma 1, g1}$ and $\omega^{\gamma 2,
  g2}$ in Eq.~\eqref{eq:partition_NNLO_prel}, contain 
overlapping collinear limits.  To disentangle them, we further split
these partitions into sectors~\cite{Behring:2020cqi}
\begin{equation}
\omega^{\gamma i,gi } 
= 
\omega^{\gamma i,g i} \, \big(\theta^{(i)}_A + \theta^{(i)}_B \big) 
\equiv
\omega^{\gamma i,g i} \, 
\Big[ 
\theta \big(\eta_{\gamma i} - \eta_{g i } \big)
+
\theta \big(\eta_{g i} - \eta_{\gamma i } \big)
\Big]
\; .
\end{equation}
We then write the soft-regulated term in Eq.~\eqref{eq:NNLO_soft_reg}
as follows
\begin{align}
\label{eq:soft_regulated_partition}
 &   \big\langle (I-S_g)  \, (I-S_\gamma)  \, F_{\rm LM}(1_q,2_{\bar{q}},3,4|5_g,6_\gamma)\big\rangle
    =
    \Big\langle (I-S_g)  \, (I-S_\gamma)  \,
    \Big[
    \omega^{\gamma1,g1} \, 
    \big(\theta^{(1)}_A + \theta^{(1)}_B\big)
    \nn
     \\
    & \qquad 
+ \,   \omega^{\gamma2,g2}  \, 
\big(\theta^{(2)}_A + \theta^{(2)}_B\big)
 +   \sum_{i=1}^4\sum_{\substack{j=1 \\ j\neq i}}^2   \omega^{\gamma i,gj}
    \Big]
    F_{\rm LM}(1_q,2_{\bar{q}},3,4|5_g,6_\gamma)\Big\rangle,
\end{align}
and note that each term that appears on  the r.h.s in Eq.~(\ref{eq:soft_regulated_partition})  is singular in one collinear
configuration only. 
To simplify the analytic computation of the corresponding limits, we
re-write Eq.~\eqref{eq:soft_regulated_partition} as follows
\begin{equation}
\begin{split}
&
\big\langle (I-S_g) (I-S_\gamma) F_{\rm LM}(1_q,2_{\bar{q}},3,4|5_g,6_\gamma)\big\rangle
=
 \\
& 
\sum_{i=1}^4
\big\langle (I-S_g) (I-S_\gamma) \, \Omega_i^{q\barq} \, F_{\rm LM}(1_q,2_{\bar{q}},3,4|5_g,6_\gamma)\big\rangle,
\end{split}
\end{equation}
where the four operators  $\Omega^{q \bar q}_i$ read 
\begin{equation}
\label{eq:Omega1}
\begin{split} 
\Omega^{\, q\bar{q}}_1 & = 
 (1-C_{g\gamma, 1})  (1-C_{g1})  \, \omega^{\gamma1,g1}  \, \theta_A^{(1)} 
+  (1-C_{g\gamma, 1})  (1-C_{\gamma1}) \, \omega^{\gamma1,g1} \, \theta_B^{(1)} 
 \\
& 
+ \,  (1-C_{g\gamma, 2})  (1-C_{g2})   \, \omega^{\gamma2,g2}  \, \theta_A^{(2)}  
+  (1-C_{g\gamma, 2})  (1-C_{\gamma2}) \, \omega^{\gamma2,g2} \, \theta_B^{(2)} 
 \\
&
+ \,  (1-C_{g2})  (1-C_{\gamma1})  \, \omega^{\gamma1,g2}
+   (1-C_{g1})  (1-C_{\gamma2}) \, \omega^{\gamma2,g1}
 \\
&
+ \,  (1-C_{g2})  (1-C_{\gamma3})  \, \omega^{\gamma3,g2}
+  (1-C_{g2})  (1-C_{\gamma4}) \, \omega^{\gamma4,g2} 
 \\
&
+ \, (1-C_{g1})  (1-C_{\gamma3})  \, \omega^{\gamma3,g1} 
+  (1-C_{g1})  (1-C_{\gamma4}) \,  \omega^{\gamma4,g1}, \\[10pt]
\Omega^{\, q\bar{q}}_2 & = 
 C_{g\gamma, 1} (1-C_{g1})  \, \omega^{\gamma1,g1}  \, \theta_A^{(1)}   
+  C_{g\gamma, 1} (1-C_{\gamma1})  \, \omega^{\gamma1,g1} \, \theta_B^{(1)} 
\\
&
+ \, C_{g\gamma, 2} (1-C_{g2})  \, \omega^{\gamma2,g2}  \, \theta_A^{(2)}  
+ C_{g\gamma, 2} (1-C_{\gamma2}) \, \omega^{\gamma2,g2} \, \theta_B^{(2)}  \; , \\[10pt]
\Omega^{\, q\bar{q}}_3 & = 
-  \, C_{g2} \, C_{\gamma1}  \, \omega^{\gamma1,g2}
-  C_{g1} \, C_{\gamma2}  \, \omega^{\gamma2,g1} 
-  C_{g2} \, C_{\gamma3} \, \omega^{\gamma3,g2} 
\\
&
- \, C_{g2}\, C_{\gamma4} \, \omega^{\gamma4,g2} 
- C_{g1} \, C_{\gamma3}  \, \omega^{\gamma3,g1} 
- C_{g1} \, C_{\gamma4} \,  \omega^{\gamma4,g1} \; , \\[10pt]
\Omega^{\, q\bar{q}}_4 & = 
C_{g1} \, 
\big[ \omega^{\gamma1,g1}  \, \theta_A^{(1)}  
+  \omega^{\gamma2,g1} 
+  \omega^{\gamma3,g1} 
+  \omega^{\gamma4,g1} \big]
 \\
&
+ \, C_{g2} \, 
\big[ \omega^{\gamma2,g2}  \, \theta_A^{(2)}  
+  \omega^{\gamma1,g2}
+  \omega^{\gamma3,g2}
+  \omega^{\gamma4,g2}
 \big]
 \\
& 
+ C_{\gamma1} \, 
\big[ \omega^{\gamma1,g1}  \, \theta_B^{(1)}  
+  \omega^{\gamma1,g2} \big]
+ C_{\gamma2} \, 
\big[ \omega^{\gamma2,g2}  \, \theta_B^{(2)}  +  \omega^{\gamma2,g1} \big] 
\\
&
+ C_{\gamma3} \, 
\big[ \omega^{\gamma3,g1} 
+  \omega^{\gamma3,g2} \big]
+ C_{\gamma4} \, 
\big[ \omega^{\gamma4,g2} 
  +  \omega^{\gamma4,g1} \big].
\end{split} 
\end{equation}
We now discuss the integrated subtraction terms for each of the four
operators separately.
The $\Omega_1^{q \bar q}$ contribution is fully regulated, i.e. all
the soft and collinear limits have been extracted. Hence,
\begin{equation}
\la (I-S_g) (I-S_\gamma)   \, 
 \Omega^{q\bar{q}}_1
 F_{\rm LM}(1_q,2_{\bar{q}},3,4|5_g,6_\gamma)\ra
 \label{eq2.40}
\end{equation}
can be numerically integrated in four space-time dimensions and does
not require further discussion.

The operator  $\Omega_{2}^{q \bar q}$ contains all triple-collinear
limits.  The corresponding integrated counterterm can be found in
Refs.~\cite{Delto:2019asp,Delto:2019ewv, Behring:2020cqi} and yields
\begin{equation}
\label{eq:triple_coll_integrated_NNLO}
\begin{split}
& \la (I-S_g) (I-S_\gamma)   \, 
 \Omega^{q\bar{q}}_2
F_{\rm LM}(1_q,2_{\bar{q}},3,4|5_g,6_\gamma)\ra 
=
 \\
& 
-2 [\alpha_s] [\alpha] \, Q_q^2 \,  C_F 
\sum_{i=1}^2 (2 E_i)^{-4\eps}
\int\limits_0^1 dz \, P_{qq}^{TC}(z) \; 
\la F_{\rm LM}^{(i)}(1_{q},2_{\bar{q}},3,4; z)
\ra \, ,
\end{split}
\end{equation}
where the function  $P_{qq}^{TC}(z) $ is defined as 
\begin{align}
P_{qq}^{TC}(z) 
&=
\frac1\eps
\bigg[
\frac32 (1-z)
+ z \log(z)
+ \frac{3+z^2}{4(1-z)} \, \log^2(z)
\bigg]
+ (1-z) \, \bigg[
\frac{11}2
-6\log(1-z)
\bigg]
\nn \\
&
-\frac{2\pi^2 \, z}{3}
-\frac z2 \, \log^2(z)
- \frac{19+9z^2}{12(1-z)} \, \log^3(z)
+ 4 z \, {\rm Li}_2(z)
\nn \\
&
- \log(z) \bigg[
z
+\frac{\pi^2 \, (5+3z^2)}{3(1-z)}
+\frac{2(1+z^2)}{1-z} \, {\rm Li}_2(z)
\bigg] 
\\
& + \frac{2(5+3z^2)}{1-z} \, \big( {\rm Li}_3(z)-\zeta_3\big)
+
\mathcal{O}(\eps) \, .
\nn
\end{align}

We continue with the discussion of double-collinear terms which
are contained in the operator $\Omega^{\, q\bar{q}}_3$.  There are two types of such
contributions that need to be considered separately: a contribution
where a photon and a gluon are emitted by two different initial-state
particles and a contribution where a gluon is emitted by one of the
initial-state quarks and a photon is emitted by one of the final-state
leptons.  In both cases collinear limits are described by leading
order splitting functions; the main difference between the two cases
is the kinematics of the underlying Born process.  We obtain
\begin{align}
\label{eq:double_coll_int_NNLO}
& 
\la (I-S_g) (I-S_\gamma)   \, 
\Omega^{\, q\bar{q}}_3 
F_{\rm LM}(1_q,2_{\bar{q}},3,4|5_g,6_\gamma)\ra
=
\frac{[\alpha_s] [\alpha] C_F}{\eps^2} \, 
\bigg\{ - 2 Q_q^2  \, 
\big(4E_{1} E_{2}\big)^{-2\eps}
\nn
 \\
&
\times \, 
\int\limits_0^1 dz_1 \, dz_2\, P_{qq}^{\rm NLO}(z_1,L_1)
\, P_{qq}^{\rm NLO}(z_2,L_2) \; 
\Big\langle
\frac{F_{\rm LM}(z_1 \cdot 1_q, z_2 \cdot 2_\barq \cdot 2,3,4)}{z_1 \, z_2}
\Big\rangle
\\
&
+ \, 
Q_e^2 
 \sum_{\substack{i = 1, 2 \\ j = 3,4 } }
 \Big\langle
  \big(2E_{j}\big)^{-2\eps}
 P_{qq}^{\rm NLO}(L_j) 
 \big(2E_{i}\big)^{-2\eps}
\int\limits_0^1 dz \,
P_{qq}^{\rm NLO}(z, L_i) \, 
 \, 
F_{\rm LM}^{(i)}(1_{q},2_{\bar{q}},3,4; z)
\Big\rangle
\bigg\} \; .
\nn
\end{align}

Finally we consider the operator $\Omega^{q\barq}_4$  which
contains all single-collinear singularities.  It is convenient to
split it into two terms
\begin{equation}
\Omega^{\, q\bar{q}}_4 \equiv
\Omega^{\, q\bar{q}, \, \rm IS}_4
+
\Omega^{\, q\bar{q}, \, \rm IS\times FS}_4\,,
\end{equation}
defined as follows 
\begin{equation}
\label{eq:Omega4_IS_FS}
\begin{split} 
\Omega^{\, q\bar{q}, \, \rm IS}_4
& =
C_{g1} \, 
\big[ \omega^{\gamma1,g1}  \, \theta_A^{(1)} 
+  \omega^{\gamma2,g1} 
+ \, C_{g2} \, 
\big[ 
\omega^{\gamma2,g2}  \, \theta_A^{(2)} 
+  \omega^{\gamma1,g2}
 \big]
 \\
& 
+ \, 
C_{\gamma1} \, 
\big[ \omega^{\gamma1,g1}  \, \theta_B^{(1)} 
+  \omega^{\gamma1,g2} \big]
+ 
C_{\gamma2} \, 
\big[ \omega^{\gamma2,g2}  \, \theta_B^{(2)} 
+  \omega^{\gamma2,g1} \big] 
 \; ,
\\
\Omega^{\, q\bar{q}, \, \rm IS\times FS}_4
& =
C_{g1} \, 
\big[
  \omega^{\gamma3,g1} 
+  \omega^{\gamma4,g1} \big]
+ \, C_{g2} \, 
\big[ 
  \omega^{\gamma3,g2}
+  \omega^{\gamma4,g2}
 \big]
 \\
&
+ \, C_{\gamma3} \, 
\big[ \omega^{\gamma3,g1} 
+  \omega^{\gamma3,g2} \big]
+ C_{\gamma4} \, 
\big[ \omega^{\gamma4,g2} 
+  \omega^{\gamma4,g1} \big] \; .
\end{split}
\end{equation}
The $\Omega^{\, q\bar{q}, \, \rm IS}_4$ operator  describes the
emission of collinear photons and gluons by the incoming quark and
anti-quark.  It is important that it contains partitions that only
allow for initial state singularities. For this reason this
contribution is closely related to similar contributions studied
earlier in the context of NNLO QCD computations.  The result can be
extracted from Refs.~\cite{Caola:2017dug, Caola:2019nzf}. After
obvious modifications that account for the fact that we deal with
mixed \QCDEW rather than NNLO QCD corrections, we find
\begin{align}
\label{eq:single_coll_int_term1_NNLO}
&
\la 
(I-S_g) (I-S_\gamma)   \, 
\Omega^{\, q\bar{q}, \, \rm IS}_4 \, 
F_{\rm LM}(1_q,2_{\bar{q}},3,4|5_g,6_\gamma)
\ra =
\nn \\
&
- \frac{1}\eps \, 
\sum_{i=1}^2
\Big\langle
\big(2 E_i\big)^{-2\eps}
\int\limits_0^1 dz \, P_{qq}^{\rm NLO}(z,L_i) \, 
\bigg\{
[\alpha_s] \, C_F \, \mathcal{O}_{\rm nlo}^\gamma \, 
\Big[
\Delta_{\gamma i} \, 
F_{\rm LM}^{(i)}(1_{q},2_{\bar{q}},3,4| 6_\gamma ; z)
\, 
\Big]
\nn \\
& \qquad \qquad \qquad 
+
[\alpha] \, Q_q^2 \,
\mathcal{O}_{\rm nlo}^g
\Big[
\Delta_{g i} \, 
F_{\rm LM}^{(i)}(1_{q},2_{\bar{q}},3,4| 5_g ; z)
\, 
\Big]
\bigg\}
\Big\rangle
\nn \\
& +
[\alpha_s] [\alpha] \, \frac{C_F \, Q_q^2}{\eps^2} \, 
 \frac{\Gamma(1-\eps) \, \Gamma(1-2\eps)}{\Gamma(1-3\eps)} \,
  \\
 &\quad  \qquad  \times \, 
 \sum_{i=1}^2
\Big\langle
 \big(2E_i\big)^{-4\eps}
\int\limits_0^1 dz \, 
\big[P_{qq}^{\rm NLO}
\otimes
P_{qq}^{\rm NLO}
\big] (z, E_i) \; 
F_{\rm LM}^{(i)}(1_{q},2_{\bar{q}},3,4; z)
\Big\rangle
\nn \\
& + 4 \, \frac{ [\alpha_s] [\alpha]  C_F \, Q_q^2}{\eps^2} \; 
\frac{\Gamma^2(1-\eps)}{\Gamma(1-2\eps)} \, 
\big(2 E_1\big)^{-2\eps}
\big(2 E_2\big)^{-2\eps} \; 
\nn  \\ 
& \quad  \qquad \times \, 
\Big\langle
\int\limits_0^1 dz_1 dz_2 \, P_{qq}^{\rm NLO}(z_1,L_1) \,
P_{qq}^{\rm NLO}(z_2,L_2)  \, 
\frac{F_{\rm LM}(z_1\cdot 1, z_2 \cdot 2, 3, 4)}{z_1 z_2}
\Big\rangle \; .
\nn
\end{align}
The convolution $\big[P_{qq}^{\rm NLO}
\otimes P_{qq}^{\rm NLO}
\big]$ that appears in Eq.~(\ref{eq:single_coll_int_term1_NNLO}) 
is defined as 
\begin{equation}
\big[P_{qq}^{\rm NLO}
\otimes
P_{qq}^{\rm NLO}
\big] (z, E_i)
=
\int \limits_{0}^{1}
dz_1  dz_2 \, 
z_1^{-2\eps} \, 
P_{qq}^{\rm NLO} (z_1, L_i) \, 
P_{qq}^{\rm NLO}(z_2, L_{iz_i}) \, 
\delta(z-z_1 z_2) \, , \; \;
\end{equation}
with
$L_{iz_i} = \log \left(E_{\rm max}/(z_i E_i)\right)$.
The quantities $\Delta_{\gamma(g) i}$ are
remnants of the  partition functions
and the phase-space measure in relevant collinear limits.
They read
\begin{equation}
\Delta_{\gamma i}
=
\tilde{\omega}^{\gamma i,gi}_{g \parallel i} \, 
\eta_{\gamma i}^{-\eps}
+
\tilde{\omega}^{\gamma j,gi}_{g \parallel i}
 \; , 
\qquad
\Delta_{g i}
=
\tilde{\omega}^{\gamma i,gi}_{\gamma \parallel i} \, \eta_{g i}^{-\eps}
+
\tilde{\omega}^{\gamma i,gj}_{\gamma \parallel i} \; , 
\end{equation}
where we have introduced
\begin{equation}
\tilde{\omega}^{\gamma i,g j}_{g \parallel j} \equiv
C_{g j} \, \omega^{\gamma i,g j},
\qquad
\tilde{\omega}^{\gamma i,g j}_{\gamma \parallel i} \equiv
C_{\gamma i} \, \omega^{\gamma i,g j}.
\end{equation}

The operator  $\Omega^{\, q\bar{q}, \, \rm IS\times FS}_4$ contains
partition functions that only allow for initial-final state
singularities.  They can be computed following the steps discussed in the
context of NLO computations in Section~\ref{sec:NLO}.  We find
\begin{align}
\label{eq:single_coll_int_term2_NNLO}
&
\la (I-S_g) (I-S_\gamma)   
\Omega^{\, q\bar{q}, \, \rm IS\times FS}_4  F_{\rm LM}(1_q,2_{\bar{q}},3,4|5_g,6_\gamma) \ra  
=
\nn\\
& -[\alpha_s] \frac{C_F}\eps 
\sum_{i=1}^2 \Big\langle \mathcal{O}_{\rm nlo}^\gamma \, 
\Big(
\tilde{\omega}^{\gamma3,gi}_{g \parallel i}
+
\tilde{\omega}^{\gamma4,gi}_{g \parallel i} 
\Big) \, 
\big(2 E_i\big)^{-2\eps} 
\int\limits_0^1 dz \, P_{qq}^{\rm NLO}(z,L_i) \, 
F_{\rm LM}^{(i)}(1_{q},2_{\bar{q}},3,4| 6_\gamma ; z)
\Big\rangle
\nn\\
&
+
[\alpha] \frac{Q_e^2}\eps \, 
\sum_{i=3}^4
\Big\langle  \mathcal{O}_{\rm nlo}^g \, 
\Big(
\tilde{\omega}^{\gamma i,g1}_{\gamma \parallel i}
+
\tilde{\omega}^{\gamma i,g2}_{\gamma \parallel i}
\Big) \, 
\big(2 E_i\big)^{-2\eps}
P_{qq}^{\rm NLO}(L_i) \, 
F_{\rm LM}(1_q, 2_\barq, 3, 4| 5_g)
\Big\rangle
 \\
 & 
- 2 \, \frac{ [\alpha_s] [\alpha]  C_F \, Q_e^2}{\eps^2}\; 
\frac{\Gamma^2(1-\eps)}{\Gamma(1-2\eps)}  \, 
\nn \\
& \qquad  \times
\sum_{\substack{i=1, 2 \\ j=3,4}}
\Big\langle
\big(2 E_i\big)^{-2\eps}
\big(2 E_j\big)^{-2\eps}
P_{qq}^{\rm NLO} (L_j) 
\int\limits_0^1 dz \, P_{qq}^{\rm NLO}(z,L_i) \, 
F_{\rm LM}^{(i)}(1_{q},2_{\bar{q}},3,4; z)
\Big\rangle
\; .
\nn
\end{align}
To compute the double-real emission contribution to the partonic cross
section $d\hat{\sigma}_{{\rm rr},g\gamma}^{q \bar{q}}$ we add
Eqs.~(\ref{eq:double_soft_integ_count_NNLO}-\ref{eq:single_soft_term2_NNLO},
\ref{eq2.40}, \ref{eq:triple_coll_integrated_NNLO},
\ref{eq:double_coll_int_NNLO}, \ref{eq:single_coll_int_term1_NNLO},
\ref{eq:single_coll_int_term2_NNLO}) and expand in $\epsilon$.  It is
straightforward to do this since there are no \emph{implicit}
singularities left. We do not show such
a result here since it is not very illuminating.

We now proceed with the calculation of real-virtual contributions to
mixed \QCDEW corrections.  As we have mentioned earlier, these
contributions are generated in two different ways, either as QCD
corrections to the process $q\bar{q} \rightarrow \ell^- \ell^+ +
\gamma$ or as electroweak corrections to the process $q\bar{q}
\rightarrow \ell^- \ell^+ + g$.  We write
\begin{equation}
2s\cdot \sum \limits_{f=g,\gamma}
d\hat{\sigma}_{{\rm rv},f}^{q \bar{q}}
=
\la F_{\rm LRV}^{\rm (EW)} (1_q,2_{\barq},3,4| 5_g) \ra
 +
  \la F_{\rm LRV}^{\rm (QCD)} (1_q,2_{\barq},3,4| 5_\gamma) \ra , 
\end{equation}
where the superscript on the r.h.s. specifies whether the loop
correction involves a gluon or an electroweak boson.  Since gluons and
photons do not interact with each other, soft limits of loop
corrections are trivial. Collinear limits can be dealt with by adapting
analogous QCD results~\cite{Caola:2017dug}. At the end, we find
\begin{align}
   \label{eq:rvint}
  & 2s\cdot\sum \limits_{f=g,\gamma}
d\hat{\sigma}_{{\rm rv},f}^{q \bar{q}}
= 
2 C_F
\frac{[\alpha_s]}{\eps^2} \big(2 E_{\rm max}\big)^{-2\eps}
\big\langle
\eta_{12}^{-\eps} \, \mathcal{F}(\eta_{12})  \,
F_{\rm LV}^{({\rm EW})}(1_q,2_\barq,3,4) 
\big\rangle
\nn\\
& 
-2 \frac{[\alpha]}{\eps^2} \big(2 E_{\rm max}\big)^{-2\eps} 
\sum_{j>i = 1}^4   
\lambda_{ij} \; Q_iQ_j \; 
\Big\langle \eta_{ij}^{-\eps} \, \mathcal{F}(\eta_{ij}) \, 
F_{\rm LV}^{({\rm QCD})}(1_q,2_\barq,3,4)
\Big\rangle
\nn \\
&
-
C_F \, 
\frac{[\alpha_s]}{\epsilon} \,
\frac{\Gamma^2(1-\eps)}{\Gamma(1-2\eps)}
\sum_{i=1}^2 \;(2 E_i)^{-2\eps}  \int \limits_{0}^{1} dz \, P_{qq}^{\rm NLO}(z, L_i) \, 
\, \Big\langle 
F_{\rm LV}^{(i), ({\rm EW}) }(1_{q},2_{\bar{q}},3,4; z) \Big\rangle
\nn \\
& 
+ 2
 C_F \, Q_q^2 \, 
\frac{[\alpha_s][\alpha]}{\eps} \, 
\frac{\Gamma^4(1-\eps) \, \Gamma(1+\eps)}{\Gamma(1-3\eps)} \, 
\nn \\
& \qquad \times \, 
\sum_{i=1}^2 (2 E_i)^{-4\eps}  \, 
 \int \limits_{0}^{1} dz \, 
\mathcal{P}_{qq}^{{\rm loop}, \, RV}(z)  \, 
\la
F_{\rm LM}^{(i)}(1_{q},2_{\bar{q}},3,4; z)
\ra
 \\
&
-
\frac{[\alpha]}{\epsilon} \, Q_q^2  \,
\frac{\Gamma^2(1-\eps)}{\Gamma(1-2\eps)} \sum_{i=1}^2  (2 E_i)^{-2\eps} 
\int \limits_{0}^{1} dz \, 
P_{qq}^{\rm NLO}(z, L_i) 
\Big\langle 
F_{\rm LV}^{(i), ({\rm QCD}) }(1_{q},2_{\bar{q}},3,4; z)
\Big\rangle
\nn\\
&
+
\frac{[\alpha]}{\eps}\, Q^2_{e} \, 
  \frac{\Gamma^2(1-\eps)}{\Gamma(1-2\eps)}
\Big\langle\sum_{i=3}^4(2E_i)^{-2\ep}
 \, P_{qq}^{\rm NLO}(L_i)
F_{\rm LV}^{({\rm QCD})}(1_q,2_\barq, 3,4)
\Big\rangle
\nn \\
 & 
 +
\langle \mathcal{O}_{\rm nlo}^g  \, F_{\rm LRV}^{({\rm EW})}(1_q,2_\barq,3,4 | 5_g) \rangle
+
 \langle \mathcal{O}_{\rm nlo}^\gamma  \, F_{\rm LRV}^{({\rm QCD}) }(1_q,2_\barq,3,4 | 5_\gamma) \rangle \; ,
 \nn
\end{align}
where $\mathcal{P}_{qq}^{\mathrm{loop},\,RV}(z)$ is defined in
Eq.~\eqref{eq:Pqq_RV}. In Eq.~\eqref{eq:rvint}, we have used the
following parametrization for the explicit infrared $1/\ep$ poles that are
present in both QCD and electroweak virtual amplitudes
\begin{equation}
  \label{eq2.55}
\la F_{\rm LRV}^{({\rm X})}(1_{q},2_{\bar q},3,4 | 5_i) \ra =
    [\alpha_X] \la I^{(1)}_{X}F_{\rm LM}
    (1_q,2_\barq,3,4|5_i) \ra + \la F_{\rm LV}^{\rm (X), \, fin}
    (1_q,2_\barq,3,4|5_i) \ra\,,
\end{equation}
where $\{X,\alpha_X,i\} = \{{\rm EW},\alpha,g\}$ for EW and
$\{X,\alpha_X,i\}=\{{\rm QCD},\alpha_s,\gamma\}$ for QCD corrections.
In
Eq.~(\ref{eq2.55}), $F_{\rm LV}^{\rm (X), \, fin}$ are one-loop
finite remainders, $I^{(1)}_{EW}$ is the electroweak Catani's operator
given in Eq.~(\ref{eq2.40a}) and $I_{ QCD}^{(1)}$ is the QCD one,
defined as~\cite{Catani:1998bh}
\begin{equation}
I_{ QCD}^{(1)}
=
-2 C_F\; 
 \bigg( \frac1{\eps^{2}}+\frac{3}{2\ep}\bigg)
 \cos(\pi \eps)\; \Big(\frac{\mu^2}{s_{12}}\Big)^{\eps}.
 \end{equation}

Next, we consider the double-virtual mixed QCD-electroweak
corrections.  Their infrared singularities can be derived by
abelianizing the NNLO QCD case in Ref.~\cite{Catani:1998bh}.  We find
\begin{align}
  \label{eq:twoloopfinrem}
  2 s \cdot d\hat{\sigma}_{\rm vv}^{q \bar{q}} &= [\alpha_s][\alpha]
  \bigg[ I^{(1)}_{QCD}\cdot I^{(1)}_{EW} +2 C_F \, Q_q^2 \,
    \frac{e^{\eps \, \gamma_E}}{\Gamma(1-\eps)} \frac1\eps \, \Big(
    \frac{\pi^2}{2} - 6 \zeta_3 -\frac38 \Big) \bigg] \la F_{\rm
    LM}(1_q,2_\barq,3,4) \ra
  \nn \\
  & + [\alpha_s] I^{(1)}_{QCD} \,
  \la F_{\rm LV}^{\rm (EW), \, fin} (1_q,2_\barq,3,4) \ra + [\alpha]
  I^{(1)}_{EW} \, \la F_{\rm LV}^{\rm (QCD), \, fin} (1_q,2_\barq,3,4)
  \ra
   \\
  & + \la F^{\rm (QCD\times EW), \, fin}_{\rm LVV+LV^2}
   (1_q,2_\barq,3,4) \ra \, .
   \nn
\end{align}
The quantity $F^{\rm (QCD\times EW), \, fin}_{\rm LVV+LV^2}$
represents the finite remainder of two- and one-loop virtual
corrections to the process $q \bar q \to \ell^- \ell^+$.  It was
recently calculated  in Ref.~\cite{Heller:2020owb}, and we briefly
discuss its computation  in the next section.

The last ingredient that we require to obtain a finite partonic cross
section comes from the renormalization of parton distribution
functions. It can be obtained from the results reported in
Ref.~\cite{Behring:2021adr}.  We find
\begin{equation}
  \label{eq:pdfren}
\begin{split}
2 s \cdot  d\hat{\sigma}_{{\rm pdf},g\gamma}^{q \bar{q}}
= \, &
-
2 [\alpha][\alpha_s]\frac{\Gamma^2(1-\ep)}{\mu^{4\ep}e^{2\ep\gamma_E}}
\bigg\{
\frac{Q_q^2\, C_F}{2\eps^2} \; 
\\
& \qquad \times \, 
\sum_{i=1}^2
\int \limits_{0}^{1} dz \, 
\big[
\bar{P}_{qq}^{\rm AP,0} 
\otimes 
\bar{P}_{qq}^{\rm AP,0}
\big] (z) \, 
\la F_{\rm LM}^{(i)}(1_{q},2_{\bar{q}},3,4; z) \ra
\\
&
+ \frac{Q_q^2 \, C_F}{\eps^2}
\int \limits_{0}^{1} dz_1 \, dz_2  \, 
\bar{P}_{qq}^{\rm AP,0}(z_1) \, 
\bar{P}_{qq}^{\rm AP,0}(z_2) \, 
\Big\langle
\frac{F_{\rm LM}(z_1\cdot 1, z_2\cdot 2, 3, 4)}{z_1 \, z_2}
\Big\rangle
\bigg\}
\\
&
+\frac{2s}\eps\cdot \, 
\bigg\{
C_F \,
[\alpha_s]\frac{\Gamma(1-\ep)}{\mu^{2\ep}e^{\ep\gamma_E}}
\Big[
\bar{P}_{qq}^{\rm AP,0} 
\otimes 
d\hat\sigma_{\rm nlo,EW}^{q\bar q}
+
d\hat\sigma_{\rm nlo,EW}^{q\bar q}
\otimes 
\bar{P}_{qq}^{\rm AP,0} 
\Big]
\\
& \qquad \; \; 
+
Q_q^2 \,
[\alpha]\frac{\Gamma(1-\ep)}{\mu^{2\ep}e^{\ep\gamma_E}}
\Big[
\bar{P}_{qq}^{\rm AP,0} 
\otimes
d\hat\sigma_{\rm nlo,QCD}^{q\bar q}
+
d\hat\sigma_{\rm nlo,QCD}^{q\bar q}
\otimes 
\bar{P}_{qq}^{\rm AP,0} 
\Big]
\bigg\}
\\
&
+
[\alpha][\alpha_s]\frac{\Gamma^2(1-\ep)}{\mu^{4\ep}e^{2\ep\gamma_E}}\,
\frac{Q_q^2\, C_F}{\eps}
\sum_{i=1}^2
\int \limits_{0}^{1} dz \, 
\bar{P}_{qq}^{\rm AP,1}(z)\, 
\la F_{\rm LM}^{(i)}(1_{q},2_{\bar{q}},3,4; z) \ra \; ,
\end{split}
\end{equation}
where the NLO corrections $d\hat\sigma^{q\bar q}_{\rm nlo,EW/QCD}$
have been discussed in the previous section.  The explicit expressions
for the various Altarelli-Parisi splitting functions and their
convolutions appearing in Eq.~\eqref{eq:pdfren} can be found in 
Appendix~\ref{sec:splittings}.

\subsection{Analytic results for mixed \QCDEW corrections in the $q\barq$ channel}
\label{subsec:analytic_qqbar}
Following the discussion in the previous section, we obtain a
manifestly finite expression for the partonic cross section
$d\hat\sigma_{{\rm mix},g\gamma}^{q\bar q}$ defined in
Eq.~\eqref{eq:mix_qqb_split}. We find it convenient to write it as a
combination of four terms that describe processes with different
multiplicities of resolved final-state particles and/or distinct
kinematic configurations.  We write
\begin{equation}
d\hat{\sigma}_{{\rm mix},g\gamma}^{q \bar{q}}
=
d\hat{\sigma}_{{\rm el},g\gamma}^{q \bar{q}}
+
d\hat{\sigma}_{{\rm bt},g\gamma}^{q \bar{q}}
+
d\hat{\sigma}_{{\mathcal{O}_{\rm nlo}},g\gamma}^{q \bar{q}}
+
d\hat{\sigma}_{{\rm reg},g\gamma}^{q \bar{q}}\,.
\label{eq2.59}
\end{equation}
For the sake of simplicity, we present results
in the center-of-mass frame of the colliding
partons and choose $E_{\max}=E_1=E_2\equiv E_c = \sqrt{s}/2$. 

The last term in Eq.~(\ref{eq2.59}) corresponds to the fully-regulated
contribution
\begin{equation}
2s \cdot d\hat{\sigma}_{{\rm reg},g\gamma}^{q \bar{q}} = \la
(I-S_g) (I-S_\gamma) \, \Omega^{\, q\bar{q}}_1 \, F_{\rm
  LM}(1_q,2_\barq,3,4|5_g,6_\gamma)\ra \, ,
\end{equation}
with $\Omega^{\, q\bar{q}}_1$ defined in Eq.~\eqref{eq:Omega1}.
It can be computed numerically in four dimensions without further ado. 

The \emph{elastic} cross section $d\hat{\sigma}_{{\rm el},g\gamma}^{q
  \bar{q}} $ contains all
contributions with Born kinematics. It reads 
\begin{align}
\label{eq:elastic_xsect_ag}
2s \cdot d\hat{\sigma}_{{\rm el}, g\gamma}^{q \bar{q}}
& = 
\la F^{\rm (QCD\times EW), \, fin}_{\rm LVV+LV^2} (1_q,2_\barq,3,4)\ra
\nn \\ 
  & +
[\alpha] \, 
\Big\langle
\Big[ \calG_{\EW} + 3 Q_q^2\, \log\Big(\frac{s}{\mu^2}\Big)\Big] \,  
F_{\rm LV}^{\rm (QCD), \, fin} (1_q,2_\barq,3,4)\Big\rangle
\nn \\
 & +
 [\alpha_s] \, C_F \, \Big[ \frac23 \pi^2+ 3 \log\Big( \frac{s}{\mu^2}\Big)
 \Big]
  \la F_{\rm LV}^{\rm (EW), \, fin} (1_q,2_\barq,3,4)\ra
   \\
  & 
  +
  [\alpha] \, [\alpha_s] \, 
  C_F \, \Big\langle
  \bigg\{
  Q_q^2  \,  
  \Big[
  -\frac{4 \pi^4}{45} + \lb 2\pi^2 + 32\zeta_3\rb\log\Big(\frac{s}{\mu^2}\Big)
\nn  \\
  & +
  \Big( 9-\frac{4\pi^2}{3}\Big) \log^2\Big(\frac{s}{\mu^2}\Big) \Big]
  +
  \calG_{\EW}
  \Big( \frac{2\pi^2 }3+ 3\log\Big(\frac{s}{\mu^2}\Big) 
  \Big) 
  \bigg\}
  F_{\rm LM} (1_q,2_\barq,3,4)\Big\rangle  \; ,
  \nn
\end{align}
where the function $\calG_{\EW}$ already appeared at NLO and was
defined in Eq.~\eqref{eq:Gewk}.

The boosted contribution reads 
\begin{align}
\label{eq:boostqqb}
2s & \cdot d \sigma_{{\rm bt}, g\gamma}^{q\bar{q}}
 = 
[\alpha] \, [\alpha_s] \, 
2 C_F \, Q_q^2 \,
\nn \\
& \qquad \times 
\int_0^1 dz_1 dz_2 \, 
\tilde{P}_{qq}^{\rm NLO} (z_1, E_c) \, 
\Big\langle \frac{F_{\rm LM} (z_1 \cdot 1, z_2\cdot 2,3,4)}{z_1 \, z_2} \Big\rangle \, 
 \tilde{P}_{qq}^{\rm NLO} (z_2, E_c)
\nn \\
 & 
 +
 \sum_{i =1}^2
\int_0^1 dz \, \tilde{P}_{qq}^{\rm NLO} (z, E_c)
\bigg[
[\alpha] \, 
Q_q^2 \, 
\Big\langle F_{\rm LV}^{(i), \rm (QCD), \, fin}(1_{q},2_{\bar{q}},3,4; z) \Big\rangle
\nn \\
& \hspace{30mm}
+
[\alpha_s] \, 
C_F\, 
\Big\langle F_{\rm LV}^{(i), \rm (EW), \, fin}(1_{q},2_{\bar{q}},3,4; z) \Big\rangle
 \bigg]
\nn \\
 &
 +
 [\alpha] \, 
Q_q^2 \, 
\sum_{i =1}^2 \, 
\int_0^1 dz \, 
 \Big\langle
  \mathcal{O}_{\rm nlo}^g \,  
  \Big[
  \tilde{P}_{qq}^{\rm NLO} (z, E_c) 
\nn   \\
  & \hspace{30mm}
  + \tilde{\omega}_{\gamma \parallel i}^{\gamma i, g i} \, 
  \log \eta_{i5} \; \bar{P}_{qq, R}^{\rm AP, 0}(z)
  \Big]
 F_{\rm LM}^{(i)}(1_{q},2_{\bar{q}},3,4| 5_g; z) 
 \Big\rangle
\nn \\
 &
 +
 [\alpha_s] \, 
C_F \, 
\sum_{i =1}^2 \, 
\int_0^1 dz \, 
 \Big\langle
  \mathcal{O}_{\rm nlo}^\gamma \,  
  \Big[
  \tilde{P}_{qq}^{\rm NLO} (z, E_c) 
\\ 
    & \hspace{30mm}
  + \tilde{\omega}_{g\parallel i}^{\gamma i, g i} \, 
  \log \eta_{i5} \; \bar{P}_{qq, R}^{\rm AP, 0}(z)
  \Big]
 F_{\rm LM}^{(i)}(1_{q},2_{\bar{q}},3,4| 5_\gamma ; z) \Big\rangle
\nn \\
 & 
 +
 [\alpha] \, [\alpha_s] \, 
 C_F \,
 \sum_{i =1}^2 \, 
\int_0^1 dz \, 
\Big\langle
\bigg\{
Q_q^2 \, P_{qq}^{\rm NNLO}(z, E_c)
+  \tilde{P}_{qq}^{\rm NLO} (z, E_c)
\nn \\
& \times \Big[
 Q_e^2 \, G_{e^2} 
 + \, 2 Q_qQ_e \, \Big(
G^{(1,2)}_{eq} + (-1)^{i} \log \Big( \frac{s_{i3}}{s_{i4}}\Big) \log(z)
\Big) \Big]
\bigg\} \, 
F_{\rm LM}^{(i)}(1_{q},2_{\bar{q}},3,4; z)
\Big\rangle \; .
\nn
\end{align}

The $\mathcal{O}_{\rm nlo}$ term reads
\begin{align}
\label{eq:ONLOqqb}
2s & \cdot d \sigma_{\mathcal{O}_{\rm nlo},g\gamma}^{q\bar{q}}
 = 
\big\langle
\mathcal{O}_{\rm nlo}^g \, 
 F_{\rm LV}^{\rm (EW), \, fin} (1_q, 2_\barq, 3, 4| 5_g)
 \big\rangle
 \nn \\
 & +
\big\langle
 \mathcal{O}_{\rm nlo}^\gamma \, 
 F_{\rm LV}^{\rm (QCD), \, fin} (1_q, 2_\barq, 3, 4| 5_\gamma)
 \big\rangle
 \\
 & 
 +  [\alpha] \,  \Big\langle
 \mathcal{O}_{\rm nlo}^g \, 
 \Big[
Q_q^2 \, 
 \bigg(
 \frac23 \pi^2
 + 3 \log\Big( \frac{s}{\mu^2}\Big) \bigg) 
 +  2 Q_q Q_e \, G^{(1,2)}_{eq} +  Q_e^2\, G_{e^2}
\Big] 
 F_{\rm LM} (1_q, 2_\barq, 3, 4| 5_g)
\Big \rangle  
\nn\\
 & 
 +  [\alpha_s] \, C_F \, 
 \Big[\,
 \frac23 \pi^2
 + 3 \log\Big( \frac{s}{\mu^2}\Big) \Big] \, 
 \la
 \mathcal{O}_{\rm nlo}^\gamma \, 
 F_{\rm LM} (1_q, 2_\barq, 3, 4| 5_\gamma) \ra
 \; .
 \nn
\end{align}
Here we have introduced
\begin{equation}
\label{eq:PqqNLOtilde}
\begin{split}
&
\tilde{P}_{qq}^{\rm NLO} (z, E) = 4 \mathcal{D}_1(z) -2 (1+z) \log(1-z) +(1-z) + \bar{P}_{qq, R}^{\rm AP, 0}(z) \, \log\Big(\frac{4E^2}{\mu^2} \Big) \, ,
\\
&
\bar{P}_{qq, R}^{\rm AP, 0}(z) = 2 \mathcal{D}_0(z)-(1+z)
 \, , 
\end{split}
\end{equation}
while $P_{qq}^{\rm NNLO}(z, E_c)$ is defined in
Eq. \eqref{pqqnnlo}. Finally, the functions $G_{eq}$ and $G_{e^2}$ in
Eq.~\eqref{eq:boostqqb} and Eq.~\eqref{eq:ONLOqqb} are given by
\begin{equation}
\label{eq:Gesq_qqb}
\begin{split}
G^{(i,j)}_{eq} = \, &
{\litwo}(1-\eta_{i3}) - {\litwo}(1-\eta_{i4})
-{\litwo}(1-\eta_{j3}) + {\litwo}(1-\eta_{j4})
\\
&
+\left[ \frac32 - \log\lb\frac{E_3}{E_c}\rb \right]
\log\lb\frac{\eta_{i3}}{\eta_{j3}}\rb
-\left[ \frac32 - \log\lb\frac{E_4}{E_c}\rb \right]
\log\lb\frac{\eta_{i4}}{\eta_{j4}}\rb \, ,
\\
G_{e^2} 
& =
13 
-\frac23 \pi^2
+\log^2 \Big( \frac{E_3}{E_4}\Big)
+ \left[ 3
- 2 \log \Big(\frac{E_3  E_4}{E_c^2} \Big) \right]  \log(\eta_{34})
+2\litwo(1-\eta_{34}) \; .
\end{split}
\end{equation}
Similar analytic expressions for all the remaining partonic channels
are collected in Appendix~\ref{sec:finite_parts}.

\section{Virtual corrections}
\label{sec:VV}

In the previous section we have described the extraction and cancellation
of infrared singularities in mixed \QCDEW corrections to DY production
within the framework of the nested soft-collinear subtraction scheme. In
doing so, we discussed the infrared singularity structure of virtual
corrections but did not explain  how to obtain the finite remainders,
cf. Eqs.~(\ref{eqa19}, \ref{eq2.55}, \ref{eq:twoloopfinrem}). In this section
we briefly outline their computation, focusing especially on how the
two-loop amplitudes presented in
Ref.~\cite{Heller:2019gkq,Heller:2020owb} can be adapted to our
subtraction framework and implemented in a numerical code.

The complete calculation of mixed \QCDEW corrections to dilepton
production requires the computation of various one- and two-loop
contributions.  We need one-loop QCD and electroweak corrections to
the partonic process $q \bar q \to \ell^-\ell^+$, one-loop QCD
corrections to the process $q \bar q \to \ell^-\ell^+ + \gamma$ and
one-loop electroweak corrections to the process $q \bar q \to
\ell^-\ell^+ + {\rm jet}$ (and their crossings), as well as two-loop
mixed \QCDEW corrections to the $q\bar{q}\to\ell^-\ell^+$ amplitude.

We first discuss one-loop contributions.  Using the definition of
infrared divergent and finite contributions described in
Section~\ref{sec:NNLOsubtraction}, we obtain the finite part of the
one-loop QCD correction
\begin{equation}
\label{eq:olqcdff}
    \la F^{(\QCD), \, \fin}_{\LV} (1_q,2_{\bar{q}},3,4) \ra = \calC^{\QCD} \la F_{\LM}(1_q,2_{\bar{q}},3,4)\ra \, ,
     \quad \text{with} \quad \calC^{\QCD} = -8 C_F \,  \frac{\alpha_s(\mu)}{2\pi}\, .
\end{equation}
The one-loop QCD amplitudes for the process $q \bar q \to \ell^-\ell^+
+ \gamma$ are obtained from the well-known QCD amplitudes for the $q
\bar q \to Z+j$ process \cite{Bern:1997sc}, which we borrow
from $\mbox{MCFM}$~\cite{Campbell:1999ah}. The
one-loop electroweak corrections to the processes $q \bar q \to
\ell^-\ell^+$ and $q \bar q \to \ell^-\ell^+ + g$ are instead computed
using \OLtwo~\cite{Cascioli:2011va,Buccioni:2017yxi,Buccioni:2019sur}.
It is simple to obtain the infrared finite part of all these
amplitudes, following the discussion in
Section~\ref{sec:NNLOsubtraction}.

\begin{figure}[t]
\begin{center}
\includegraphics[scale=0.9]{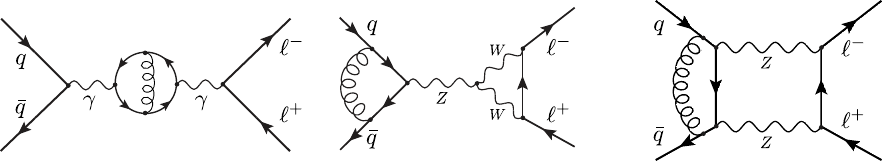}
\caption{Examples of Feynman diagrams that contribute to the two-loop
  amplitude. Left: two-loop fermionic non-factorizable corrections,
  middle: factorizable corrections, right: bosonic non-factorizable 
  corrections.}
\label{fig:twoloopfd}
\end{center}
\end{figure}
The double-virtual corrections to the process $q \bar q \to
\ell^-\ell^+$ are calculated starting from the two-loop amplitudes
presented in Refs.~\cite{Heller:2019gkq,Heller:2020owb}.  We note,
however, that in that reference only bosonic contributions to the
amplitudes were considered, see middle and right diagrams in
Fig.~\ref{fig:twoloopfd} for examples.  Thus, we have calculated the
additional terms arising from closed fermionic loops, see the left diagram in
\reffi{fig:twoloopfd} for an example.
We note that fermionic corrections to dilepton production in both the
charged- and neutral-current cases were studied earlier in
Ref.~\cite{Dittmaier:2020vra}.  We have performed an independent
calculation and checked our analytic results against those in
Refs.~\cite{Djouadi:1993ss,Dittmaier:2020vra}.  We also note that
these contributions are the only ones relevant for the on-shell
renormalization of the electroweak coupling $\alpha$, and they are the
only diagrams that make the extension of the complex mass scheme to
$\calO(\alpha\alpha_s)$ non-trivial, see Ref.~\cite{Dittmaier:2020vra}
for further details.

In our implementation, we find it convenient to separate virtual
corrections into a factorizable and a non-factorizable part. We
define the former as the product of the one-loop EW contribution and the
QCD $K$-factor from Eq.~\eqref{eq:olqcdff}:
\begin{equation}
  \left\langle
  F^{\rm fact}_{\rm LVV+LV^2} (1_q,2_\barq,3,4)
  \right\rangle \equiv
  \mathcal C^{\rm QCD}\left\langle F^{\rm(EW),fin}_{\rm LV}(1_q,2_{\bar q},3,4)
  \right\rangle.
\end{equation}
Also, we separate the non-factorizable contribution into a bosonic part
-- extracted from Ref.~\cite{Heller:2020owb} -- and
a fermionic part which accounts for closed fermion loops. 
In summary, we write the finite two-loop contribution to the cross section
as
\begin{equation}
  \begin{split}
    \label{eq:twoloopsplit}
    \la F^{\rm (QCD\times EW), \, fin}_{\rm LVV+LV^2} (1_q,2_\barq,3,4) \ra 
    &= 
    \la F^{\rm fact}_{\rm LVV+LV^2} (1_q,2_{\bar{q}},3,4) \ra
    + \la F^{{\rm non-fact,bos}}_{\rm LVV}(1_q,2_{\bar{q}},3,4)\ra
    \\
    &
    + \la F^{{\rm non-fact,ferm}}_{\rm LVV}(1_q,2_{\bar{q}},3,4)\ra \,.
  \end{split}
\end{equation}
To avoid confusion, we note that the non-factorizable fermionic term
only contains 1PI contributions similar to  the leftmost diagram in
Fig.~\ref{fig:twoloopfd}. Indeed, it is easy to convince oneself that
all reducible terms involving closed fermion loops are
included in the factorizable part. A representative diagram for each
of the three terms on the right-hand side of Eq.~\eqref{eq:twoloopsplit}
is shown in Fig.~\ref{fig:twoloopfd}.

The reason for separating the two-loop virtual corrections into
factorizable and non-factorizable parts is that the former should be
dominant at high energy since it contains leading Sudakov
logarithms. Indeed, we have checked that the non-factorizable
contribution to the cross section is typically an order of magnitude
smaller than the factorizable one. This happens across the entire
phase space that we have investigated. The practical advantage of this
observation is that the non-factorizable contribution -- whose
numerical evaluation is CPU expensive -- can be determined to a much
lower accuracy to obtain the cross section with a target
precision.
We also note that the separation of two-loop virtual corrections shown
in Eq.~\eqref{eq:twoloopsplit} allows us to capture the bulk of the
contribution coming from virtual top quarks, as we now
explain. Computing such contributions exactly for the full two-loop
amplitude is beyond the reach of current technology. As a consequence,
they were dropped in Ref.~\cite{Heller:2020owb}. Here, we neglect them
in the finite part of the bosonic non-factorizable term but include
them in all the other contributions. Since bosonic non-factorizable
contributions should be subdominant, this approach indeed allows us to
capture the leading top-quark effects in a relatively simple way.

We now discuss how to obtain the bosonic non-factorizable
contributions from Ref.~\cite{Heller:2020owb}. This
reference presents the result in terms of infrared subtracted finite
helicity remainders, referred to as ``hard functions''. Hard functions
that describe $\mathcal O(\alpha^i \alpha_s^j)$ corrections to the
scattering amplitude are denoted as $\mathcal H^{(i,j)}$.\footnote{We
note that compared to Ref.~\cite{Heller:2020owb} we have dropped the
helicity labels to simplify the notation.} We stress that the
$\mathcal H^{(i,j)}$ finite remainders do not include contributions
arising from closed fermion loops.
We also note that in
Ref.~\cite{Heller:2020owb} wave functions and masses are renormalized in the
on-shell scheme but both the QCD and EW couplings are renormalized in
the $\overline {\rm MS}$ scheme. In contrast, in this paper we renormalize
the EW coupling on-shell, so in principle we should perform a scheme change.
However, it is easy to convince oneself that such a change does not affect
the bosonic non-factorizable contribution. Hence, we can take the amplitudes
from Ref.~\cite{Heller:2020owb} as they are.

To obtain
$F_{\rm LVV}^{\rm non-fact,bos}$, we first define the non-factorizable
hard function
\begin{equation}
  \mathcal H^{(1,1)}_{\rm non-fact} = \mathcal H^{(1,1)} -
  \frac{\mathcal H^{(1,0)} \mathcal H^{(0,1)}}{\mathcal H^{(0,0)}},
  \label{eq:H11nonfact}
\end{equation}
and then use it to compute a non-factorizable $K$-factor
\begin{equation}
\begin{split}
  K_{\rm LVV}^{\rm non-fact,bos} = \frac{1}{2}\frac{ \sum \limits_{\rm
      spin,color}^{} \mathrm{Re}\left[{\mathcal H^{(0,0)*}}\mathcal
      H^{(1,1)}_{\rm non-fact}\right] }{\sum \limits_{\rm
      spin,color}^{} |\mathcal H^{(0,0)}|^2} &+ \Delta\mathcal
  H(\mu^2,s),
\end{split}
\end{equation}
where
\begin{equation}
    \Delta \mathcal H(\mu^2,s) = Q_q^2\, C_F\, \left[-\frac{1}{8} + 29 \zeta_2
      -30\zeta_3 - 22 \zeta_4 - \left(\frac{3}{2} - 12\zeta_2 + 24
      \zeta_3 \right)\log\Big(\frac{\mu^2}{s}\Big) \right]\, .
    \label{eq:dH}
\end{equation}
The non-factorizable bosonic contribution to the cross section then
reads
\begin{equation}
\begin{split} 
  \la F^{\rm non-fact,bos}_{\rm LVV}(1_q,2_{\bar{q}},3,4)\ra &=
  \frac{\alpha_s(\mu)}{2\pi} \frac{\alpha}{2\pi} \; \la K_{\rm
    LVV}^{\rm non-fact,bos} F_{\LM}(1_q,2_{\bar{q}},3,4)\ra \; .
\end{split}
\label{eq:bosnonf}
\end{equation}
We note that the $\Delta\mathcal H$ term in Eq.~\eqref{eq:dH} appears
because the definition of the two-loop finite remainders in
Ref.~\cite{Heller:2020owb} is slightly different from ours, cf.
Eq.~\eqref{eq:twoloopfinrem}. We also note that the $\Delta\mathcal H$
term is the only source of explicit scale dependence in the double-virtual
finite contribution to the cross section
$\la F^{\rm (QCD\times EW), \, fin}_{\rm LVV+LV^2}\ra$. 

We conclude this section by briefly discussing the numerical
implementation of these results. We have developed an efficient
\texttt{C++} code for the evaluation  of  the finite remainders of
the non-factorizable
two-loop bosonic corrections~\cite{Heller:2020owb}. These 
 are  given  in terms of complicated
rational functions multiplying Goncharov polylogarithms.  We have
minimized the set of rational functions by finding $\mathbb{Q}$-linear
relations among them \cite{Abreu:2019odu,
  Chawdhry:2019bji,Agarwal:2021grm}, performed a partial fraction
decomposition with minimal denominator
powers~\cite{Agarwal:2021grm} using the package \textsc{
  MultivariateApart}~\cite{Heller:2020owb}, and identified common
subexpressions to optimize the performance.  For the evaluation of the
Goncharov polylogarithms we employ the \textsc{handyG}
library~\cite{Naterop:2019xaf}.\footnote{ As a cross-check, we have
also used the \textsc{PolyLogTools} package of
Ref.~\cite{Duhr:2019tlz}, which employs
\textsc{GiNaC}~\cite{Bauer:2000cp,Vollinga:2004sn} for the numerical
evaluation of Goncharov polylogarithms, to compute the two-loop
amplitudes. We found perfect agreement with the results obtained with
\textsc{handyG}.}  The total evaluation time of the double-virtual
contributions for a single phase-space point is, on average, about
$0.7$~s. We have also performed an independent \textsc{Mathematica}
implementation of Eq.~\eqref{eq:bosnonf} and found perfect agreement
with the \texttt{C++} result for a random kinematic configuration.

\section{Phenomenological results}
\label{sec:pheno}
We are now in position to perform a phenomenological study of dilepton
production at high invariant mass.  We begin by specifying the
renormalization scheme and the input parameters. As we have mentioned, wave
functions, masses and the electric charge are renormalized
on-shell. The strong coupling and parton distribution functions are
instead renormalized in the $\overline{\rm MS}$ scheme.
We use the so-called $G_\mu$ input scheme for the EW parameters.  We
also employ the complex-mass scheme~\cite{Denner:2005fg} and its
extension to $\calO(\alpha \alpha_s)$ corrections as described in
Ref.~\cite{Dittmaier:2020vra}.

We consider proton-proton collisions at 13.6~$\TeV$ center-of-mass
energy.  We use the
\texttt{NNPDF31\_nnlo\_as\_0118\_luxqed}~\cite{NNPDF:2017mvq} parton
distribution functions for \emph{all}
computations reported in this paper, including leading and
next-to-leading order ones.
We use the strong coupling constant
$\alpha_s$ as provided by the PDF set; numerically, $\alpha_s(m_Z) =
0.118$.
In our numerical code, we have used both the $\texttt{LHAPDF}$
library~\cite{Buckley:2014ana} and
\textsc{Hoppet}~\cite{Salam:2008qg} to deal with PDFs. 
For the electroweak input parameters, the following values are used: 
$m_Z = 91.1876~\GeV$, $\Gamma_Z = 2.4952~\GeV$,
$m_W = 80.398~\GeV$,  $\Gamma_W = 2.1054~\GeV$,
$m_H = 125~\GeV$, $\Gamma_H = 4.165 ~{\rm MeV}$,
$m_t = 173.2~\GeV$ and
$G_F = 1.16639 \cdot 10^{-5}~{\rm GeV}^{-2}$.
With these input parameters, the fine structure constant reads $\alpha = 1/132.277$.

We note that since we work with
massless leptons, their momenta are not collinear-safe
quantities.  For this reason, we cluster photons and leptons into
``lepton jets'', often referred to as ``dressed leptons'' in
the literature, if the separation between leptons and photons
$R_{\ell\gamma}=\sqrt{(y_\ell-y_\gamma)^2 +
  (\varphi_\ell-\varphi_\ell)^2}$ is smaller than $R_{\rm{cut}}$.
We choose $R_{\rm cut}$ to be $0.1$. We recombine momenta in the
so-called $E$ scheme, i.e. to obtain the dressed-lepton momentum
we sum the four-momenta of the clustered leptons and photons.
For numerical computations, we take the renormalization scale $\mu_R$
and the factorization scale $\mu_F$ to be equal, and we choose the
invariant mass of the (dressed) dilepton system divided by two \ie
$\mu_F = \mu_R = \mu = m_{\ell\ell}/2$ as the central value. Scale
uncertainty is estimated  by increasing or decreasing the scale $\mu$ by a factor of two.

Following the ATLAS analysis in the high invariant
mass region \cite{ATLAS:2016gic}, we define the fiducial region by
requiring
\begin{equation}
\label{eq:fidspaceone}
m_{\ell\ell} > 200 \, \GeV \, , \quad
p_{T,\ell^{\pm}} > 30  \, \GeV \, , \quad
\sqrt{p_{T,\ell^-}p_{T,\ell^+}} > 35 \, \GeV \, , \quad
|y_{\ell^\pm}| < 2.5 \, .
\end{equation}
We note, however, that at variance with Ref.~\cite{ATLAS:2016gic}
we do not impose asymmetric cuts on the lepton transverse momenta
but we adopt the product
cuts recently proposed in Ref.~\cite{Salam:2021tbm}.  We also note that all
quantities that appear in Eq.~\eqref{eq:fidspaceone}, are defined in
terms of dressed leptons. This applies to leptons' transverse momenta
and rapidities $p_{T,\ell}$ and $y_{\ell}$, respectively, as well as
to the dilepton invariant mass $m_{\ell\ell}$.

To discuss the  impact of the various higher-order corrections, we find it convenient to introduce
the following notation for the differential cross section $d\sigma$
and its integrated counterpart $\delta\sigma$
\begin{equation}
\label{eq:pertexpansion}
d\sigma = \sum_{i,j=0} d\sigma^{(i,j)}\, , \quad\quad
\delta\sigma^{(i,j)} = \int  d\sigma^{(i,j)}\quad \text{with} \quad \sigma^{(0,0)}\equiv \delta\sigma^{(0,0)}\,.
\end{equation}
In the above equation, $d\sigma^{(0,0)}$ and $\sigma^{(0,0)}$
represent the LO cross sections while  $d\sigma^{(i,j)}$ and 
$\delta\sigma^{(i,j)}$ with $i,j>0$ stand for contributions to cross
sections at order ${\cal O}(\alpha_s^i\alpha^j)$.

\begin{table}[t]
\begin{center}
\renewcommand{\arraystretch}{1.1}
\begin{tabular}{| l | r | r | r | r | r |}
\hline
$\sigma$[fb]
& \multicolumn{1}{|c|}{$\sigma^{(0,0)}$}
& \multicolumn{1}{|c|}{$\delta\sigma^{(1,0)}$}
& \multicolumn{1}{|c|}{$\delta\sigma^{(0,1)}$}
& \multicolumn{1}{|c|}{$\delta\sigma^{(2,0)}$}
& \multicolumn{1}{|c|}{$\delta\sigma^{(1,1)}$} \\
\hline\hline
$q\bar{q}$     & $1561.42$     & $340.31$ &  $-49.907$  & $44.60$      & $-16.80$  \\
\hline
$\gamma\gamma$ & $59.645$      &          &  $3.166$    &              &           \\
\hline
$q g$          &               & $0.060$  &             & $-32.66$     & $1.03$    \\
\hline
$q \gamma$     &               &          &  $-0.305$   &              & $-0.207$  \\
\hline
$g \gamma$     &               &          &             &              & $0.2668$  \\
\hline
$g g$          &               &          &             & $1.934$      &           \\
\hline
\hline
sum            & $1621.06$     & $340.37$ & $-47.046$   & $13.88$      & $-15.71$  \\
\hline
\end{tabular}
\caption{Results for fiducial cross sections for central value of the
  renormalization and factorization scales
  $\mu_R=\mu_F=m_{\ell\ell}/2$.  Contributions are separated by
  partonic channels.  Selection cuts for
  final-state leptons and jets are given in
  Eq.~(\ref{eq:fidspaceone}). Here, $\delta\sigma^{(i,j)}$ denotes
  the correction of relative order $\alpha_s^i \alpha^j$. We note that
  the numerical precision on the correction $\delta\sigma^{(1,1)}$ is about
  1\%.}
\label{tab:integratedXSsetup1}
\end{center}
\end{table}

The results for fiducial cross sections are summarized in
\refta{tab:integratedXSsetup1}.  We note that we have compared NLO
QCD and EW results against \texttt{Sherpa}~\cite{Sherpa:2019gpd} and
\texttt{MoCaNLO}+\texttt{Recola}~\cite{Actis:2016mpe,Denner:2016kdg,Denner:2021hqi,Denner:2021csi}. We have found perfect agreement for
all the channels listed in Table ~\ref{tab:integratedXSsetup1}.
We observe that NLO QCD corrections
increase the leading-order cross section by about twenty percent, the
NNLO QCD corrections change it by about $0.9\%$, and the NLO electroweak
corrections reduce it by about $3\%$. 
We note that numerical results reported
in Table~\ref{tab:integratedXSsetup1} are consistent with expectations
based on the magnitude of the respective coupling constants, although
the NNLO QCD corrections are slightly smaller than could have been
anticipated.
In particular, NLO EW corrections are compatible with
a naive power counting $\delta^{\rm EW}\sim\alpha/\sin^2\theta_W\sim 0.03$,
where $\theta_W$ is the weak mixing angle.

An interesting feature of the results shown in
\refta{tab:integratedXSsetup1} is that the contribution of the
diphoton channel at leading order, where dileptons are produced
directly in collisions of photon ``partons'' that originate from the
proton, is quite large, about $3.6\%$ of the total cross section.  The
reason for this is the enhancement of this contribution by a logarithm
$\ln (m_{\ell\ell}^2/p_{T,\ell^{\pm}}^2) \sim 5$.  We also note
that there is a strong cancellation between this contribution and the
NLO electroweak corrections.

We observe that the NLO QCD correction does not show the cancellation
between $q \bar q$ and $qg $ partonic channels that is observed in the
resonant region; in fact, we see that at high invariant masses, QCD
corrections to the $q \bar q$ channel are the dominant ones with the
$qg$ channel playing only a minor role.  The picture changes if we
consider scales $\mu=m_{\ell \ell}/4$ or $\mu=m_{\ell \ell}$, in which
case the contributions of the $qg$ channels are of the same order as
the $q\bar q$ ones.  At NNLO QCD, there is a strong cancellation
between these two partonic channels, making this correction even
smaller than the NLO EW one.  This, of course, illustrates the somewhat
unphysical nature of individual partonic channels at higher orders
since they require collinear subtractions to be well-defined.

It follows from \refta{tab:integratedXSsetup1} that mixed \QCDEW
corrections are quite large and decrease the fiducial cross section by
about $1\%$, whereas an estimate based on power counting suggests
that $\mathcal{O}(\alpha \alpha_s)$ corrections should be at the per
mille level.  In fact, the mixed \QCDEW corrections are about $30\%$
of the electroweak corrections and \emph{larger} than the NNLO QCD
ones. These corrections receive the dominant contribution from the $q \bar
q$ partonic channel; all other channels affect the fiducial cross section
by a much smaller amount.

It is also instructive to compare the magnitude of the mixed \QCDEW
corrections with the theoretical uncertainty. To estimate it, we
increase and decrease the central scale $\mu = m_{\ell \ell}/2$ by a
factor of two and also choose a different input scheme for the electroweak
parameters. In particular, we consider the so-called
$\alpha(m_Z)$-scheme where $\alpha(m_Z) =1/128$ is an input parameter,
and the other input parameters are kept fixed.\footnote{For a comprehensive
discussion of electroweak input schemes see
Ref.~\cite{Denner:2019vbn}.}  We then take the envelope of these
results as an estimate of the theoretical uncertainty.  We find that the
(asymmetric) uncertainty of the leading-order cross section is $+12\%$
and $-6\%$.  Instead, if the cross section is computed through NNLO QCD and NLO
EW, but the mixed \QCDEW corrections are neglected, we find
\begin{equation}
  \sigma^{(0,0)}+\delta\sigma^{(1,0)}+\delta\sigma^{(0,1)}+\delta\sigma^{(2,0)} = 1928.3^{+1.8\%}_{-0.15\%}~{\rm fb}.
  \label{eq4.3}
\end{equation}
We note that the main source of the theoretical uncertainty in
Eq.~(\ref{eq4.3}) is the input-scheme change which, however, is
reduced from about 6\% at leading order to about two percent when NLO
EW contributions are accounted for.
The mixed QCD-electroweak corrections are about $-1\%$ and, thus,
comparable in size to the theoretical uncertainty in
Eq.~(\ref{eq4.3}).  Upon including them, the central value of the
fiducial cross section and its uncertainty decrease. We obtain
\begin{equation}
\label{eq:bestpredictionxs}
    \sigma_{\QCD\times\EW}\equiv \sigma^{(0,0)}+\delta\sigma^{(1,0)}+\delta\sigma^{(0,1)}+\delta\sigma^{(2,0)} +\delta \sigma^{(1,1)}= 1912.6^{+0.65\%}_{-0\%}~{\rm fb}.
\end{equation}
The main reason behind the reduction of uncertainty with respect to
Eq.~\eqref{eq4.3} is that now the mixed \QCDEW corrections remove a
large source of input-scheme dependence coming from the NLO QCD
contribution.\footnote{Indeed, we note that the pure EW scheme uncertainty is
reduced from about 1\% to about 0.5\% after the inclusion of mixed
corrections.}
We note that the above error estimates do not include uncertainties
from PDFs, which are known to be significant. Indeed, the uncertainty on the
$q\bar q$ luminosity ranges from  about
2\% for $m_{\ell\ell}\lesssim 1~\TeV$ to about $\sim 5\%$ for
$m_{\ell\ell}\sim 2~\TeV$.

It is well-known that at high invariant masses, EW corrections are
dominated by the universal Sudakov logarithms.  This implies that the
mixed QCD-electroweak corrections should be well described by the
product of QCD and electroweak corrections, at least inasmuch as the
leading logarithms are concerned.  Although it is not clear when this
``factorized'' approximation becomes a good representation of the
full result, it is easy to check its efficacy  by comparing exact
and approximate results for mixed \QCDEW corrections at various values
of $m_{\ell\ell}$.  To this end, we consider four invariant-mass
windows defined as follows
\be
\begin{split} 
\label{eq:invmasswindows}
\Phi^{(1)}: &~~200~\GeV <  m_{\ell\ell} <  300~\GeV,  \\
\Phi^{(2)}: &~~300~\GeV <  m_{\ell\ell} < 500~\GeV,  \\
\Phi^{(3)}: &~~500~\GeV <  m_{\ell\ell} <  1.5~\TeV, \\
\Phi^{(4)}: &~~1.5~\TeV < m_{\ell\ell} < \infty.
\end{split}
  \ee For each of these windows, we apply the
  $m_{\ell\ell}$-independent kinematic cuts described in
  Eq.~\eqref{eq:fidspaceone}.
\begin{table}[t]
\begin{center}
\renewcommand{\arraystretch}{1.2}
\begin{tabular}{|c|r|r|r|r|r|r|c|}
\hline
  $\sigma\lsb\fb\rsb$  & $\sigma^{(0,0)}$  & $\delta \sigma^{(1,0)}$  & $\delta \sigma^{(0,1)}$  &$\delta \sigma^{(2,0)}$   & $\delta \sigma^{(1,1)}$      & $\delta \sigma^{(1,1)}_{\mathrm{fact.}}$ & $\sigma_{\QCD\times\EW}$ \\ \hline\hline
$\Phi^{(1)}$           & $1169.8$ & $254.3$  & $-30.98$  &  $10.18$  & $-10.74$  & $-6.734$  & $1392.6^{+0.75\%}_{-0\%}$ \\ \hline
$\Phi^{(2)}$           & $368.29$ & $71.91$  & $-11.891$ & $2.85$    & $-4.05$   & $-2.321$  & $427.1^{+0.41\%}_{-0.02\%}$  \\ \hline
$\Phi^{(3)}$           & $82.08$  & $14.31$  & $-4.094$  &  $0.691$  & $-1.01$  & $-0.7137$ & $91.98^{+0.22\%}_{-0.14\%}$  \\ \hline
$\Phi^{(4)} \times 10$ & $9.107$  & $1.577$  & $-1.124$  &  $0.146$  & $-0.206$ & $-0.1946$ & $9.500^{+0\%}_{-0.97\%}$  \\ \hline
\end{tabular}
\caption{Fiducial cross sections in the invariant mass windows
  specified through Eq.~\refeq{eq:invmasswindows}.  We show the LO
  predictions, $\sigma^{(0,0)}$ and the higher-order 
  ones, $\delta\sigma^{(i,j)}$.  In order to improve readability, we
  multiply the fiducial cross sections in the $\Phi^{(4)}$ phase space by
  a factor 10.  In the next-to-last column we quote the result of the
  factorized approximation defined in Eq.~\eqref{eq:factapprox}.  In
  the last column we show our best predictions obtained by including
  all the higher-order corrections considered in this paper,
  cf. Eq.~\eqref{eq:bestpredictionxs}.}
\label{tab:masswindows}
\end{center}
\end{table}
To compare the quality of the factorized approximation in each of
the four mass regions, we define approximate mixed corrections as
follows \be
\label{eq:factapprox}
\delta \sigma_{\rm fact}^{(1,1)} = \delta_{\rm NLO}^{(1,0)} \; \delta_{\rm NLO}^{(0,1)} \; \sigma^{(0,0)},
\ee
where 
\begin{equation}
  \delta_{\rm NLO}^{(1,0)}  = \frac{\delta \sigma^{(1,0)}}{\sigma^{(0,0)}}, \;\;\;\;\;\; \delta_{\rm NLO}^{(0,1)}  = \frac{\delta \sigma^{(0,1)}}{\sigma^{(0,0)}}.
\end{equation}
The approximate mixed corrections are compared to their exact
counterparts in \refta{tab:masswindows}.  We find that $\delta
\sigma^{(1,1)}_{\rm fact}$ captures the main features of the mixed
corrections but underestimates them for lower invariant masses.
At high invariant masses $m_{\ell\ell} > 1~{\rm TeV}$ the situation
changes and the quality of the factorized approximation improves. For
the highest invariant-mass window the factorized approximation
captures more than $90\%$ of the exact result.  This behavior is not
surprising since, as we already mentioned, the factorized
approximation correctly reproduces the leading Sudakov logarithms that
are expected to provide the dominant contribution at large invariant
masses.  In this table, we also show our predictions for the quantity
$\sigma_{\rm QCD \times EW}$ defined in
Eq.~\eqref{eq:bestpredictionxs}, i.e. including NLO QCD, NLO EW, NNLO
QCD and mixed \QCDEW corrections, in the four invariant mass
windows. We observe that the theoretical uncertainty, estimated by a
simultaneous variation of scales and input scheme, is below the
percent level across the different windows considered.

\begin{figure}[t]
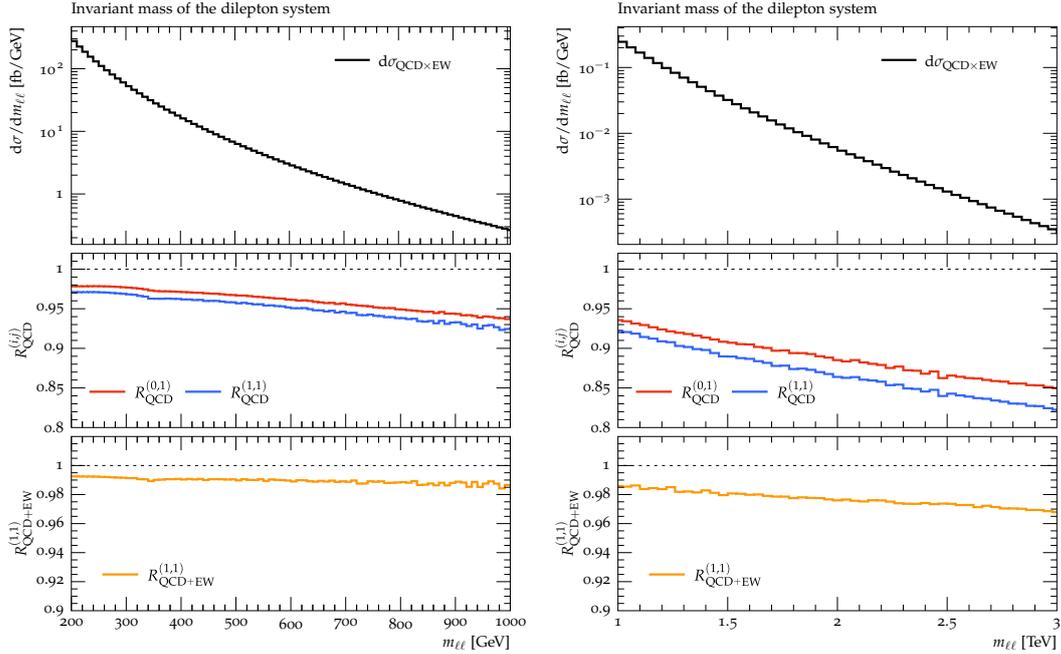

\showtwoobs{mll}{mll_high}
\caption{Dilepton invariant mass distribution for the Drell-Yan
  process $pp \to \ell^- \ell ^+$ at the 13.6 TeV LHC. The upper pane
  shows our best prediction for $d\sigma$ which included NLO
  QCD, NNLO QCD, NLO EW, and mixed \QCDEW corrections. The middle pane
  shows the ratio of the NLO EW and mixed \QCDEW corrections to the
  full NLO QCD result. The lower pane shows the ratio of mixed \QCDEW
  corrections to a result which includes both QCD and EW NLO
  corrections. The left plot
  shows results in the range $200~{\rm GeV} < m_{\ell \ell} < 1~{\rm
    TeV}$, the right plot shows the range $1~{\rm TeV} < m_{\ell \ell}
  < 3~{\rm TeV}$. See text for details.}
\vspace*{3ex}
\label{fig:mll}
\end{figure}

We now turn to the discussion of kinematic distributions. The dilepton
invariant mass case is shown in Fig.~\ref{fig:mll}.  There,
the distributions in the upper panes include all corrections
considered in this paper
\begin{equation}
 d\sigma_{\QCD\times\EW} = d\sigma^{(0,0)} + d\sigma^{(1,0)} + d\sigma^{(0,1)} + d\sigma^{(2,0)} + d\sigma^{(1,1)},
\end{equation}
the middle panes show the impact of the NLO EW and mixed \QCDEW
corrections on the results computed through NLO QCD, and the lower panes
show the impact of the mixed \QCDEW corrections on cross sections
computed through NLO QCD and NLO EW accuracy.  To this end, we define
the following quantities
\begin{equation}
    R_{\rm QCD}^{(0,1)} = \frac{d\sigma^{(0,0)} + d\sigma^{(1,0)} + d\sigma^{(0,1)}}{d\sigma^{(0,0)}+d\sigma^{(1,0)}}, 
    \quad 
    R_{\rm QCD}^{(1,1)} = \frac{d\sigma^{(0,0)} + d\sigma^{(1,0)} + d\sigma^{(0,1)} +
    d\sigma^{(1,1)}}{d\sigma^{(0,0)}+d\sigma^{(1,0)}},
\end{equation}
\begin{equation}
     R^{(1,1)}_{\rm QCD+EW} = R_{\rm QCD}^{(1,1)}/R_{\rm QCD}^{(0,1)} = \frac{d\sigma^{(0,0)} + d\sigma^{(1,0)} +
     d\sigma^{(0,1)} + d\sigma^{(1,1)}}
     {d\sigma^{(0,0)}+d\sigma^{(1,0)}+d\sigma^{(0,1)}},
\end{equation}
and plot them in Fig.~\ref{fig:mll} as a function of the dilepton
invariant mass.
It follows from Fig.~\ref{fig:mll} that NLO EW corrections grow from
${\cal O}(-2\%)$ at $m_{\ell\ell}~=~200~{\rm GeV}$ to ${\cal
  O}(-15\%)$ at $3$~TeV, and that the mixed \QCDEW corrections follow
the shape of the NLO EW ones. Nevertheless, $R_{\rm \QCDEW}$ is not
entirely flat over the range of invariant masses that we consider;
indeed, the magnitude of \QCDEW corrections slowly increases from
${\calO}(1.5\%)$ at $m_{\ell\ell} \approx 200~{\rm GeV}$ to
${\calO}(3\%)$ at $m_{\ell\ell} \approx 3~{\rm TeV}$.  These results
are consistent with those presented in Table~\ref{tab:masswindows} and
are indicative of the presence of Sudakov logarithms in the virtual EW
corrections, as mentioned previously. We note that the small dip in
the middle pane at $m_{\ell \ell} \approx 340$ GeV originates from the
$t \bar t$ thresholds in closed fermion loops that modify the
propagators of the electroweak bosons.

While the magnitude of mixed QCD-electroweak corrections at large
invariant masses is fairly easy to understand, their size at lower
values of $m_{\ell\ell}$ is more puzzling as they seem to be enhanced
relative to naive expectations.  Indeed, it follows both from
Table~\ref{tab:integratedXSsetup1} and Table~\ref{tab:masswindows}
that mixed \QCDEW corrections are only three times smaller than the EW
corrections themselves and it is unclear why this is the case, given
that one does not expect large Sudakov logarithms at such energy
scales.  However, one should also keep in mind that the NLO QCD
corrections to the leading-order cross section are twenty percent whereas
the mixed \QCDEW corrections are thirty percent of the NLO EW contribution
which implies that the difference is not too large.  Hence, it can
also be that these fairly large effects at small invariant masses are
just the result of a numerical interplay of various contributions and
that the observed enhancement is more or less accidental.

\begin{figure}[t]
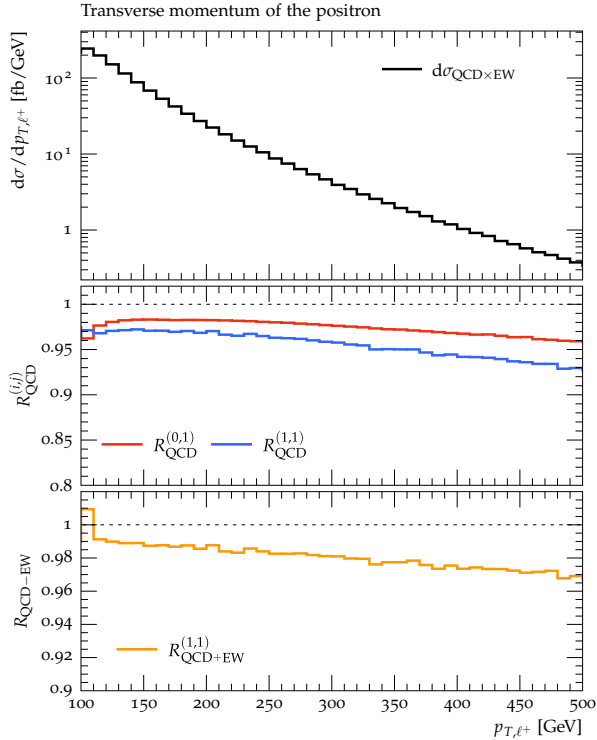

\showobs{ptlp}
\caption{Transverse momentum distribution of the positively charged lepton.
See the caption in Fig.~\ref{fig:mll} and the text for details.}
\vspace*{3ex}
\label{fig:ptylept}
\end{figure}

We continue with the discussion of other observables.  In
Fig.~\ref{fig:ptylept} we show the transverse momentum distribution of
the positively-charged lepton.  Since at leading order the lepton transverse momentum 
$p_{T,\ell}$ is always smaller than $m_{\ell\ell}/2$, there is a
correlation between the $p_{T,\ell}$ and the $m_{\ell\ell}$
distributions.  Indeed, it follows from Fig.~\ref{fig:ptylept} that 
corrections to the transverse momentum distribution are similar to the
ones to the $m_{\ell\ell}$ distribution, in that NLO EW corrections
are negative and quite large, while the relative \QCDEW corrections
are unusually large at low values of $p_{T,\ell}$, which give the
largest contribution to the fiducial cross section.  Mixed \QCDEW
corrections largely follow the pattern of the NLO EW corrections.
Nevertheless, the impact of the \QCDEW corrections does become
slightly more important at higher values of $p_{T,\ell}$, amounting to
around -3\% on top of the NLO QCD and EW result at $p_{T,\ell} =
500\,\GeV$.

\begin{figure}[t]
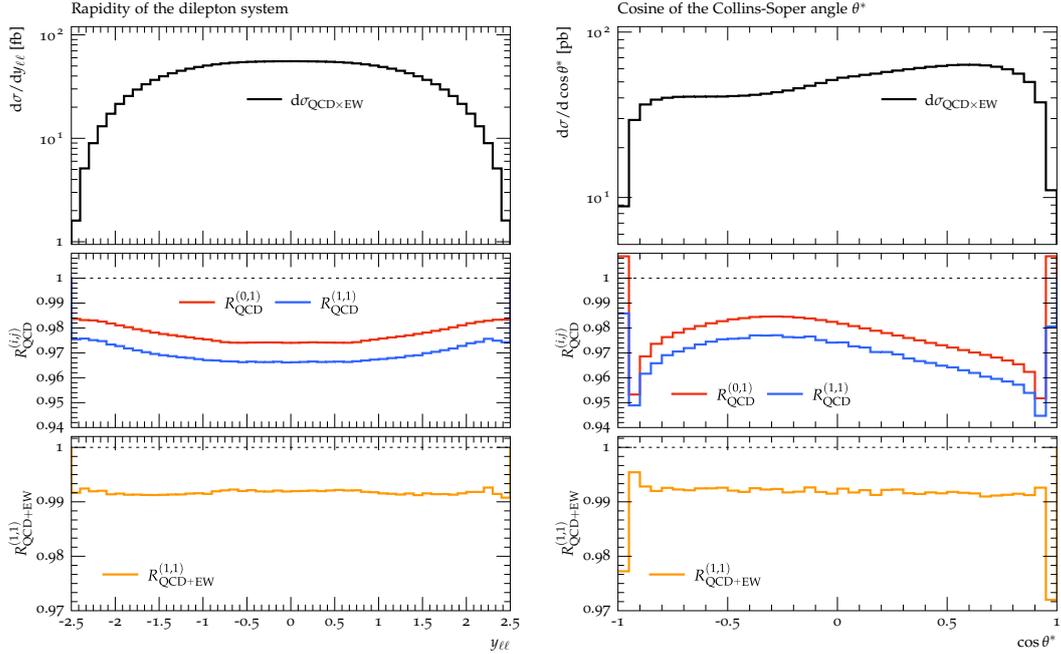

\showtwoobs{yll}{costhCS}
\caption{Distributions of the rapidity of the dilepton system and the cosine 
  of the Collins-Soper angle. See the caption in Fig.~\ref{fig:mll} and the
  text for details.}
\vspace*{3ex}
\label{fig:yllcosthCS}
\end{figure}

Another interesting class of observables are rapidity and angular
distributions. In left panes of Fig.~\ref{fig:yllcosthCS} we show the
rapidity of the dilepton system.  We observe that both the NLO EW
corrections and the mixed \QCDEW corrections are fairly flat over the
considered rapidity range and amount to ${\cal O}(-3\%)$ and ${\cal
  O}(-1\%)$, respectively.  As the fiducial cross sections are
dominated by low values of $m_{\ell\ell} \approx 200$~GeV, the
corrections that we see in the rapidity distribution correspond to
those shown in Table~\ref{tab:integratedXSsetup1}.

Angular distributions can be used to analyze the structure of the
currents that facilitate the transition from quarks to leptons both
within and beyond the Standard Model. Although these angular
distributions can be computed in full generality, it is simpler to
discuss an integrated quantity -- the forward-backward asymmetry of a
lepton relative to the direction of an incoming quark.
A convenient variable that allows one to study such an asymmetry is
the Collins-Soper angle~\cite{Collins:1977iv}, defined as follows
\begin{equation}
\cos \theta^* = \frac{p_{\ell^-}^+ p^-_{\ell^+} - p_{\ell^-}^- p^+_{\ell^+}}{m_{\ell\ell} \, \sqrt{m^2_{\ell\ell} + p_{\ell\ell,\perp}^2}} \times {\rm sgn}(p_{\ell\ell,z}) \; .
\label{eq4.11}
\end{equation}
In Eq.~(\ref{eq4.11}), $p_i^{\pm} = E_i \pm p_{z,i}$ and $p_{\ell\ell}
= p_{\ell^-} + p_{\ell^+}$.  We show the $\cos \theta^*$ distribution
for events in the fiducial volume Eq.~\eqref{eq:fidspaceone} in the
right panel of Fig.~\ref{fig:yllcosthCS}.  Similar to the dilepton
rapidity distribution, both the NLO EW and mixed \QCDEW corrections to
the Collins-Soper angle are fairly flat and are comparable to corrections to
the fiducial cross section, cf.  Table~\ref{tab:integratedXSsetup1}.

It is clear from Fig.~\ref{fig:yllcosthCS} that the distribution of
the Collins-Soper angle is not symmetric and that there are more
events with $\cos\theta^* > 0$ than with $\cos \theta^* < 0$.  To
quantify this effect, we consider the forward-backward
asymmetry
\begin{equation}
  A_{\rm FB} = \frac{\sigma_F -
    \sigma_B}{\sigma_F + \sigma_B},
\end{equation}
where
\begin{equation}
\label{eq:sigmafb}
\sigma_F = \int \limits_{0}^{1} {\rm d} \cos \theta^* \; \frac{{\rm d} \sigma(pp \to \ell^- \ell^+) }{{\rm d} \cos \theta^*},
\;\;\;\;\;
\sigma_B = \int \limits_{-1}^{0} {\rm d} \cos \theta^* \;  \frac{{\rm d} \sigma(pp \to \ell^- \ell^+)}{{\rm d} \cos \theta^*}.
\end{equation}
We calculate the forward-backward asymmetry for the fiducial phase space defined in Eq.~\refeq{eq:fidspaceone}
including all corrections computed in this paper and find 
\begin{equation}
A_{\rm FB} = 0.1580^{+0.15\%}_{-0.07\%} \; ,
\label{eq4.5}
\end{equation}
where the uncertainties are estimated from a simultaneous scale and
input-scheme variations as described above.  Omitting the mixed \QCDEW
corrections changes the prediction in Eq.~(\ref{eq4.5}) by about 2 per
mille which is again comparable with the uncertainty on the central
value.

It is well-known that the forward-backward asymmetry increases with
the invariant mass of dileptons. For this reason, it is instructive to
study the forward-backward asymmetry and mixed QCD-electroweak
corrections to it in the four $m_{\ell \ell}$ windows defined in
Eq.~(\ref{eq:invmasswindows}).  The results are shown in
Table~\ref{tab:CS}. There we display predictions for the
forward-backward asymmetry that include all corrections considered in
this paper ($A_{\rm FB}$) as well as the prediction for the
forward-backward asymmetry without the mixed \QCDEW correction
($\tilde A_{\rm FB}$).  We observe that the mixed \QCDEW corrections
impact the value of $A_{\rm FB}$ below the percent level in the lower
invariant mass windows, and reach $-1.3\%$ at high $m_{\ell\ell}$.
Such percent-level shifts above $m_{\ell \ell} \sim \TeV$ should
become observable at the high-luminosity LHC, provided that systematic
uncertainties can be controlled.

\begin{table}[t]
\begin{center}
\renewcommand{\arraystretch}{1.4}
\begin{tabular}{|l|c|c|}
\hline
$$    & $\tilde A_{\rm FB}$   &  $A_{\rm FB}$   \\ \hline\hline
$\Phi^{(1)}$         & $0.1442^{+0.05\%}_{-0.31\%}$  & $0.1440^{+0.11\%}_{-0.09\%}$ \\ \hline
$\Phi^{(2)}$         & $0.1852^{+0.08\%}_{-0.40\%}$  & $0.1847^{+0.10\%}_{-0.19\%}$ \\ \hline
$\Phi^{(3)}$         & $0.2401^{+0.13\%}_{-0.64\%}$  & $0.2388^{+0.06\%}_{-0.47\%}$ \\ \hline
$\Phi^{(4)}$         & $0.3070^{+0.49\%}_{-1.5\%}$  & $0.3031^{+0.19\%}_{-1.2\%}$ \\ \hline
\end{tabular}
\caption{Forward-backward asymmetry in the invariant
  mass windows specified through Eq.~\refeq{eq:invmasswindows}.  We
  label as $\tilde A_{\rm FB}$ the predictions including the LO
  contribution and higher order corrections from NLO-QCD, NLO-EW and
  NNLO-QCD, whereas $A_{\rm FB}$ further includes the mixed
  \QCDEW correction computed here.}
\label{tab:CS}
\end{center}
\end{table}

\section{Conclusions}

We presented a computation of mixed QCD-electroweak corrections
to the production of dilepton pairs in proton-proton collisions, $pp
\to \ell^- \ell^+$, focusing on the high-invariant mass region.  We
have used the two-loop amplitudes computed in
Ref.~\cite{Heller:2020owb} and the nested soft-collinear subtraction
scheme~\cite{Caola:2017dug} to extract and regulate the real-emission
contributions. Our results are fully differential with respect to the
kinematics of resolved final-state particles and can be used to compute
any infrared safe observable. 

We applied our result to the study of high-mass dilepton production at
the 13.6$~\TeV$ LHC.  We presented results for fiducial cross sections
and distributions defined by kinematic cuts applied to final-state
leptons in typical experimental analyses. We have selected the
high-mass region by requiring that the dilepton invariant mass is
larger than $200~{\rm GeV}$.  With these cuts, the mixed \QCDEW
corrections amount to about $-1\%$ of the LO cross section. They are
therefore larger than what could have been expected based on the
magnitudes of the coupling constants. In fact, in this setup they are
larger than the NNLO QCD ones. The remaining uncertainty coming from
scale and input-scheme variation is reduced to the sub-percent level.

For
even higher invariant masses, above $1~{\rm TeV}$, the mixed
corrections become even larger and appear to be driven by Sudakov
logarithms.  For this reason, the exact mixed \QCDEW corrections can
be reliably approximated by a product of NLO QCD and EW
contributions.  We have checked that this factorized approximation
reproduces the size of mixed corrections at $m_{\ell \ell} \sim 1~{\rm
  TeV}$ to within thirty percent and the accuracy of this
approximation increases at higher invariant masses.
We have also found that mixed \QCDEW corrections may affect the
forward-backward asymmetry in the process $pp \to \ell^- \ell^+$ at
the percent level for dilepton invariant masses above $1~{\rm
  TeV}$. This region is especially interesting for searching for New
Physics effects in dilepton production. Hopefully,
 measurements with such a precision can be
performed at the HL-LHC.

\label{sec:conclusions}

\section*{Acknowledgements}
We are grateful to Konstantin Asteriadis for useful discussions about
developing the multi-processor interface for the numerical code that
we used to compute the results reported in this paper. We are indebted
to Giovanni Pelliccioli for providing
results for NLO EW
corrections in our setup, and for discussions on the structure of NLO
EW corrections.  We gratefully acknowledge Robert M. Schabinger for help
with the two-loop amplitudes used in this paper. 
We thank Marek Schoenherr for useful correspondence
and help with \texttt{Sherpa}. This
research was partially supported by the Deutsche
Forschungsgemeinschaft (DFG, German Research Foundation) under grants
396021762-TRR 257, 204404729-SFB 1044 and 39083149-EXC 2118/1,
by the UK Science and Technology Facilities
Council (STFC) under grant ST/T000864/1,
by the ERC Starting Grant 804394 \textsc{hipQCD}, and by
the National Science Foundation (NSF) under grant 2013859.

\appendix
\section{Analytic results for mixed \QCDEW corrections for other
  partonic channels}
\label{sec:finite_parts}

\subsection{The $g\bar{q}$ and $q g$ channels}
\label{sec:finite_parts_gqbar}
In this section we present the finite partonic cross sections for the
$g\barq(gq)$ and $qg(\bar{q}g)$ channels.  For the sake of simplicity
we only present results for the $g\barq(gq)$ case, results for the
$qg(\bar{q}g)$ channel can be obtained with straightforward
modifications that will be mentioned below.  Similar to the $q\barq$
channel (see Sec.~\ref{subsec:analytic_qqbar}), we isolate four
different structures
\begin{equation}
d\hat{\sigma}_{\rm nnlo}^{g \bar{q}}
=
d\hat{\sigma}_{\rm bt}^{g \bar{q}}
+
d\hat{\sigma}_{\mathcal{O}_{\rm nlo}}^{g \bar{q}}
+
d\hat{\sigma}_{\rm reg}^{g \bar{q}}.
\label{eq:gqbar_finite_contributions}
\end{equation}
The regulated term reads
\begin{equation}
  2s\cdot d\hat{\sigma}_{\rm reg}^{g \bar{q}} = \left\langle
  \left(I-S_6\right)
  \Omega_{g\bar q} F_{\rm LM}(1_g,2_{\bar q},3_{\ell^-},4_{\ell^+},5_{\bar
    q},6_\gamma)\right\rangle, 
\end{equation}
with
\begin{equation}
  \begin{split}
  \Omega_{g\bar{q}} &= (1-C_{56, 1})
  \,\omega^{65} \, \tilde\theta_A
 +
(1-C_{56, 1}) (1-C_{56}) \,
 \omega^{65} \, \tilde\theta_B
 \\ & + \, (1-C_{56, 1}) (1-C_{51}) \, \omega^{65} \,
 \tilde\theta_C
+
(1-C_{56, 1}) (1-C_{56}) \,
\omega^{65} \, \tilde\theta_D
\\
& + \, \sum_{i=2}^{4}(1-C_{51})
(1-C_{6i}) \, \omega^{6 i},
  \end{split}
  \label{eq:omega_gq}
\end{equation}
and
\begin{equation}
  \begin{gathered}
    \tilde\theta_A = \theta\left(\eta_{16}<\frac{\eta_{15}}{2}\right)
    ~~~~
    \tilde\theta_B = \theta\left(\frac{\eta_{15}}{2}<\eta_{16}<\eta_{15}\right)
    \\
    \tilde\theta_C = \theta\left(\eta_{15}<\frac{\eta_{16}}{2}\right)
    ~~~~
    \tilde\theta_D = \theta\left(\frac{\eta_{16}}{2}<\eta_{15}<\eta_{16}\right).
  \end{gathered}
  \label{eq:thetatilde}
\end{equation}
In Eq.~\eqref{eq:omega_gq} we have also introduced the damping factors
\begin{equation}
\omega^{6 i} = \frac{1/\eta_{6 i}}{\sum_{a=2}^5 1/\eta_{6 a}} \; .
\end{equation}
We note that Eq.~\eqref{eq:omega_gq} is finite, and can be evaluated
numerically.

The other contributions are reported below assuming
$E_{\rm max}=E_1=E_2\equiv E_c$.  We have
\begin{equation}
\label{eq:gqboost}
\begin{split}
2s \cdot d\hat{\sigma}_{\rm bt}^{g \bar{q}} 
= \, &
[\alpha] \, [\alpha_s] \, 
 T_R \, Q_q^2 \, 
\int_0^1 dz_1 dz_2 \,\tilde{P}^{\rm NLO}_{\bar{q}g} (z_1, E_c)
\\ 
& \hspace{10mm} \times 
 \Big\langle \frac{F_{\rm LM} (z_1 \cdot 1_{q},z_2 \cdot 2_{\bar{q}},3,4))}{z_1 \, z_2} \Big\rangle \, 
 \tilde{P}_{qq}^{\rm NLO} (z_2, E_c)
 \\
 &
 +
[\alpha_{s}] \,T_R 
\int_{0}^1 dz \, 
\tilde{P}^{\rm NLO}_{\bar{q}g} (z, E_c) \, \Big\langle
 \frac{F_{\rm LV}^{\rm (EW), \, fin}(z \cdot 1_{q},2_{\bar{q}},3,4)}{z} \Big\rangle
 \\
 & +
[\alpha] [\alpha_{s}] \,T_R 
\int_{0}^1 dz \, 
 \Big\langle
\Big[
\tilde{P}^{\rm NLO}_{\bar{q}g} (z, E_c) \,
\Big(
2 \, Q_e Q_q \, 
G^{(1,2)}_{eq} 
\\
& \hspace{10mm} 
- \log \Big( \frac{s_{13}}{s_{14}}\Big) \log(z)
+
Q_e^2 \, G_{e^2} 
\Big)
+
Q_q^2 \, P_{\barq g}^{\rm NNLO}(z, E_c)
\Big] \; 
\\
& \hspace{10mm} \times \; 
 \frac{F_{\rm LM}(z \cdot 1_{q},2_{\bar{q}},3,4)}{z} \Big\rangle
 \; ,
 \end{split}
\end{equation}
where $P_{\barq g}^{\rm NNLO}(z, E_c)$ is reported in Eq.~\eqref{eq:PqgNNLO}.
The $\mathcal{O}_{\rm nlo}$ term reads
\begin{align}
\label{eq:gqOnlo}
2s \cdot d\hat{\sigma}_{\mathcal{O}_{\rm nlo}}^{g \bar{q}}
= \, &
 \Big\langle
 \mathcal{O}^{\barq}_{\rm nlo} \; 
F_{\rm LV}^{\rm (EW), \, fin}(1_g,2_{\bar{q}},3,4| 5_\barq) \Big\rangle
\nn \\
 &
 + \, [\alpha] 
  \Big\langle
 \mathcal{O}^{\barq}_{\rm nlo} \, 
 \bigg[
 Q_q^2\lb G^{(5,2)}_{q^2} -\Big(\frac32 - 2 \log \Big( \frac{E_5}{E_c}\Big) \Big) \, \log \Big(\frac{\eta_{51}}{1-\eta_{51}}\Big)\rb
\nn\\
&
+ \, Q_e^2 \;  G_{e^2} 
+ 2 \, Q_e Q_q \lb G^{(5,2)}_{eq} - \log \Big(\frac{E_5}{E_c}\Big) \log\Big(\frac{s_{35}}{s_{45}}\Big) \rb
\bigg]
 \nn\\
& \hspace{35mm}\times F_{\rm LM}(1_{g},2_{\bar{q}},3,4| 5_\barq) \Big\rangle
\\
 &
 + \, 
[\alpha] \,Q_q^2 
\int_{0}^1 dz \, 
\tilde{P}_{qq}^{\rm NLO}(z,E_c) \, 
\Big\langle
\mathcal{O}^{\bar{q}}_{\rm nlo}\, 
\frac{F_{\rm LM}(1_g,z\cdot2_{\bar{q}},3,4 | 5_{\bar{q}})}{z}
\Big\rangle
\nn\\
&
+
[\alpha_{s}] \,T_R 
\int_{0}^1 dz \, 
\Big\langle  
 \mathcal{O}^{\gamma}_{\rm nlo}\, 
 \Big[
 \tilde{P}^{\rm NLO}_{\bar{q}g} (z, E_c)
 +
 \tilde{\omega}_{5 \parallel 1}^{65}  \,
 \log \Big(\frac{\eta_{16}}{2}\Big) \, 
\bar{P}_{\barq g}^{\rm AP, 0}(z)
\Big] 
 \nn
 \\
& \hspace{35mm} \times
 \,  \frac{F_{\rm LM}(z \cdot 1_{q},2_{\bar{q}},3,4 |6_\gamma)}{z} \Big\rangle \; .
 \nn
\end{align}
Here, $\mathcal{O}^{\barq}_{\rm nlo} = 1 -C_{15}$,
$\mathcal O^\gamma_{\rm nlo}$ is defined in Eq.~\eqref{eq:onloa}
and $\tilde{\omega}_{5 \parallel 1}^{65} = C_{5 1} \, \omega^{65}$.
The relevant splitting functions are 
\begin{equation}
\label{eq:Ptildegq}
\begin{split}
&
\tilde{P}^{\rm NLO}_{\bar{q}g} (z, E_c) 
=
2 \bar{P}_{\barq g}^{\rm AP, 0}(z) \log(1-z)
+2 z (1-z) 
+ \bar{P}_{\barq g}^{\rm AP, 0}(z) \log \Big(\frac{4 E_c^2}{\mu^2} \Big) \; ,
\\
&
\bar{P}_{\barq g}^{\rm AP, 0}(z) =(1-z)^2+z^2 \, .
\end{split}
\end{equation}
Finally, we have
\begin{equation}
\label{eq:Gqsq_ij}
\begin{split}
G^{(i,j)}_{q^2}
 = &
 \frac{13}{2}
 - \pi^2
 + 2 {\rm Li}_2\big( 1-\eta_{ij}\big)
  +\log^2 \Big( \frac{E_i}{E_c} \Big)
 + \frac32 \log \Big( \frac{E_c^2}{\mu^2}\Big) \\
 &
 +\bigg(
 3
 -2  \log \Big( \frac{E_i}{E_c} \Big)
 \bigg) \, \log\big( 4 \, \eta_{ij}\big) \; .
 \end{split}
 \end{equation}
The functions $G_{e^2}$ and $G_{eq}$ have already been defined for the
$q\barq$ channel in Eq.~\eqref{eq:Gesq_qqb}.

In Eqs.~\refeqs{eq:gqboost}{eq:gqOnlo} we define $Q_q$ 
as minus the electric charge of the initial state anti-quark, i.e. $Q_q = \{-1/3, 2/3\}$
for the down and up quarks respectively.
In order to obtain the results for the 
$g q$ channel it is sufficient to 
flip the sign of the quark electric charge, \ie $Q_q \rightarrow -Q_q$, 
and apply the replacement $\bar{q}\to q$ 
inside the relevant matrix elements. 
As for the $q g$ channel one can start from 
Eqs.~\refeqs{eq:gqboost}{eq:gqOnlo} and follow a simple set of rules. 
In practice, for Eq.~\eqref{eq:gqboost} it is enough to consider
\begin{equation*}
F_{\rm LM(V)}(z\cdot 1_q, 2_{\bar{q}}) \to F_{\rm LM(V)}(1_q, z\cdot 2_{\bar{q}}), \quad
\tilde{P}^{\NLO}_{ij}(z_1,E_c) \leftrightarrow \tilde{P}^{\NLO}_{ij}(z_2,E_c), \quad
s_{1k} \to s_{2k} ,
\end{equation*}
whereas for Eq.~\eqref{eq:gqOnlo}, one has
\begin{gather*}
F_{\rm LM(V)}(z\cdot 1_q, 2_{\bar{q}}) \to F_{\rm LM(V)}(1_q, z\cdot 2_{\bar{q}}), \quad
F_{\rm LM(V)}(1_g, z\cdot 2_{\bar{q}}) \to F_{\rm LM(V)}(z\cdot 1_q, 2_g), \quad \\
G^{(5,2)}_{q^2} \to G^{(5,1)}_{q^2} \,, \quad
\eta_{k1} \to \eta_{k2}\,, \quad
\tilde{\omega}_{5 \parallel 1}^{65} \to \tilde{\omega}_{5 \parallel 2}^{65} \, .
\end{gather*}

\subsection{The $\gamma\bar{q}$ and $q \gamma$ channels}
\label{sec:finite_parts_aqbar}
The partonic channels induced by photon-(anti)quark scattering receive
contributions from two different configurations: one where an
intermediate $Z/\gamma$ decays into leptons, and one where the leptons are
produced directly from the initial state photon.  The IR structure of
the first configuration is similar to the $g q (\barq)$ case.  We then
expect the final formulas to be similar to
Eqs.~\refeqs{eq:gqboost}{eq:gqOnlo}, upon setting $Q_e \rightarrow 0$,
$T_R \rightarrow N_c C_F$, and replacing the gluon with the photon in
all the relevant matrix elements.  The contribution of the direct
lepton production results in additional terms proportional to Born-
and NLO-level boosted matrix elements.  For the sake of simplicity we
focus on the $\gamma(1) \barq(2) \rightarrow \ell^-\ell^+g(5)\barq(6)$
process.  In order to disentangle all the relevant collinear
singularities, we introduce the partition
\begin{equation}
1 = \omega^{51,61}+ \omega^{52,62} + \omega^{52,61} + \omega^{51,62} \, ,
\label{eq:partition_aq_temp}
\end{equation}
where the definition of the $\omega^{5i,6j}$ functions can be found in 
Ref.~\cite{Caola:2017dug}.
We then write the final result as 
\begin{equation}
d\hat{\sigma}_{\rm nnlo}^{\gamma \bar{q}}
=
d\hat{\sigma}_{\rm bt}^{\gamma \bar{q}}
+
d\hat{\sigma}_{\mathcal{O}_{\rm nlo}}^{\gamma \bar{q}}
+
d\hat{\sigma}_{\rm reg}^{\gamma \bar{q}}\,.
\label{eq:gammaqbar_finite_contributions}
\end{equation}
The regulated contribution reads
\begin{equation}
  2s\cdot d\hat{\sigma}_{\rm reg}^{\gamma \bar{q}} =
  \left\langle(1-S_5)\Omega_{\gamma\barq} F_{\rm LM}(1_\gamma,2_{\bar
    q},3_{\ell^-},4_{\ell^+},5_{g},6_{\bar q})\right\rangle,
\end{equation}
with
\begin{equation}
  \begin{split}
  \Omega_{\, \gamma\bar{q}} 
  &=
(1-C_{56, 1})  (1-C_{61}) \, \omega^{51,61}  \, \tilde\theta_A 
+ 
 (1-C_{56, 1})  (1-C_{56}) \,  \omega^{51,61} \, \tilde\theta_B
  \\
 & 
 + \,
 (1-C_{56, 1})   \,  \omega^{51,61} \, \tilde\theta_C
 +
  (1-C_{56, 1})  (1-C_{56})  \,  \omega^{51,61} \, \tilde\theta_D
  \\
  &
  + \, 
  (1-C_{56,2})\, \omega^{52,62}  \, \tilde\theta_A 
  +
  (1-C_{56,2})  (1-C_{56}) \,  \omega^{52,62} \, \tilde\theta_B 
   \\
  & 
  + \, 
  (1-C_{56,2})  (1-C_{52}) \,  \omega^{52,62} \, \tilde\theta_C  
  +
  (1-C_{56,2})  (1-C_{56})  \,  \omega^{52,62} \, \tilde\theta_D
   \\
  & 
   + \omega^{51,62} +
   (1-C_{52})(1-C_{61})\omega^{52,61} \;,
   \end{split}
\end{equation}
and $\tilde\theta_I$ defined in Eq.~\eqref{eq:thetatilde}.

The \it{boost} contribution reads
\begin{equation}
\begin{split}
2s \, \cdot \, & d\hat{\sigma}_{\rm bt}^
{\gamma \bar{q}}
= \,
[\alpha] \, [\alpha_s] \, 
Q_q^2 \, N_c \, C_F \,  
\int_0^1 dz_1 dz_2 \,\tilde{P}^{\rm NLO}_{\bar{q}g} (z_1, E_c)
\\
& \hspace{20mm} \times 
\Big\langle \frac{F_{\rm LM} (z_1 \cdot 1_{q},z_2 \cdot 2_{\bar{q}},3,4))}{z_1 \, z_2} \Big\rangle \, 
\tilde{P}_{qq}^{\rm NLO} (z_2, E_c)
\\
&
+ [\alpha] \,Q_q^2 \, N_c \, 
\int_{0}^1 dz \, 
\tilde{P}^{\rm NLO}_{\bar{q}g} (z, E_c) \, \Big\langle
 \frac{F_{\rm LV}^{\rm (QCD), \, fin}(z \cdot 1_{q},2_{\bar{q}},3,4)}{z} \Big\rangle
\\
&
+  \, 
[\alpha] \, [\alpha_s]  \,Q_q^2 \, C_F
\int_{0}^1 dz \, 
\bigg[ N_c \,  
P^{{\rm NNLO}, \, s}_{\gamma\bar{q}}(z, E_c)
\Big\langle
\frac{F_{LM}(z \cdot 1_{q},2_{\bar{q}},3,4)}{z} \Big\rangle
\\
& \hspace{20mm}
+  \, 
P^{{\rm NNLO}, \, t}_{\gamma\bar{q}}(z, E_c)
\Big\langle
\frac{F_{\rm LM}(1_{\gamma},z\cdot2_{\gamma},3,4)}{z} \Big\rangle
 \bigg] \; ,
 \end{split}
 \end{equation}
where the first term in squared brackets is the same as for the
$g\barq$ channel, i.e. $P^{{\rm NNLO}, \, s}_{\gamma\bar{q}} =
P_{\barq g}^{\rm NNLO}$, while $P^{{\rm NNLO}, \, t}_{\gamma\bar{q}}$
stems from the direct lepton production and is reported in
Eq.~\eqref{eq:PNNLO_t_aq}.  The $\mathcal{O}_{\rm nlo}$ contributions
is
\begin{equation}
\label{eq:ONLO_gammaqbar}
\begin{split}
2s \, \cdot \, & d\hat{\sigma}_{\mathcal{O}_{\rm nlo}}^{\gamma \bar{q}}
= 
\Big\langle
 \mathcal{O}_{\rm nlo}^\barq \;  F_{\rm LV}^{\rm (QCD), \, fin}(1_{\gamma},2_{\bar{q}},3,4|6_\barq)
 \Big\rangle \\
 &
 + [\alpha_s] \,C_F \;
 \Big\langle
 \mathcal{O}_{\rm nlo}^\barq \; 
 \Big[
  G^{(6,2)}_{q^2}
 - \sum_{i=1}^2 \bigg(
 \frac32
 - 2  \log \Big( \frac{E_6}{E_c} \Big)
 \bigg) \, \log \Big( \frac{\eta_{i6}}{1-\eta_{i6}}\Big) \, 
  \tilde{\omega}^{5i,6i}_{5\parallel6}  
  \Big]
  \\
  &  \hspace{25mm}
\; \times
  F_{\rm LM}(1_{\gamma},2_{\bar{q}},3,4|6_\barq)
 \Big\rangle
 \\
 &  
 +
 [\alpha] \, Q_q^2 \, N_c 
 \int_0^1 dz \, 
  \Big\langle
  \mathcal{O}_{\rm nlo}^{g} \; 
  \Big[
  \tilde{P}^{\rm NLO}_{\bar{q}g} (z, E_c) 
 \\
& \hspace{25mm}
  + \, 
  \tilde{\omega}^{51,61}_{6\parallel1} \,
  \log \Big( \frac{\eta_{51}}{2}\Big) \; 
  \bar{P}_{g\barq}^{\rm AP,0} (z)
  \Big] \, 
 \frac{F_{\rm LM}(z \cdot 1_{q},2_{\bar{q}},3,4|5_g)}{z} \Big\rangle
 \\
 &
 +
 [\alpha_s] \, C_F
 \int_0^1 dz \, 
 \Big\langle
   \mathcal{O}_{\rm nlo}^{\barq} \; 
   \Big[
  \tilde{P}_{qq}^{\rm NLO} (z, E_c) 
 \\
  & \hspace{25mm}
  + \, 
    \tilde{\omega}^{52,62}_{5\parallel2}  \; 
   \log \Big( \frac{\eta_{62}}{2}\Big) \,
\bar{P}_{qq, \rm R}^{\rm AP,0} (z)
\Big]
 \frac{F_{\rm LM}(1_\gamma,z\cdot2_{\bar{q}},3,4|6_\barq)}{z} \Big\rangle \; ,
\end{split}
\end{equation}
where we have introduced $\tilde{P}^{\rm NLO}_{\bar{q}g}$ and
$\bar{P}_{\barq g}^{\rm AP, 0}$ in Eq.~\eqref{eq:Ptildegq},
$G^{(6,2)}_{q^2}$ in Eq.~\eqref{eq:Gqsq_ij}, and $\bar{P}_{qq, \, \rm
  R}^{\rm AP, 0}$ in Eq.~\eqref{eq:PqqNLOtilde}.  One can obtain
results for the $\gamma q$, $q\gamma$ and $\barq \gamma$ channels
following the discussion at the end of
Sec.~\ref{sec:finite_parts_gqbar}.

\subsection{The $\gamma g$ and $g \gamma$ channels}
\label{sec:finite_parts_ag}
The NNLO corrections to the cross sections in the 
$\gamma g$ and $g \gamma$ partonic channels are affected only
by collinear singularities that cancel upon combining
real corrections and PDF renormalization. In order
to regularize real radiations, we introduce the same
phase space partition as in Eq.~\eqref{eq:partition_aq_temp}.
The subtraction then proceeds as usual.
The final result for the $\gamma g$ channel
can be cast in the following form
\begin{equation}
d\hat{\sigma}_{\rm nnlo}^{\gamma g}
=
d\hat{\sigma}_{\rm bt}^{\gamma g}
+
d\hat{\sigma}_{\mathcal{O}_{\rm nlo}}^{\gamma g}
+
d\hat{\sigma}_{\rm reg}^{\gamma g}\,.
\label{eq:gammag_finite_contributions}
\end{equation}
The regulated part reads
\begin{equation}
  2s\cdot d\hat{\sigma}_{\rm reg}^{\gamma g} = \left\langle
  \Omega_{\gamma g} F_{\rm
    LM}(1_\gamma,2_g,3_{\ell^-},4_{\ell^-},5_q,6_{\bar q})\right\rangle,
\end{equation}
where
\begin{equation}
  \begin{split}
    \Omega_{\gamma g} & =
    (1-C_{51}-C_{61}) \, \omega^{51,61}  
     \\
     &
  + \, 
  (1-C_{56,2})\,  (1-C_{62})\, \omega^{52,62}  \, \tilde\theta_A 
  +
   (1-C_{56,2})  (1-C_{52}) \,  \omega^{52,62} \, \tilde\theta_C 
   \\
  & 
  + (1-C_{62}) \,(1-C_{51}) \, \omega^{51,62} 
   + (1-C_{61}) \,(1-C_{52}) \, \omega^{52,61}
    \\
   &
   +  \, 
   (1-C_{56,2})\, \omega^{52,62}  \, \tilde\theta_B 
  +
   (1-C_{56,2})\, \omega^{52,62} \, \tilde\theta_D 
     \; , 
  \end{split}
\end{equation}
with $w^{5i,6j}$ and $\tilde\theta$ as in Sec.~\ref{sec:finite_parts_aqbar}.

The other relevant definitions are
\begin{equation}
\label{eq:finite_ag}
\begin{split}
2s \, \cdot  \, & d\hat{\sigma}_{\rm bt}^{\gamma g}
 = 
[\alpha] \, [\alpha_s] \, 
Q_q^2 \, T_R \, N_c \,  
\int_0^1 dz_1 dz_2 \,
\tilde{P}^{\rm NLO}_{\bar{q}g} (z_1, E_c) \, 
\tilde{P}_{\bar{q}g}^{\rm NLO} (z_2, E_c)
\\
& \quad \times 
\bigg[
 \Big\langle \frac{F_{\rm LM} (z_1 \cdot 1_{q},z_2 \cdot 2_{\bar{q}},3,4))}{z_1 \, z_2} \Big\rangle
 + 
  \Big\langle \frac{F_{\rm LM} (z_1 \cdot 1_{\barq},z_2 \cdot 2_q,3,4))}{z_1 \, z_2} \Big\rangle 
 \bigg] 
\\
& 
 + \,
[\alpha] \, [\alpha_s] \,Q_q^2 \, T_R \, 
\int_{0}^1 dz \, 
P^{\rm NNLO}_{\gamma g}(z, E_c) \, 
  \Big\langle
 \frac{F_{\rm LM}(1_{\gamma},z\cdot2_{\gamma},3,4)}{z} \Big\rangle \; ,
 \\
2s \, \cdot \, & d\hat{\sigma}_{\mathcal{O}_{\rm nlo}}^{\gamma g}
  =
  [\alpha] \, Q_q^2 \, N_c 
 \int_0^1 dz \, 
  \tilde{P}^{\rm NLO}_{\bar{q}g} (z, E_c) 
\\
& \qquad \times \, 
  \bigg[
  \Big\langle
  \mathcal{O}_{\rm nlo}^{q} \; 
  \frac{F_{\rm LM}(z \cdot 1_{q},2_g,3,4|5_q)}{z} \Big\rangle
  +
    \Big\langle
  \mathcal{O}_{\rm nlo}^{\barq} \; 
  \frac{F_{\rm LM}(z \cdot 1_{\barq},2_g,3,4|6_\barq)}{z} \Big\rangle
  \bigg]
  \\
   & \quad
   + \, 
 [\alpha_s] \,T_R
 \int_0^1 dz \, 
  \Big\langle
  \mathcal{O}_{\rm nlo}^{\barq} \; 
  \Big[
  \tilde{P}^{\rm NLO}_{\bar{q}g} (z, E_c) +
  \\
 & \hspace{25mm}
  + \, 
  \tilde{\omega}^{52,62}_{6\parallel2} \,
  \log \Big( \frac{\eta_{52}}{2}\Big) \; 
  \bar{P}_{g\barq}^{\rm AP,0} (z)
  \Big] \, 
 \frac{F_{\rm LM}(1_\gamma,z\cdot2_q,3,4|5_q)}{z} \Big\rangle
 \\
   & \quad
   + \, 
 [\alpha_s] \,T_R
 \int_0^1 dz \, 
  \Big\langle
  \mathcal{O}_{\rm nlo}^{\barq} \; 
  \Big[
  \tilde{P}^{\rm NLO}_{\bar{q}g} (z, E_c) +
   \\
  & \hspace{25mm}
  + \, 
  \tilde{\omega}^{52,62}_{5\parallel2} \,
  \log \Big( \frac{\eta_{62}}{2}\Big) \; 
  \bar{P}_{g\barq}^{\rm AP,0} (z)
  \Big] \, 
 \frac{F_{\rm LM}(1_\gamma,z\cdot2_\barq,3,4|6_\barq)}{z} \Big\rangle \; , 
 \end{split}
\end{equation}
where $P^{\rm NNLO}_{\gamma g}(z, E_c)$ is defined in
Eq.~\eqref{eq:PNNLO_ag}. One can deduce the result for the $g \gamma$
channel by taking Eq.~\eqref{eq:finite_ag} and swapping the indices 1 and 2.

\subsection{The $q \barq \rightarrow \ell^- \ell^+ q \barq$ and  $q q$ channels}
\label{sec:finite_parts_qq_qq}

The $q \barq \rightarrow \ell^- \ell^+ q \barq$ and $q q \rightarrow
\ell^- \ell^+ q q$ partonic processes are only affected by triple
collinear singularities, that are compensated by the PDF
renormalization contribution.  The phase space partition that we
choose for both channels reads
\begin{equation}
1 = w^{51,61}+w^{52,62} = \frac{\eta_{52}}{\eta_{51}+\eta_{52}}
+
\frac{\eta_{51}}{\eta_{51}+\eta_{52}} \, .
\end{equation}
For both channels we write
\begin{equation}
  \label{eq:xsect_qqb_into_qqb_decomp}
  \begin{split}
    d\hat{\sigma}_{\rm nnlo}^{qq}
    =
    d\hat{\sigma}_{\rm bt}^{qq}
    +
    d\hat{\sigma}_{\rm reg}^{qq},
    ~~~~~~~~~~~~
    d\hat{\sigma}_{{\rm nnlo},q\bar q}^{q\bar q}
    =
    d\hat{\sigma}_{{\rm bt},q\bar q}^{q\bar q}
    +
    d\hat{\sigma}_{{\rm reg},q\bar q}^{q\bar q}\,.
  \end{split}
\end{equation}
In the $q\bar q\to \ell^-\ell^+q\bar q$ channel we have
\begin{equation}
  2s \cdot d\hat{\sigma}_{{\rm reg},q\bar q}^{q\bar q} =
  \left\langle
  \sum_{i=1}^2 (1-C_{56,i})\omega^{5i,6i}
  F_{\rm LM}(1_q,2_{\bar q},3_{\ell^-},4_{\ell^+},5_{q},6_{\bar q})
  \right\rangle\,,
\end{equation}
and
\begin{equation}
\label{eq:qqb_qqb_channel_dec}
\begin{split}
2s \cdot d\hat{\sigma}_{{\rm bt},q\bar q}^{q\bar q}
= \, &
 [\alpha] \, 
[\alpha_s] \,
2 \, Q_q^2\, C_F \, 
\int_0^1 dz \, 
P^{\rm NNLO}_{q\bar q \rightarrow q\bar q}(z, E_c) 
\\ & \times \, 
\bigg[
\Big\langle
\frac{F_{\rm LM}(z\cdot1_q,
2_{\barq},3,4 )}{z}
\Big\rangle
+
\Big\langle
\frac{F_{\rm LM}(1_q,z\cdot2_\barq,3,4)}{z}
\Big\rangle
\bigg] \; ,
\end{split}
\end{equation}
with $P^{\rm NNLO}_{q\bar q\to q\bar q}$ defined in
Eq.~\eqref{eq:PNNLO_qqbar_qqbar}.

For the $qq\to\ell^-\ell^+qq$ channel, we have instead
\begin{equation}
  2s \cdot d\hat{\sigma}_{{\rm reg}}^{qq} =
  \left\langle
  \big[(1-C_{56,1})\omega^{51,61} + \omega^{52,62}\big]
  F_{\rm LM}(1_q,2_{q},3_{\ell^-},4_{\ell^+},5_{q},6_{q})
  \right\rangle\,,
\end{equation}
and
\begin{equation}
\label{eq:qq_qq_channel_dec}
\begin{split}
2s \cdot d\hat{\sigma}_{\rm bt}^{qq}
= \, &
 [\alpha] \, 
[\alpha_s] \,
2 \, Q_q^2\, C_F \, 
\int_0^1 dz \, 
P^{\rm NNLO}_{qq \rightarrow qq}(z, E_c) 
\Big\langle
\frac{F_{\rm LM}(z\cdot1_{\barq},2_q,3,4 )}{z}
\Big\rangle\,,
\end{split}
\end{equation}
with $P^{\rm NNLO}_{qq \rightarrow qq}$ given in Eq.~\eqref{eq:PNNLO_qq_qq}.

\section{Splitting functions}
\label{sec:splittings}
In this section we collect the splitting functions
that we have introduced in the main text. 
We begin by defining the functions that 
describe the hard-collinear integrated counterterm. 
For initial state and final state emission we have, respectively,
\begin{equation}
\label{eq:PqqNLO}
\begin{split}
P_{qq}^{\rm NLO}(z, L_i) 
= \, &
\bigg[
(1-z)^{-2\eps} 
P_{qq} (z)
+ \frac1\eps \, 
e^{-2\eps L_i}   \,
 \delta(1-z)
\bigg] \; , 
\\ 
P_{qq}^{\rm NLO}(L_i)
= \, &
\gamma_{ee}^{22}
+ \frac{1}{\eps} 
\bigg(1 - e^{- 2 \eps L_i} \bigg)
\; ,
\end{split}
\end{equation}
where 
\beq
\label{eq:gamma_ee_22}
\gamma_{ee}^{22} 
=
- \int_0^1 dz \; 
\bigg\{
\big[ z(1-z)\big]^{-2\eps}
P_{qq} \big(z\big) 
- 2 \,  \frac{ (1-z)^{-2\eps}}{1-z}
\bigg\} \; .
\eeq
The LO Altarelli-Parisi splitting functions we have used are
\begin{equation}
\label{eq:PbarqgAP0}
\begin{split}
\bar{P}_{qq}^{\rm AP,0} (z) 
= \, &
2 \mathcal{D}_0(z)
-(1+z)
+\frac32 \, \delta(1-z) \, ,
\\
\bar{P}_{\bar{q} g}^{\rm AP, 0}(z)
= \, & 
(1-z)^2+z^2 \, ,
\\ 
\bar{P}_{q\gamma}^{\rm AP, 0} (z)
= \, & 
\frac{1+(1-z)^2}{z} \, ,
\end{split}
\end{equation}
where the regular part of $\bar{P}_{qq}^{\rm AP,0}$ is equal to
\beq
\bar{P}_{qq, R}^{\rm AP, 0}(z) = 2 \mathcal{D}_0(z)-(1+z) \, .
\eeq
In order to describe the single-collinear limits of the 
$q\barq \rightarrow \ell^-\ell^+(g\gamma)$ we compute
\begin{equation}
\begin{split}
\big[
\bar{P}_{qq}^{\rm AP,0} 
\otimes 
\bar{P}_{qq}^{\rm AP,0}
\big] (z)
= \, &
6 \mathcal{D}_0(z)
+ 8 \mathcal{D}_1(z)
+ \Big(\frac94 -\frac{2\pi^2}{3} \Big) \, \delta(1-z)
- \frac{\big(3z^2+1\big) \log(z)}{1-z}
\\ &
-z-4(z+1) \, \log(1-z)
-5 \, .
\end{split}
\end{equation}
At one-loop level, the Altarelli-Parisi splitting function 
for process $q\rightarrow qg$ reads
\begin{equation}
\begin{split}
\bar{P}_{qq}^{\rm AP,1} (z) 
=& 3 - 2 z + 
2 \left[1 - \frac{1 + z^2}{1 - z} \log(1 - z)\right] \log z +
\frac{1 + 3 z^2}{2 (1 - z)} \log^2 z
\\
&+ \frac{2 (1 + z^2)}{1 - z} \,{\rm Li}_2(1 - z) + 
   \delta(1 - z) \left( \frac{3}{8} - \frac{\pi^2}{2} + 6 \zeta_3\right).
\end{split}
\end{equation}
The real-virtual splitting is 
\begin{equation}
\label{eq:Pqq_RV}
\begin{split}
\mathcal{P}_{qq}^{{\rm loop}, \, RV}(z) 
= \, &
\frac1\eps \frac{1+z^2}{1-z} \, \log z
\\ &
- \frac{1+z^2}{1-z}
	\Big( {\rm Li}_2(1-z) 
	+3 \log(1-z) \log z \Big)
- \frac z2
- (1-z) \log z
\\ &
- \bigg[
 \frac12 
	\Big(
	1
	+ z
	- 3 z \log(1-z)
	\Big)
- (1-z) \Big(
	3 \log(1-z) \, \log z
\\ & \quad 
+ {\rm Li}_2(1-z)
	\Big)
- \frac12 \, \frac{1+z^2}{1-z}
	\Big(
	9 \log^2(1-z) \, \log z 
\\ & \qquad 
	+ 6 \log (1-z) \,{\rm Li}_2(1-z)
	- 2 {\rm Li}_3(1-z) 
	\Big)
\bigg] \eps 
+ \mathcal{O}(\eps^2)\, .
\end{split}
\end{equation}
To express the finite contributions to the $q\barq \rightarrow
\ell^-\ell^+ (\gamma g)$ channel 
partonic cross sections we have introduced 
the NLO splitting function 
\begin{equation}
\begin{split}
\tilde{P}_{qq}^{\rm NLO} (z, E_c) &=\, 
4 \mathcal{D}_1(z) -2 (1+z) \log(1-z) +(1-z) 
\\ 
& + 2 \log\Big(\frac{2E_c}{\mu} \Big) \big( 2 \mathcal{D}_0(z)-(1+z)\big) \; .
\end{split}
\end{equation}
The finite contributions to the different partonic 
channels depend on collinear functions that multiply 
boosted matrix elements.
For the $q\barq \rightarrow
\ell^-\ell^+ (\gamma g)$ channel we have
\begin{align}
\label{pqqnnlo}
P_{qq}^{\rm NNLO}(z, E_c) 
\, & =   
\mathcal{D}_0(z) \Big[48 \log ^2\Big(\frac{2 E_c}{\mu }\Big)-\frac{16}{3} \log ^3(z)+32 \zeta_3 \Big]
 + \log \Big(\frac{2 E_c}{\mu }\Big)\Big[48
   \mathcal{D}_1(z)
\nn\\
  &-8 (z+1) \text{Li}_2(z)
  -\frac{8(z^2+1) \text{Li}_2(1-z)}{z-1}
  -20 z+\frac{4}{3} \pi ^2
   (z+1)+24\Big]
\nn   \\
   & \, 
   -16 (-2 \mathcal{D}_1(z)+z+2) \log ^2\Big(\frac{2
   E_c}{\mu }\Big)
   +\mathcal{D}_2(z) \Big(48 \log \Big(\frac{2 E_c}{\mu }\Big)-16 \log (z)\Big)
\nn   \\
   &
   +\log ^2(z) \Big[-8 \mathcal{D}_1(z)+2 (z+1) \log \Big(\frac{2 E_c}{\mu
   }\Big)-z+2\Big]
\nn   \\
   &
   +16 \mathcal{D}_3(z)
   +\log (1-z) 
   \Big[-\frac{\log (z)}{1-z} \Big(8 (z^2+1) \log \Big(\frac{2 E_c}{\mu
   }\Big)
\nn   \\
   &
   -2 (5 z^2+2 z+5)\Big)
   -16 (z+1) \log ^2\Big(\frac{2 E_c}{\mu }\Big)
   -16 (z+2) \log \Big(\frac{2 E_c}{\mu }\Big)
\nn   \\
   &
    -\frac{6 (z^2+1) \text{Li}_2(1-z)}{z-1}
   -8 (z+1)
   \text{Li}_2(z)-9 z+\frac{4}{3} \pi ^2 (z+1)
   \\
   &  +4 (z+1) \log ^2(z)+12\Big]
   +\log (z) \Big[\frac{4 (3 z^2+1)
   \log ^2\Big(\frac{2 E_c}{\mu }\Big)}{z-1}
\nn   \\
   &
   -\frac{8 (z^2+z+1) \log \Big(\frac{2 E_c}{\mu
   }\Big)}{z-1}
   -\frac{4 (z^2+1) \text{Li}_2(z)}{z-1}
   +\frac{\pi ^2 (8 z^2+8)}{3-3 z}-9
   z+5\Big]
\nn   \\
   &
   +\log ^2(1-z) \Big[-24 (z+1) \log \Big(\frac{2 E_c}{\mu }\Big)+\frac{(5 z^2+13) \log
   (z)}{1-z}+4 (z-1)\Big]
\nn   \\
   &
   +\frac{(10-6 z^2) \text{Li}_3(1-z)}{z-1}+\frac{4 (3 z^2+5)
   \text{Li}_3(z)}{z-1}
    +(2-2 z) \text{Li}_2(1-z)
\nn   \\
   &
   +4 (z-1)\text{Li}_2(z)-\frac{4 (7 z^2+1) \zeta_3}{z-1}
   +\frac{(13 z^2+47) \log ^3(z)}{6-6 z}-\frac{10}{3} \pi ^2 (z-1)
\nn   \\
   &
   +2 (8 z-9)
   -8 (z+1) \log ^3(1-z)
   \; .
\nn   
\end{align}
When discussing the finite remainders for the $g\bar{q}$ and $q g$ channels we introduced
\begingroup
\begin{align}
\label{eq:PqgNNLO}
P_{\barq g}^{\rm NNLO}(z, E_c) 
= \, &
\log \Big(\frac{2 E_c}{\mu }\Big) \bigg[
(8 z-4) \text{Li}_2(z)
+\frac{4}{3} \pi ^2 (z-1)^2
+z (30 z-41)
\nn \\
&
\quad 
+\big(12 (z-1) z+6\big) \log ^2(1-z)
+(1-2 z) \log ^2(z)
\nn \\
& 
\quad
-(8 (z-2) z+2) \log (1-z)
+(20 (z-1) z
\nn \\
&
\quad 
-(8 (z-1) z+4) \log (1-z)+3) \log (z)
+16\bigg]
\nn \\
& + \log^2 \Big(\frac{2 E_c}{\mu }\Big) \bigg[
4 z (3 z-2)-(8 z^2-4 z+2) \log (z) 
\\ 
& +(8 (z-1) z+4) \log (1-z)+5
\bigg] +\big(2 (9-5 z) z-9\big) \text{Li}_3(1-z)
\nn \\
& +\big(2 (9-7 z) z-9\big) \text{Li}_3(z) -49 z^2 +\text{Li}_2(z) \Big(\big(2 (z-1) z+1\big) \log (z)
\nn \\
& - (2 (z-5) z+5) \log (1-z) +3\Big) + \log ^3(z) \bigg[\frac{7 z^2}{3}-\frac{5 z}{2}+\frac{5}{4}\bigg]
\nn \\
& +\log ^2(z) \bigg[z^2 -\frac{z}{2} -\Big(5 (z-1)   z+\frac{5}{2}\Big) \log (1-z)-\frac{3}{8}\bigg]
\nn \\
& +\log (z) \bigg[ \frac{35 z}{4} +\frac{4}{3} \pi ^2 (2 (z-1) z+1)
-\Big((z-5) z+\frac{5}{2}\Big) \log ^2(1-z)
\nn \\
&  +(14 (z-1) z+6) \log (1-z)-8 z^2+1\bigg] 
\nn \\
   &
   +\log (1-z) \bigg[44 z^2+\pi ^2 \Big(z^2-\frac{7
   z}{3}+\frac{7}{6}\Big)-\frac{109 z}{2}+16\bigg] 
\nn \\
   &
   +8 (2 (z-1) z+1) \zeta_3
   +(4 z (4 z-5)+10) \zeta_3
   +\frac{255z}{4}
\nn \\
   &
   +\frac{1}{12} \pi ^2 \big(2 z (9 z-13)+11\big)
   +\frac{11}{6} \big(2 (z-1) z+1\big) \log ^3(1-z)
\nn \\
   &
   +\big((21-17 z) z-7\big) \log ^2(1-z)
   -\frac{69}{4} \; \nn , 
\end{align}
\endgroup
To present the results for the $\gamma\bar{q}$ and $q \gamma$ channels
we have defined the function 
\begin{align}
\label{eq:PNNLO_t_aq}
P^{{\rm NNLO}, \,  t}_{\gamma\bar{q}}(z, E_c)
=&
\log\Big(\frac{2 E_c}{\mu }\Big)
 \bigg[
 4 (z-2) \text{Li}_2(z)
 +\pi ^2 \Big(-2 z-\frac{8}{3 z}+4\Big)
 -4 z
 +\frac{10}{z}
\nn\\
& \qquad \qquad 
 +6 \Big(z+\frac{2}{z}-2\Big) \log ^2(1-z)
 +(2-z) \log ^2(z)
\nn\\
& \qquad \qquad  
 +\Big(-8 z-\frac{12}{z}+20\Big) \log (1-z)
 +(5 z+8) \log (z)
 -15
   \bigg]
\nn \\
 &
   +\log^2\Big(\frac{2 E_c}{\mu }\Big)
   \bigg[-z+\Big(4 z+\frac{8}{z}-8\Big) \log (1-z)+(4-2 z) \log
   (z)+4\bigg]
\nn \\
 &
   -(5 z+8) \text{Li}_2(z)
   +4 (z-2) \text{Li}_3(1-z)
   +(2 z-4) \text{Li}_3(z)
    \\
 &
  +\Big(6 z+\frac{16}{z}-12\Big) \zeta_3
   +\frac{9 \pi ^2z}{4}
   -\frac{29 z}{4}
   +\frac{5 \pi ^2}{2 z}
   -\frac{27}{z}
   -\frac{11 \pi ^2}{6}
   +\frac{73}{2}
\nn \\
 &
   + \frac{\log (1-z)}{6 z}
   	\bigg[24 (z-2) z  \, \text{Li}_2(z)
	-3 z (3 z+46)
	-2 \pi ^2 \big(7 (z-2) z+10\big)
\nn \\
	& +108\bigg]
   +\frac{13 \big((z-2) z+2\big) \log ^3(1-z)}{6 z}
   +\frac{1}{12} (2-z) \log ^3(z)
\nn   \\
   &
   +\frac{1}{4} \bigg(-27z-\frac{42}{z}+58\bigg) \log ^2(1-z)
   +\Big(\frac{17 z}{8}+\frac{5}{2}\Big) \log ^2(z)
\nn   \\
   &
+\Big(\frac{1}{4} (5 z-3)+2 (z-2) \log^2(1-z)\Big) \log (z) \; .
\nn
\end{align}
For the $\gamma g$ and $g \gamma$ channels we need
\begin{align}
\label{eq:PNNLO_ag}
P^{\rm NNLO}_{\gamma g}(z, E_c)
 = \, &
 \log (1-z) \bigg[-16 (z+1) \text{Li}_2(z)-\frac{8 (z-1) \big(4 z^2+7 z+4\big)}{3z} \log \Big( \frac{2 E_c}{\mu}\Big)
\nn \\
 & \quad 
   +\frac{4}{9 z} \big(38 z^3+6 \pi ^2 z(z+1))+39 z^2-57 z-20\big)\bigg]
\nn \\
 &
   +\text{Li}_2(z) \bigg[4 (3 z+1)-16 (z+1) \log \Big( \frac{2 E_c}{\mu}\Big)\bigg]
   -16 (z+1) \text{Li}_3(1-z)
\nn \\
 &  -8 (z+1) \text{Li}_3(z)
   -\frac{4 (z-1) \big(4 z^2+7 z+4\big)}{3 z} \,  \log^2 \Big( \frac{2 E_c}{\mu}\Big)
\nn \\
 & 
   +\frac{4}{9 z} \big(38 z^3+6 \pi ^2 z(z+1)+39 z^2-57
   z-20\big)  \log \Big( \frac{2 E_c}{\mu}\Big)
  \\
 &  
   +\log ^2(z) \Big[4 (z+1) \log \Big( \frac{2 E_c}{\mu}\Big)
   +\frac{1}{2} (-11 z-5)\Big]
\nn  \\
 & 
   +\log (z) \bigg[8 (z+1)
   \log^2 \Big( \frac{2 E_c}{\mu}\Big)
   -4 (3 z+1) \Big( \frac{2 E_c}{\mu}\Big)
   +2 (3 z+1)
\nn   \\
   & -8 (z+1) \log ^2(1-z)\bigg]
      -\frac{4 (z-1) \big(4 z^2+7 z+4\big) \log^2(1-z)}{3 z}
\nn \\
 &
   +8 (z+1) \zeta_3+\frac{1}{3} (z+1) \log ^3(z) +\frac{2}{27 z} \Big(12 \pi ^2 (z^3-1)-211 z^3
\nn \\
 &
-18 \pi ^2 z(z+1)-420 z^2+528 z+103\Big) \; .
\nn
\end{align}
Finally, we report the collinear functions appearing in the final
results for the $q \barq \rightarrow \ell^- \ell^+ (q \barq)$ and $q q
\rightarrow \ell^- \ell^+ (q q)$ channels.  We define respectively
\begin{align}
\label{eq:PNNLO_qqbar_qqbar}
P^{\rm NNLO}_{q\barq \rightarrow q\barq}(z, E_c)
= \, &
\frac{1+z^2}{1-z} \bigg[\log \Big(\frac{2 E_c}{\mu}\Big) \bigg(4 \text{Li}_2(z)-2 \log ^2(z)+4 \log (1-z) \log (z)-\frac{2 \pi ^2}{3}\bigg)
\nn\\
&
+6 \text{Li}_3(1-z)
+8 \text{Li}_3(-z)
+9 \text{Li}_3(z)
+4 \text{Li}_2(z) \log (1-z)
\nn   \\
   &
   -\log (z) \bigg(4 \text{Li}_2(-z)+3 \text{Li}_2(z)-4 \log ^2(1-z)+\frac{2 \pi ^2}{3}\bigg)
\nn   \\
   &
   -\frac{7}{6} \log ^3(z)-\frac{1}{2} \log (1-z) \log
   ^2(z)-\frac{2}{3} \pi ^2 \log (1-z)\bigg]
   \\
   &
   +\bigg(\frac{(10-4 z^2) \log (z)}{z-1}
   +2 (7 z-8)\bigg) \log \Big(\frac{2 E_c}{\mu}\Big)
   +\frac{(z^2+6 z-13) \, \text{Li}_2(z)}{z-1}
\nn   \\
   &
   +4 (z+1) \text{Li}_2(-z)
   +\frac{z^2 (6 \zeta_3+7)-6 z+6 \zeta_3-1}{2 (z-1)}
   +\frac{\pi ^2 \big(z^2-6 z+11\big)}{6 (z-1)}
\nn   \\
   &
   -\frac{\big(4 z^2+12 z-25\big) \log
   ^2(z)}{4 (z-1)}
   +2 (7 z-8) \log (1-z)
   -\Big(
   13 z^2 +19 z
\nn   \\
   &
   -8 (z^2-1) \log (z+1)
   +6 (z-1)^2 \log (1-z)
   -22
   \Big) \, 
   \frac{ \log (z)}{2 (z-1)}
   \; ,
\nn   
\end{align}
and 
\begin{align}
\label{eq:PNNLO_qq_qq}
P^{\rm NNLO}_{qq \rightarrow qq}(z, E_c)
= \, &
\log \Big(\frac{2 E_c}{\mu }\Big) 
\bigg[
\frac{1+z^2}{1+z}
\bigg(-8 \text{Li}_2(-z)
+2 \log ^2(z)
-8 \log (z+1) \log (z)\bigg)
\nn\\
& \qquad \qquad \qquad
-\frac{2 \pi ^2 \big(z^2+1\big)}{3(z+1)}
-8 (z-1)+4 (z+1) \log (z)\bigg]
\nn\\
&
+\frac{1+z^2}{1+z} \bigg[
-4 \text{Li}_3(1-z^2)
+8 \text{Li}_3(1-z)
-18 \text{Li}_3(-z)
-8 \text{Li}_3(z)
\nn\\
& \hspace{17mm}
-12 \text{Li}_3\Big(\frac{z}{z+1}\Big)
+\log (z) \Big(2 \text{Li}_2(-z)
+2 \text{Li}_2(z)
-6 \log ^2(z+1)\Big)
\nn\\
&  \hspace{17mm}
-8\log (1-z) \Big(\text{Li}_2(-z)+\log (z) \log
   (z+1)\Big) 
   +\frac{7 \log ^3(z)}{6}
\nn   \\
   &  \hspace{17mm}
   +2 \log ^3(z+1)
   -\log (z+1) \log ^2(z)
   -\pi ^2 \log (z+1)\bigg]  
   \\
   &
   -6 (z+1) \text{Li}_2(-z)
   -2 (z+3) \text{Li}_2(z)
   +\frac{5 (z^2+1) \zeta_3}{z+1}
\nn   \\
   &
   +\bigg(\frac{2 \pi ^2 (z^2+1)}{3 (z+1)} 
   +\frac{1}{2} (11 z+19)
   -6 (z+1) \log (z+1)\bigg) \log (z)
\nn   \\
   &
   -2\log (1-z)
   \bigg(\frac{\pi ^2 (z^2+1)}{3 (z+1)}
   +4 (z-1)
   - (z-1) \log (z)\bigg)
\nn   \\
   &
   +\frac{1}{6} \pi ^2 (3-z)
   -\frac{15 (z-1)}{2}+2 (z+2) \log ^2(z)    \; .
\nn   
\end{align}

\bibliographystyle{JHEP}
\bibliography{mixedll}

\providecommand{\href}[2]{#2}\begingroup\raggedright\begin{thebibliography}{100}

\bibitem{Drell:1970wh}
S.~D. Drell and T.-M. Yan, \emph{{Massive Lepton Pair Production in
  Hadron-Hadron Collisions at High-Energies}},
  \href{https://doi.org/10.1103/PhysRevLett.25.316}{\emph{Phys. Rev. Lett.}
  {\bfseries 25} (1970) 316}.

\bibitem{CMS:2021ctt}
{\scshape CMS} collaboration, \emph{{Search for resonant and nonresonant new
  phenomena in high-mass dilepton final states at $ \sqrt{s} $ = 13 TeV}},
  \href{https://doi.org/10.1007/JHEP07(2021)208}{\emph{JHEP} {\bfseries 07}
  (2021) 208} [\href{https://arxiv.org/abs/2103.02708}{{\ttfamily
  2103.02708}}].

\bibitem{ATLAS:2017fih}
{\scshape ATLAS} collaboration, \emph{{Search for new high-mass phenomena in
  the dilepton final state using 36 $\mathrm{fb}^{-1}$ of proton-proton
  collision data at $ \sqrt{s}=13 $ TeV with the ATLAS detector}},
  \href{https://doi.org/10.1007/JHEP10(2017)182}{\emph{JHEP} {\bfseries 10}
  (2017) 182} [\href{https://arxiv.org/abs/1707.02424}{{\ttfamily
  1707.02424}}].

\bibitem{Alioli:2017nzr}
S.~Alioli, M.~Farina, D.~Pappadopulo and J.~T. Ruderman, \emph{{Catching a New
  Force by the Tail}},
  \href{https://doi.org/10.1103/PhysRevLett.120.101801}{\emph{Phys. Rev. Lett.}
  {\bfseries 120} (2018) 101801}
  [\href{https://arxiv.org/abs/1712.02347}{{\ttfamily 1712.02347}}].

\bibitem{Buchmuller:1985jz}
W.~Buchmuller and D.~Wyler, \emph{{Effective Lagrangian Analysis of New
  Interactions and Flavor Conservation}},
  \href{https://doi.org/10.1016/0550-3213(86)90262-2}{\emph{Nucl. Phys. B}
  {\bfseries 268} (1986) 621}.

\bibitem{Grzadkowski:2010es}
B.~Grzadkowski, M.~Iskrzynski, M.~Misiak and J.~Rosiek, \emph{{Dimension-Six
  Terms in the Standard Model Lagrangian}},
  \href{https://doi.org/10.1007/JHEP10(2010)085}{\emph{JHEP} {\bfseries 10}
  (2010) 085} [\href{https://arxiv.org/abs/1008.4884}{{\ttfamily 1008.4884}}].

\bibitem{Peskin:1990zt}
M.~E. Peskin and T.~Takeuchi, \emph{{A New constraint on a strongly interacting
  Higgs sector}}, \href{https://doi.org/10.1103/PhysRevLett.65.964}{\emph{Phys.
  Rev. Lett.} {\bfseries 65} (1990) 964}.

\bibitem{Falkowski:2015krw}
A.~Falkowski and K.~Mimouni, \emph{{Model independent constraints on
  four-lepton operators}},
  \href{https://doi.org/10.1007/JHEP02(2016)086}{\emph{JHEP} {\bfseries 02}
  (2016) 086} [\href{https://arxiv.org/abs/1511.07434}{{\ttfamily
  1511.07434}}].

\bibitem{Farina:2016rws}
M.~Farina, G.~Panico, D.~Pappadopulo, J.~T. Ruderman, R.~Torre and A.~Wulzer,
  \emph{{Energy helps accuracy: electroweak precision tests at hadron
  colliders}},
  \href{https://doi.org/10.1016/j.physletb.2017.06.043}{\emph{Phys. Lett. B}
  {\bfseries 772} (2017) 210}
  [\href{https://arxiv.org/abs/1609.08157}{{\ttfamily 1609.08157}}].

\bibitem{Dawson:2021ofa}
S.~Dawson and P.~P. Giardino, \emph{{New Physics Through Drell Yan SMEFT
  Measurements at NLO}},  \href{https://arxiv.org/abs/2105.05852}{{\ttfamily
  2105.05852}}.

\bibitem{LHCb:2017avl}
{\scshape LHCb} collaboration, \emph{{Test of lepton universality with $B^{0}
  \rightarrow K^{*0}\ell^{+}\ell^{-}$ decays}},
  \href{https://doi.org/10.1007/JHEP08(2017)055}{\emph{JHEP} {\bfseries 08}
  (2017) 055} [\href{https://arxiv.org/abs/1705.05802}{{\ttfamily
  1705.05802}}].

\bibitem{LHCb:2019hip}
{\scshape LHCb} collaboration, \emph{{Search for lepton-universality violation
  in $B^+\to K^+\ell^+\ell^-$ decays}},
  \href{https://doi.org/10.1103/PhysRevLett.122.191801}{\emph{Phys. Rev. Lett.}
  {\bfseries 122} (2019) 191801}
  [\href{https://arxiv.org/abs/1903.09252}{{\ttfamily 1903.09252}}].

\bibitem{LHCb:2021trn}
{\scshape LHCb} collaboration, \emph{{Test of lepton universality in
  beauty-quark decays}},  \href{https://arxiv.org/abs/2103.11769}{{\ttfamily
  2103.11769}}.

\bibitem{Bifani:2018zmi}
S.~Bifani, S.~Descotes-Genon, A.~Romero~Vidal and M.-H. Schune, \emph{{Review
  of Lepton Universality tests in $B$ decays}},
  \href{https://doi.org/10.1088/1361-6471/aaf5de}{\emph{J. Phys. G} {\bfseries
  46} (2019) 023001} [\href{https://arxiv.org/abs/1809.06229}{{\ttfamily
  1809.06229}}].

\bibitem{Greljo:2017vvb}
A.~Greljo and D.~Marzocca, \emph{{High-$p_T$ dilepton tails and flavor
  physics}}, \href{https://doi.org/10.1140/epjc/s10052-017-5119-8}{\emph{Eur.
  Phys. J. C} {\bfseries 77} (2017) 548}
  [\href{https://arxiv.org/abs/1704.09015}{{\ttfamily 1704.09015}}].

\bibitem{Hamberg:1990np}
R.~Hamberg, W.~van Neerven and T.~Matsuura, \emph{{A Complete calculation of
  the order $\alpha-s^{2}$ correction to the Drell-Yan $K$ factor}},
  \href{https://doi.org/10.1016/0550-3213(91)90064-5}{\emph{Nucl.Phys.}
  {\bfseries B359} (1991) 343}.

\bibitem{vanNeerven:1991gh}
W.~L. van Neerven and E.~B. Zijlstra, \emph{{The $O(\alpha_s^2)$ corrected
  Drell-Yan $K$ factor in the DIS and MS scheme}},
  \href{https://doi.org/10.1016/0550-3213(92)90078-P}{\emph{Nucl. Phys. B}
  {\bfseries 382} (1992) 11}.

\bibitem{Harlander:2002wh}
R.~V. Harlander and W.~B. Kilgore, \emph{{Next-to-next-to-leading order Higgs
  production at hadron colliders}},
  \href{https://doi.org/10.1103/PhysRevLett.88.201801}{\emph{Phys. Rev. Lett.}
  {\bfseries 88} (2002) 201801}
  [\href{https://arxiv.org/abs/hep-ph/0201206}{{\ttfamily hep-ph/0201206}}].

\bibitem{Anastasiou:2003yy}
C.~Anastasiou, L.~J. Dixon, K.~Melnikov and F.~Petriello, \emph{{Dilepton
  rapidity distribution in the Drell-Yan process at NNLO in QCD}},
  \href{https://doi.org/10.1103/PhysRevLett.91.182002}{\emph{Phys. Rev. Lett.}
  {\bfseries 91} (2003) 182002}
  [\href{https://arxiv.org/abs/hep-ph/0306192}{{\ttfamily hep-ph/0306192}}].

\bibitem{Anastasiou:2003ds}
C.~Anastasiou, L.~J. Dixon, K.~Melnikov and F.~Petriello, \emph{{High precision
  QCD at hadron colliders: Electroweak gauge boson rapidity distributions at
  NNLO}}, \href{https://doi.org/10.1103/PhysRevD.69.094008}{\emph{Phys. Rev.}
  {\bfseries D69} (2004) 094008}
  [\href{https://arxiv.org/abs/hep-ph/0312266}{{\ttfamily hep-ph/0312266}}].

\bibitem{Melnikov:2006kv}
K.~Melnikov and F.~Petriello, \emph{{Electroweak gauge boson production at
  hadron colliders through O(alpha(s)**2)}},
  \href{https://doi.org/10.1103/PhysRevD.74.114017}{\emph{Phys. Rev.}
  {\bfseries D74} (2006) 114017}
  [\href{https://arxiv.org/abs/hep-ph/0609070}{{\ttfamily hep-ph/0609070}}].

\bibitem{Catani:2009sm}
S.~Catani, L.~Cieri, G.~Ferrera, D.~de~Florian and M.~Grazzini, \emph{{Vector
  boson production at hadron colliders: a fully exclusive QCD calculation at
  NNLO}}, \href{https://doi.org/10.1103/PhysRevLett.103.082001}{\emph{Phys.
  Rev. Lett.} {\bfseries 103} (2009) 082001}
  [\href{https://arxiv.org/abs/0903.2120}{{\ttfamily 0903.2120}}].

\bibitem{Gavin:2010az}
R.~Gavin, Y.~Li, F.~Petriello and S.~Quackenbush, \emph{{FEWZ 2.0: A code for
  hadronic Z production at next-to-next-to-leading order}},
  \href{https://doi.org/10.1016/j.cpc.2011.06.008}{\emph{Comput. Phys. Commun.}
  {\bfseries 182} (2011) 2388}
  [\href{https://arxiv.org/abs/1011.3540}{{\ttfamily 1011.3540}}].

\bibitem{Duhr:2020seh}
C.~Duhr, F.~Dulat and B.~Mistlberger, \emph{{Drell-Yan Cross Section to Third
  Order in the Strong Coupling Constant}},
  \href{https://doi.org/10.1103/PhysRevLett.125.172001}{\emph{Phys. Rev. Lett.}
  {\bfseries 125} (2020) 172001}
  [\href{https://arxiv.org/abs/2001.07717}{{\ttfamily 2001.07717}}].

\bibitem{Duhr:2020sdp}
C.~Duhr, F.~Dulat and B.~Mistlberger, \emph{{Charged current Drell-Yan
  production at N$^{3}$LO}},
  \href{https://doi.org/10.1007/JHEP11(2020)143}{\emph{JHEP} {\bfseries 11}
  (2020) 143} [\href{https://arxiv.org/abs/2007.13313}{{\ttfamily
  2007.13313}}].

\bibitem{Duhr:2021vwj}
C.~Duhr and B.~Mistlberger, \emph{{Lepton-pair production at hadron colliders
  at N$^3$LO in QCD}},  \href{https://arxiv.org/abs/2111.10379}{{\ttfamily
  2111.10379}}.

\bibitem{Chen:2021vtu}
X.~Chen, T.~Gehrmann, N.~Glover, A.~Huss, T.-Z. Yang and H.~X. Zhu,
  \emph{{Dilepton Rapidity Distribution in Drell-Yan Production to Third Order
  in QCD}}, \href{https://doi.org/10.1103/PhysRevLett.128.052001}{\emph{Phys.
  Rev. Lett.} {\bfseries 128} (2022) 052001}
  [\href{https://arxiv.org/abs/2107.09085}{{\ttfamily 2107.09085}}].

\bibitem{Camarda:2021ict}
S.~Camarda, L.~Cieri and G.~Ferrera, \emph{{Drell-Yan lepton-pair production:
  $q_T$ resummation at N$^3$LL accuracy and fiducial cross sections at
  N$^3$LO}},  \href{https://arxiv.org/abs/2103.04974}{{\ttfamily 2103.04974}}.

\bibitem{Chen:2022cgv}
X.~Chen, T.~Gehrmann, E.~W.~N. Glover, A.~Huss, P.~Monni, E.~Re, L.~Rottoli
  et~al., \emph{{Third order fiducial predictions for Drell-Yan at the LHC}},
  \href{https://arxiv.org/abs/2203.01565}{{\ttfamily 2203.01565}}.

\bibitem{Dittmaier:2001ay}
S.~Dittmaier and M.~Kr\"amer, \emph{{Electroweak radiative corrections to W
  boson production at hadron colliders}},
  \href{https://doi.org/10.1103/PhysRevD.65.073007}{\emph{Phys. Rev. D}
  {\bfseries 65} (2002) 073007}
  [\href{https://arxiv.org/abs/hep-ph/0109062}{{\ttfamily hep-ph/0109062}}].

\bibitem{Baur:2001ze}
U.~Baur, O.~Brein, W.~Hollik, C.~Schappacher and D.~Wackeroth,
  \emph{{Electroweak radiative corrections to neutral current Drell-Yan
  processes at hadron colliders}},
  \href{https://doi.org/10.1103/PhysRevD.65.033007}{\emph{Phys.Rev.} {\bfseries
  D65} (2002) 033007} [\href{https://arxiv.org/abs/hep-ph/0108274}{{\ttfamily
  hep-ph/0108274}}].

\bibitem{Baur:2004ig}
U.~Baur and D.~Wackeroth, \emph{{Electroweak radiative corrections to $p
  \bar{p} \to W^\pm \to \ell^\pm \nu$ beyond the pole approximation}},
  \href{https://doi.org/10.1103/PhysRevD.70.073015}{\emph{Phys. Rev. D}
  {\bfseries 70} (2004) 073015}
  [\href{https://arxiv.org/abs/hep-ph/0405191}{{\ttfamily hep-ph/0405191}}].

\bibitem{Arbuzov:2005dd}
A.~Arbuzov, D.~Bardin, S.~Bondarenko, P.~Christova, L.~Kalinovskaya, G.~Nanava
  and R.~Sadykov, \emph{{One-loop corrections to the Drell-Yan process in SANC.
  I. The Charged current case}},
  \href{https://doi.org/10.1140/epjc/s2006-02505-y}{\emph{Eur. Phys. J. C}
  {\bfseries 46} (2006) 407}
  [\href{https://arxiv.org/abs/hep-ph/0506110}{{\ttfamily hep-ph/0506110}}].

\bibitem{Zykunov:2005tc}
V.~A. Zykunov, \emph{{Weak radiative corrections to Drell-Yan process for large
  invariant mass of di-lepton pair}},
  \href{https://doi.org/10.1103/PhysRevD.75.073019}{\emph{Phys. Rev. D}
  {\bfseries 75} (2007) 073019}
  [\href{https://arxiv.org/abs/hep-ph/0509315}{{\ttfamily hep-ph/0509315}}].

\bibitem{CarloniCalame:2006zq}
C.~M. Carloni~Calame, G.~Montagna, O.~Nicrosini and A.~Vicini, \emph{{Precision
  electroweak calculation of the charged current Drell-Yan process}},
  \href{https://doi.org/10.1088/1126-6708/2006/12/016}{\emph{JHEP} {\bfseries
  12} (2006) 016} [\href{https://arxiv.org/abs/hep-ph/0609170}{{\ttfamily
  hep-ph/0609170}}].

\bibitem{Zykunov:2006yb}
V.~A. Zykunov, \emph{{Radiative corrections to the Drell-Yan process at large
  dilepton invariant masses}},
  \href{https://doi.org/10.1134/S1063778806090109}{\emph{Phys. Atom. Nucl.}
  {\bfseries 69} (2006) 1522}.

\bibitem{CarloniCalame:2007cd}
C.~Carloni~Calame, G.~Montagna, O.~Nicrosini and A.~Vicini, \emph{{Precision
  electroweak calculation of the production of a high transverse-momentum
  lepton pair at hadron colliders}},
  \href{https://doi.org/10.1088/1126-6708/2007/10/109}{\emph{JHEP} {\bfseries
  0710} (2007) 109} [\href{https://arxiv.org/abs/0710.1722}{{\ttfamily
  0710.1722}}].

\bibitem{Arbuzov:2007db}
A.~Arbuzov, D.~Bardin, S.~Bondarenko, P.~Christova, L.~Kalinovskaya et~al.,
  \emph{{One-loop corrections to the Drell--Yan process in SANC. (II). The
  Neutral current case}},
  \href{https://doi.org/10.1140/epjc/s10052-008-0531-8}{\emph{Eur.Phys.J.}
  {\bfseries C54} (2008) 451}
  [\href{https://arxiv.org/abs/0711.0625}{{\ttfamily 0711.0625}}].

\bibitem{Dittmaier:2009cr}
S.~Dittmaier and M.~Huber, \emph{{Radiative corrections to the neutral-current
  Drell-Yan process in the Standard Model and its minimal supersymmetric
  extension}}, \href{https://doi.org/10.1007/JHEP01(2010)060}{\emph{JHEP}
  {\bfseries 01} (2010) 060} [\href{https://arxiv.org/abs/0911.2329}{{\ttfamily
  0911.2329}}].

\bibitem{Kuhn:1999nn}
J.~H. Kuhn, A.~A. Penin and V.~A. Smirnov, \emph{{Summing up subleading Sudakov
  logarithms}}, \href{https://doi.org/10.1007/s100520000462}{\emph{Eur. Phys.
  J. C} {\bfseries 17} (2000) 97}
  [\href{https://arxiv.org/abs/hep-ph/9912503}{{\ttfamily hep-ph/9912503}}].

\bibitem{Ciafaloni:2001vu}
M.~Ciafaloni, P.~Ciafaloni and D.~Comelli, \emph{{Enhanced electroweak
  corrections to inclusive boson fusion processes at the TeV scale}},
  \href{https://doi.org/10.1016/S0550-3213(01)00375-3}{\emph{Nucl. Phys. B}
  {\bfseries 613} (2001) 382}
  [\href{https://arxiv.org/abs/hep-ph/0103316}{{\ttfamily hep-ph/0103316}}].

\bibitem{Denner:2000jv}
A.~Denner and S.~Pozzorini, \emph{{One loop leading logarithms in electroweak
  radiative corrections. 1. Results}},
  \href{https://doi.org/10.1007/s100520100551}{\emph{Eur. Phys. J. C}
  {\bfseries 18} (2001) 461}
  [\href{https://arxiv.org/abs/hep-ph/0010201}{{\ttfamily hep-ph/0010201}}].

\bibitem{Denner:2001gw}
A.~Denner and S.~Pozzorini, \emph{{One loop leading logarithms in electroweak
  radiative corrections. 2. Factorization of collinear singularities}},
  \href{https://doi.org/10.1007/s100520100721}{\emph{Eur. Phys. J. C}
  {\bfseries 21} (2001) 63}
  [\href{https://arxiv.org/abs/hep-ph/0104127}{{\ttfamily hep-ph/0104127}}].

\bibitem{Barze:2013fru}
L.~Barze, G.~Montagna, P.~Nason, O.~Nicrosini, F.~Piccinini and A.~Vicini,
  \emph{{Neutral current Drell-Yan with combined QCD and electroweak
  corrections in the POWHEG BOX}},
  \href{https://doi.org/10.1140/epjc/s10052-013-2474-y}{\emph{Eur. Phys. J.}
  {\bfseries C73} (2013) 2474}
  [\href{https://arxiv.org/abs/1302.4606}{{\ttfamily 1302.4606}}].

\bibitem{Frederix:2018nkq}
R.~Frederix, S.~Frixione, V.~Hirschi, D.~Pagani, H.~S. Shao and M.~Zaro,
  \emph{{The automation of next-to-leading order electroweak calculations}},
  \href{https://doi.org/10.1007/JHEP11(2021)085}{\emph{JHEP} {\bfseries 07}
  (2018) 185} [\href{https://arxiv.org/abs/1804.10017}{{\ttfamily
  1804.10017}}].

\bibitem{Delto:2019ewv}
M.~Delto, M.~Jaquier, K.~Melnikov and R.~Roentsch, \emph{{Mixed QCD$\otimes$QED
  corrections to on-shell $Z$ boson production at the LHC}},
  \href{https://arxiv.org/abs/1909.08428}{{\ttfamily 1909.08428}}.

\bibitem{Buccioni:2020cfi}
F.~Buccioni, F.~Caola, M.~Delto, M.~Jaquier, K.~Melnikov and R.~R\"ontsch,
  \emph{{Mixed QCD-electroweak corrections to on-shell Z production at the
  LHC}}, \href{https://doi.org/10.1016/j.physletb.2020.135969}{\emph{Phys.
  Lett. B} {\bfseries 811} (2020) 135969}
  [\href{https://arxiv.org/abs/2005.10221}{{\ttfamily 2005.10221}}].

\bibitem{Bonciani:2020tvf}
R.~Bonciani, F.~Buccioni, N.~Rana and A.~Vicini, \emph{{Next-to-Next-to-Leading
  Order Mixed QCD-Electroweak Corrections to on-Shell Z Production}},
  \href{https://doi.org/10.1103/PhysRevLett.125.232004}{\emph{Phys. Rev. Lett.}
  {\bfseries 125} (2020) 232004}
  [\href{https://arxiv.org/abs/2007.06518}{{\ttfamily 2007.06518}}].

\bibitem{Behring:2020cqi}
A.~Behring, F.~Buccioni, F.~Caola, M.~Delto, M.~Jaquier, K.~Melnikov and
  R.~R\"ontsch, \emph{{Mixed QCD-electroweak corrections to W-boson production
  in hadron collisions}},  \href{https://arxiv.org/abs/2009.10386}{{\ttfamily
  2009.10386}}.

\bibitem{Behring:2021adr}
A.~Behring, F.~Buccioni, F.~Caola, M.~Delto, M.~Jaquier, K.~Melnikov and
  R.~R\"ontsch, \emph{{Estimating the impact of mixed QCD-electroweak
  corrections on the $W$-mass determination at the LHC}},
  \href{https://doi.org/10.1103/PhysRevD.103.113002}{\emph{Phys. Rev. D}
  {\bfseries 103} (2021) 113002}
  [\href{https://arxiv.org/abs/2103.02671}{{\ttfamily 2103.02671}}].

\bibitem{Bonciani:2021iis}
R.~Bonciani, F.~Buccioni, N.~Rana and A.~Vicini, \emph{{On-shell Z boson
  production through ${\mathcal O}(\alpha \alpha_s)$}},
  \href{https://doi.org/10.1007/JHEP02(2022)095}{\emph{JHEP} {\bfseries 02}
  (2022) 095} [\href{https://arxiv.org/abs/2111.12694}{{\ttfamily
  2111.12694}}].

\bibitem{Dittmaier:2014qza}
S.~Dittmaier, A.~Huss and C.~Schwinn, \emph{{Mixed QCD-electroweak
  $\mathcal{O}(\alpha_s\alpha)$ corrections to Drell-Yan processes in the
  resonance region: pole approximation and non-factorizable corrections}},
  \href{https://doi.org/10.1016/j.nuclphysb.2014.05.027}{\emph{Nucl.Phys.}
  {\bfseries B885} (2014) 318}
  [\href{https://arxiv.org/abs/1403.3216}{{\ttfamily 1403.3216}}].

\bibitem{Dittmaier:2015rxo}
S.~Dittmaier, A.~Huss and C.~Schwinn, \emph{{Dominant mixed QCD-electroweak
  $\mathcal{O}(\alpha_s)$ corrections to Drell-Yan processes in the resonance
  region}}, \href{https://doi.org/10.1016/j.nuclphysb.2016.01.006}{\emph{Nucl.
  Phys.} {\bfseries B904} (2016) 216}
  [\href{https://arxiv.org/abs/1511.08016}{{\ttfamily 1511.08016}}].

\bibitem{Bonciani:2016ypc}
R.~Bonciani, S.~Di~Vita, P.~Mastrolia and U.~Schubert, \emph{{Two-Loop Master
  Integrals for the mixed EW-QCD virtual corrections to Drell-Yan scattering}},
  \href{https://doi.org/10.1007/JHEP09(2016)091}{\emph{JHEP} {\bfseries 09}
  (2016) 091} [\href{https://arxiv.org/abs/1604.08581}{{\ttfamily
  1604.08581}}].

\bibitem{vonManteuffel:2017myy}
A.~von Manteuffel and R.~M. Schabinger, \emph{{Numerical Multi-Loop
  Calculations via Finite Integrals and One-Mass EW-QCD Drell-Yan Master
  Integrals}}, \href{https://doi.org/10.1007/JHEP04(2017)129}{\emph{JHEP}
  {\bfseries 04} (2017) 129}
  [\href{https://arxiv.org/abs/1701.06583}{{\ttfamily 1701.06583}}].

\bibitem{Heller:2019gkq}
M.~Heller, A.~von Manteuffel and R.~M. Schabinger, \emph{{Multiple
  polylogarithms with algebraic arguments and the two-loop EW-QCD Drell-Yan
  master integrals}},
  \href{https://doi.org/10.1103/PhysRevD.102.016025}{\emph{Phys. Rev. D}
  {\bfseries 102} (2020) 016025}
  [\href{https://arxiv.org/abs/1907.00491}{{\ttfamily 1907.00491}}].

\bibitem{Heller:2020owb}
M.~Heller, A.~von Manteuffel, R.~M. Schabinger and H.~Spiesberger, \emph{{Mixed
  EW-QCD two-loop amplitudes for $q\bar{q} \to \ell^+\ell^-$ and $\gamma_5$
  scheme independence of multi-loop corrections}},
  \href{https://doi.org/10.1007/JHEP05(2021)213}{\emph{JHEP} {\bfseries 05}
  (2021) 213} [\href{https://arxiv.org/abs/2012.05918}{{\ttfamily
  2012.05918}}].

\bibitem{Hasan:2020vwn}
S.~M. Hasan and U.~Schubert, \emph{{Master Integrals for the mixed QCD-QED
  corrections to the Drell-Yan production of a massive lepton pair}},
  \href{https://doi.org/10.1007/JHEP11(2020)107}{\emph{JHEP} {\bfseries 11}
  (2020) 107} [\href{https://arxiv.org/abs/2004.14908}{{\ttfamily
  2004.14908}}].

\bibitem{Armadillo:2022bgm}
T.~Armadillo, R.~Bonciani, S.~Devoto, N.~Rana and A.~Vicini, \emph{{Two-loop
  mixed QCD-EW corrections to neutral current Drell-Yan}},
  \href{https://arxiv.org/abs/2201.01754}{{\ttfamily 2201.01754}}.

\bibitem{Caola:2017dug}
F.~Caola, K.~Melnikov and R.~R\"ontsch, \emph{{Nested soft-collinear
  subtractions in NNLO QCD computations}},
  \href{https://doi.org/10.1140/epjc/s10052-017-4774-0}{\emph{Eur. Phys. J. C}
  {\bfseries 77} (2017) 248}
  [\href{https://arxiv.org/abs/1702.01352}{{\ttfamily 1702.01352}}].

\bibitem{Bonciani:2021zzf}
R.~Bonciani, L.~Buonocore, M.~Grazzini, S.~Kallweit, N.~Rana, F.~Tramontano and
  A.~Vicini, \emph{{Mixed strong$-$electroweak corrections to the Drell$-$Yan
  process}},  \href{https://arxiv.org/abs/2106.11953}{{\ttfamily 2106.11953}}.

\bibitem{Buonocore:2021rxx}
L.~Buonocore, M.~Grazzini, S.~Kallweit, C.~Savoini and F.~Tramontano,
  \emph{{Mixed QCD-EW corrections to $\boldsymbol{pp\!\to\!\ell\nu_\ell\!+\!X}$
  at the LHC}}, \href{https://doi.org/10.1103/PhysRevD.103.114012}{\emph{Phys.
  Rev. D} {\bfseries 103} (2021) 114012}
  [\href{https://arxiv.org/abs/2102.12539}{{\ttfamily 2102.12539}}].

\bibitem{Heinrich:2020ybq}
G.~Heinrich, \emph{{Collider Physics at the Precision Frontier}},
  \href{https://doi.org/10.1016/j.physrep.2021.03.006}{\emph{Phys. Rept.}
  {\bfseries 922} (2021) 1} [\href{https://arxiv.org/abs/2009.00516}{{\ttfamily
  2009.00516}}].

\bibitem{TorresBobadilla:2020ekr}
W.~J. Torres~Bobadilla et~al., \emph{{May the four be with you: Novel
  IR-subtraction methods to tackle NNLO calculations}},
  \href{https://doi.org/10.1140/epjc/s10052-021-08996-y}{\emph{Eur. Phys. J. C}
  {\bfseries 81} (2021) 250}
  [\href{https://arxiv.org/abs/2012.02567}{{\ttfamily 2012.02567}}].

\bibitem{Caola:2018pxp}
F.~Caola, M.~Delto, H.~Frellesvig and K.~Melnikov, \emph{{The double-soft
  integral for an arbitrary angle between hard radiators}},
  \href{https://doi.org/10.1140/epjc/s10052-018-6180-7}{\emph{Eur. Phys. J. C}
  {\bfseries 78} (2018) 687}
  [\href{https://arxiv.org/abs/1807.05835}{{\ttfamily 1807.05835}}].

\bibitem{Caola:2019nzf}
F.~Caola, K.~Melnikov and R.~R\"ontsch, \emph{{Analytic results for
  color-singlet production at NNLO QCD with the nested soft-collinear
  subtraction scheme}},
  \href{https://doi.org/10.1140/epjc/s10052-019-6880-7}{\emph{Eur. Phys. J. C}
  {\bfseries 79} (2019) 386}
  [\href{https://arxiv.org/abs/1902.02081}{{\ttfamily 1902.02081}}].

\bibitem{Delto:2019asp}
M.~Delto and K.~Melnikov, \emph{{Integrated triple-collinear counter-terms for
  the nested soft-collinear subtraction scheme}},
  \href{https://doi.org/10.1007/JHEP05(2019)148}{\emph{JHEP} {\bfseries 05}
  (2019) 148} [\href{https://arxiv.org/abs/1901.05213}{{\ttfamily
  1901.05213}}].

\bibitem{Altarelli:1977zs}
G.~Altarelli and G.~Parisi, \emph{{Asymptotic Freedom in Parton Language}},
  \href{https://doi.org/10.1016/0550-3213(77)90384-4}{\emph{Nucl. Phys. B}
  {\bfseries 126} (1977) 298}.

\bibitem{Catani:1999ss}
S.~Catani and M.~Grazzini, \emph{{Infrared factorization of tree level QCD
  amplitudes at the next-to-next-to-leading order and beyond}},
  \href{https://doi.org/10.1016/S0550-3213(99)00778-6}{\emph{Nucl. Phys. B}
  {\bfseries 570} (2000) 287}
  [\href{https://arxiv.org/abs/hep-ph/9908523}{{\ttfamily hep-ph/9908523}}].

\bibitem{Bern:1999ry}
Z.~Bern, V.~Del~Duca, W.~B. Kilgore and C.~R. Schmidt, \emph{{The infrared
  behavior of one loop QCD amplitudes at next-to-next-to leading order}},
  \href{https://doi.org/10.1103/PhysRevD.60.116001}{\emph{Phys. Rev. D}
  {\bfseries 60} (1999) 116001}
  [\href{https://arxiv.org/abs/hep-ph/9903516}{{\ttfamily hep-ph/9903516}}].

\bibitem{Catani:2000pi}
S.~Catani and M.~Grazzini, \emph{{The soft gluon current at one loop order}},
  \href{https://doi.org/10.1016/S0550-3213(00)00572-1}{\emph{Nucl. Phys. B}
  {\bfseries 591} (2000) 435}
  [\href{https://arxiv.org/abs/hep-ph/0007142}{{\ttfamily hep-ph/0007142}}].

\bibitem{Kosower:1999rx}
D.~A. Kosower and P.~Uwer, \emph{{One loop splitting amplitudes in gauge
  theory}}, \href{https://doi.org/10.1016/S0550-3213(99)00583-0}{\emph{Nucl.
  Phys. B} {\bfseries 563} (1999) 477}
  [\href{https://arxiv.org/abs/hep-ph/9903515}{{\ttfamily hep-ph/9903515}}].

\bibitem{Ermolaev:1981cm}
B.~I. Ermolaev and V.~S. Fadin, \emph{{Log - Log Asymptotic Form of Exclusive
  Cross-Sections in Quantum Chromodynamics}}, {\emph{JETP Lett.} {\bfseries 33}
  (1981) 269}.

\bibitem{Bassetto:1983mvz}
A.~Bassetto, M.~Ciafaloni and G.~Marchesini, \emph{{Jet Structure and Infrared
  Sensitive Quantities in Perturbative QCD}},
  \href{https://doi.org/10.1016/0370-1573(83)90083-2}{\emph{Phys. Rept.}
  {\bfseries 100} (1983) 201}.

\bibitem{Dokshitzer:1987nm}
Y.~L. Dokshitzer, V.~A. Khoze, S.~I. Troian and A.~H. Mueller, \emph{{QCD
  Coherence in High-Energy Reactions}},
  \href{https://doi.org/10.1103/RevModPhys.60.373}{\emph{Rev. Mod. Phys.}
  {\bfseries 60} (1988) 373}.

\bibitem{Frixione:1995ms}
S.~Frixione, Z.~Kunszt and A.~Signer, \emph{{Three jet cross-sections to
  next-to-leading order}},
  \href{https://doi.org/10.1016/0550-3213(96)00110-1}{\emph{Nucl. Phys. B}
  {\bfseries 467} (1996) 399}
  [\href{https://arxiv.org/abs/hep-ph/9512328}{{\ttfamily hep-ph/9512328}}].

\bibitem{Czakon:2010td}
M.~Czakon, \emph{{A novel subtraction scheme for double-real radiation at
  NNLO}}, \href{https://doi.org/10.1016/j.physletb.2010.08.036}{\emph{Phys.
  Lett. B} {\bfseries 693} (2010) 259}
  [\href{https://arxiv.org/abs/1005.0274}{{\ttfamily 1005.0274}}].

\bibitem{Czakon:2011ve}
M.~Czakon, \emph{{Double-real radiation in hadronic top quark pair production
  as a proof of a certain concept}},
  \href{https://doi.org/10.1016/j.nuclphysb.2011.03.020}{\emph{Nucl. Phys. B}
  {\bfseries 849} (2011) 250}
  [\href{https://arxiv.org/abs/1101.0642}{{\ttfamily 1101.0642}}].

\bibitem{Caola:2019pfz}
F.~Caola, K.~Melnikov and R.~R\"ontsch, \emph{{Analytic results for decays of
  color singlets to $gg$ and $q \bar q$ final states at NNLO QCD with the
  nested soft-collinear subtraction scheme}},
  \href{https://doi.org/10.1140/epjc/s10052-019-7505-x}{\emph{Eur. Phys. J. C}
  {\bfseries 79} (2019) 1013}
  [\href{https://arxiv.org/abs/1907.05398}{{\ttfamily 1907.05398}}].

\bibitem{Catani:1998bh}
S.~Catani, \emph{{The Singular behavior of QCD amplitudes at two loop order}},
  \href{https://doi.org/10.1016/S0370-2693(98)00332-3}{\emph{Phys. Lett. B}
  {\bfseries 427} (1998) 161}
  [\href{https://arxiv.org/abs/hep-ph/9802439}{{\ttfamily hep-ph/9802439}}].

\bibitem{Bern:1997sc}
Z.~Bern, L.~J. Dixon and D.~A. Kosower, \emph{{One loop amplitudes for e+ e- to
  four partons}},
  \href{https://doi.org/10.1016/S0550-3213(97)00703-7}{\emph{Nucl. Phys. B}
  {\bfseries 513} (1998) 3}
  [\href{https://arxiv.org/abs/hep-ph/9708239}{{\ttfamily hep-ph/9708239}}].

\bibitem{Campbell:1999ah}
J.~M. Campbell and R.~Ellis, \emph{{An Update on vector boson pair production
  at hadron colliders}},
  \href{https://doi.org/10.1103/PhysRevD.60.113006}{\emph{Phys.Rev.} {\bfseries
  D60} (1999) 113006} [\href{https://arxiv.org/abs/hep-ph/9905386}{{\ttfamily
  hep-ph/9905386}}].

\bibitem{Cascioli:2011va}
F.~Cascioli, P.~Maierhofer and S.~Pozzorini, \emph{{Scattering Amplitudes with
  Open Loops}},  \href{https://arxiv.org/abs/1111.5206}{{\ttfamily 1111.5206}}.

\bibitem{Buccioni:2017yxi}
F.~Buccioni, S.~Pozzorini and M.~Zoller, \emph{{On-the-fly reduction of open
  loops}}, \href{https://doi.org/10.1140/epjc/s10052-018-5562-1}{\emph{Eur.
  Phys. J. C} {\bfseries 78} (2018) 70}
  [\href{https://arxiv.org/abs/1710.11452}{{\ttfamily 1710.11452}}].

\bibitem{Buccioni:2019sur}
F.~Buccioni, J.-N. Lang, J.~M. Lindert, P.~Maierh\"ofer, S.~Pozzorini, H.~Zhang
  and M.~F. Zoller, \emph{{OpenLoops 2}},
  \href{https://doi.org/10.1140/epjc/s10052-019-7306-2}{\emph{Eur. Phys. J. C}
  {\bfseries 79} (2019) 866}
  [\href{https://arxiv.org/abs/1907.13071}{{\ttfamily 1907.13071}}].

\bibitem{Dittmaier:2020vra}
S.~Dittmaier, T.~Schmidt and J.~Schwarz, \emph{{Mixed NNLO
  QCD\texttimes{}electroweak corrections of $\mathcal{O}(N_f \alpha_s \alpha)$
  to single-W/Z production at the LHC}},
  \href{https://doi.org/10.1007/JHEP12(2020)201}{\emph{JHEP} {\bfseries 12}
  (2020) 201} [\href{https://arxiv.org/abs/2009.02229}{{\ttfamily
  2009.02229}}].

\bibitem{Djouadi:1993ss}
A.~Djouadi and P.~Gambino, \emph{{Electroweak gauge bosons selfenergies:
  Complete QCD corrections}},
  \href{https://doi.org/10.1103/PhysRevD.49.3499}{\emph{Phys. Rev. D}
  {\bfseries 49} (1994) 3499}
  [\href{https://arxiv.org/abs/hep-ph/9309298}{{\ttfamily hep-ph/9309298}}].

\bibitem{Abreu:2019odu}
S.~Abreu, J.~Dormans, F.~Febres~Cordero, H.~Ita, B.~Page and V.~Sotnikov,
  \emph{{Analytic Form of the Planar Two-Loop Five-Parton Scattering Amplitudes
  in QCD}}, \href{https://doi.org/10.1007/JHEP05(2019)084}{\emph{JHEP}
  {\bfseries 05} (2019) 084}
  [\href{https://arxiv.org/abs/1904.00945}{{\ttfamily 1904.00945}}].

\bibitem{Chawdhry:2019bji}
H.~A. Chawdhry, M.~L. Czakon, A.~Mitov and R.~Poncelet, \emph{{NNLO QCD
  corrections to three-photon production at the LHC}},
  \href{https://doi.org/10.1007/JHEP02(2020)057}{\emph{JHEP} {\bfseries 02}
  (2020) 057} [\href{https://arxiv.org/abs/1911.00479}{{\ttfamily
  1911.00479}}].

\bibitem{Agarwal:2021grm}
B.~Agarwal, F.~Buccioni, A.~von Manteuffel and L.~Tancredi, \emph{{Two-loop
  leading colour QCD corrections to $q \bar{q} \to \gamma \gamma g$ and $q g
  \to \gamma \gamma q$}},
  \href{https://doi.org/10.1007/JHEP04(2021)201}{\emph{JHEP} {\bfseries 04}
  (2021) 201} [\href{https://arxiv.org/abs/2102.01820}{{\ttfamily
  2102.01820}}].

\bibitem{Naterop:2019xaf}
L.~Naterop, A.~Signer and Y.~Ulrich, \emph{{handyG \textemdash{}Rapid numerical
  evaluation of generalised polylogarithms in Fortran}},
  \href{https://doi.org/10.1016/j.cpc.2020.107165}{\emph{Comput. Phys. Commun.}
  {\bfseries 253} (2020) 107165}
  [\href{https://arxiv.org/abs/1909.01656}{{\ttfamily 1909.01656}}].

\bibitem{Duhr:2019tlz}
C.~Duhr and F.~Dulat, \emph{{PolyLogTools \textemdash{} polylogs for the
  masses}}, \href{https://doi.org/10.1007/JHEP08(2019)135}{\emph{JHEP}
  {\bfseries 08} (2019) 135}
  [\href{https://arxiv.org/abs/1904.07279}{{\ttfamily 1904.07279}}].

\bibitem{Bauer:2000cp}
C.~W. Bauer, A.~Frink and R.~Kreckel, \emph{{Introduction to the GiNaC
  framework for symbolic computation within the C++ programming language}},
  \href{https://arxiv.org/abs/cs/0004015}{{\ttfamily cs/0004015}}.

\bibitem{Vollinga:2004sn}
J.~Vollinga and S.~Weinzierl, \emph{{Numerical evaluation of multiple
  polylogarithms}},
  \href{https://doi.org/10.1016/j.cpc.2004.12.009}{\emph{Comput.Phys.Commun.}
  {\bfseries 167} (2005) 177}
  [\href{https://arxiv.org/abs/hep-ph/0410259}{{\ttfamily hep-ph/0410259}}].

\bibitem{Denner:2005fg}
A.~Denner, S.~Dittmaier, M.~Roth and L.~Wieders, \emph{{Electroweak corrections
  to charged-current e+ e- to 4 fermion processes: Technical details and
  further results}}, {\emph{Nucl.Phys.} {\bfseries B724} (2005) 247}
  [\href{https://arxiv.org/abs/hep-ph/0505042}{{\ttfamily hep-ph/0505042}}].

\bibitem{NNPDF:2017mvq}
{\scshape NNPDF} collaboration, \emph{{Parton distributions from high-precision
  collider data}},
  \href{https://doi.org/10.1140/epjc/s10052-017-5199-5}{\emph{Eur. Phys. J. C}
  {\bfseries 77} (2017) 663}
  [\href{https://arxiv.org/abs/1706.00428}{{\ttfamily 1706.00428}}].

\bibitem{Buckley:2014ana}
A.~Buckley, J.~Ferrando, S.~Lloyd, K.~Nordstr\"om, B.~Page, M.~R\"ufenacht,
  M.~Sch\"onherr et~al., \emph{{LHAPDF6: parton density access in the LHC
  precision era}},
  \href{https://doi.org/10.1140/epjc/s10052-015-3318-8}{\emph{Eur. Phys. J. C}
  {\bfseries 75} (2015) 132} [\href{https://arxiv.org/abs/1412.7420}{{\ttfamily
  1412.7420}}].

\bibitem{Salam:2008qg}
G.~P. Salam and J.~Rojo, \emph{{A Higher Order Perturbative Parton Evolution
  Toolkit (HOPPET)}},
  \href{https://doi.org/10.1016/j.cpc.2008.08.010}{\emph{Comput. Phys. Commun.}
  {\bfseries 180} (2009) 120}
  [\href{https://arxiv.org/abs/0804.3755}{{\ttfamily 0804.3755}}].

\bibitem{ATLAS:2016gic}
{\scshape ATLAS} collaboration, \emph{{Measurement of the double-differential
  high-mass Drell-Yan cross section in pp collisions at $ \sqrt{s}=8 $ TeV with
  the ATLAS detector}},
  \href{https://doi.org/10.1007/JHEP08(2016)009}{\emph{JHEP} {\bfseries 08}
  (2016) 009} [\href{https://arxiv.org/abs/1606.01736}{{\ttfamily
  1606.01736}}].

\bibitem{Salam:2021tbm}
G.~P. Salam and E.~Slade, \emph{{Cuts for two-body decays at colliders}},
  \href{https://doi.org/10.1007/JHEP11(2021)220}{\emph{JHEP} {\bfseries 11}
  (2021) 220} [\href{https://arxiv.org/abs/2106.08329}{{\ttfamily
  2106.08329}}].

\bibitem{Sherpa:2019gpd}
{\scshape Sherpa} collaboration, \emph{{Event Generation with Sherpa 2.2}},
  \href{https://doi.org/10.21468/SciPostPhys.7.3.034}{\emph{SciPost Phys.}
  {\bfseries 7} (2019) 034} [\href{https://arxiv.org/abs/1905.09127}{{\ttfamily
  1905.09127}}].

\bibitem{Actis:2016mpe}
S.~Actis, A.~Denner, L.~Hofer, J.-N. Lang, A.~Scharf and S.~Uccirati,
  \emph{{RECOLA: REcursive Computation of One-Loop Amplitudes}},
  \href{https://doi.org/10.1016/j.cpc.2017.01.004}{\emph{Comput. Phys. Commun.}
  {\bfseries 214} (2017) 140}
  [\href{https://arxiv.org/abs/1605.01090}{{\ttfamily 1605.01090}}].

\bibitem{Denner:2016kdg}
A.~Denner, S.~Dittmaier and L.~Hofer, \emph{{Collier: a fortran-based Complex
  One-Loop LIbrary in Extended Regularizations}},
  \href{https://doi.org/10.1016/j.cpc.2016.10.013}{\emph{Comput. Phys. Commun.}
  {\bfseries 212} (2017) 220}
  [\href{https://arxiv.org/abs/1604.06792}{{\ttfamily 1604.06792}}].

\bibitem{Denner:2021hqi}
A.~Denner and G.~Pelliccioli, \emph{{Combined NLO EW and QCD corrections to
  off-shell $\text {t} \overline{\text {t}}\text {W} $ production at the LHC}},
  \href{https://doi.org/10.1140/epjc/s10052-021-09143-3}{\emph{Eur. Phys. J. C}
  {\bfseries 81} (2021) 354}
  [\href{https://arxiv.org/abs/2102.03246}{{\ttfamily 2102.03246}}].

\bibitem{Denner:2021csi}
A.~Denner and G.~Pelliccioli, \emph{{NLO EW and QCD corrections to polarized ZZ
  production in the four-charged-lepton channel at the LHC}},
  \href{https://doi.org/10.1007/JHEP10(2021)097}{\emph{JHEP} {\bfseries 10}
  (2021) 097} [\href{https://arxiv.org/abs/2107.06579}{{\ttfamily
  2107.06579}}].

\bibitem{Denner:2019vbn}
A.~Denner and S.~Dittmaier, \emph{{Electroweak Radiative Corrections for
  Collider Physics}},
  \href{https://doi.org/10.1016/j.physrep.2020.04.001}{\emph{Phys. Rept.}
  {\bfseries 864} (2020) 1} [\href{https://arxiv.org/abs/1912.06823}{{\ttfamily
  1912.06823}}].

\bibitem{Collins:1977iv}
J.~C. Collins and D.~E. Soper, \emph{{Angular Distribution of Dileptons in
  High-Energy Hadron Collisions}},
  \href{https://doi.org/10.1103/PhysRevD.16.2219}{\emph{Phys. Rev. D}
  {\bfseries 16} (1977) 2219}.

\end{thebibliography}\endgroup

\end{document}